\documentclass[aps,prb,twocolumn,amsfonts,showpacs,longbibliography,superscriptaddress]{revtex4-2}
\usepackage{verbatim,epsfig,amsmath,amssymb,bm,epsf,graphicx,psfrag,bbold,amsthm,amsfonts}
\usepackage[export]{adjustbox}
\usepackage[bottom]{footmisc}
\usepackage{hyperref}
\usepackage{slashed}
\usepackage{lipsum}
\usepackage{standalone}
\usepackage[capitalize]{cleveref} 
\hypersetup{
    colorlinks=true,
    linkcolor=blue,
    citecolor=blue,
    filecolor=blue,
    urlcolor=blue,
}
\usepackage{enumitem}
\usepackage{framed}
\usepackage{mathrsfs}
\usepackage{esint}
\setlist{nosep}
\usepackage{color}
\usepackage[utf8]{inputenc}
\usepackage{float}
\usepackage{natbib}
\usepackage{tikz-cd}
\usepackage{leftidx}
\usepackage[normalem]{ulem}
\usepackage[caption=false]{subfig}
\usepackage{notes2bib}

\newcommand{\ket}[1]{\mbox{$ | #1 \rangle $}}

\newcommand{\tr}{\operatorname{tr}}

\newcommand{\Ch}{\mathsf{C}}
\newcommand{\CS}{\mathsf{CS}}
\newcommand{\Ham}{\mathrm{H}}
\newcommand{\HamdBerry}[1]{\mathrm{H}^{[#1]}_{\text{Berry}}}
\newcommand{\h}{\mathrm{h}}
\newcommand{\q}{\mathrm{q}}

\newcommand{\half}{\frac{1}{2}}
\newcommand{\rmd}{\mathrm{d}}

\newcommand{\T}{\mathrm{T}}
\newcommand{\uone}{\mathrm{U}(1)}
\newcommand{\un}[1]{\mathrm{U}(#1)}

\newcommand{\cL}{\mathcal{L}}
\newcommand{\cM}{\mathcal{M}}
\newcommand{\cN}{\mathcal{N}}
\newcommand\bZ {{\mathbb Z}}

\newcommand{\sI}{\mathscr{I}}

\newcommand{\innerproduct}[2]{\langle #1| #2\rangle}
\newcommand{\outerproduct}[2]{|#1\rangle\langle #2|}

\newcommand{\snpar}[1]{S^{#1}_{\text{par}}}
\newcommand{\sonebz}{S^{1}_{\text{BZ}}}
\newcommand{\tnbz}[1]{T^{#1}_{\text{BZ}}}

\theoremstyle{plain}

\usepackage{cancel}
\theoremstyle{definition}

\theoremstyle{remark}

\makeatletter
\def\l@subsubsection#1#2{}
\makeatother
\begin{document}
\title{\large Textured phase diagrams of featureless insulators}
 \author{Sashank Singam}
 \email{singam.sashank\_chandra\_reddy@students.iiserpune.ac.in}
 \affiliation{Indian Institute of Science Education and Research (IISER), Pune 411008, India}
\affiliation{Harish-Chandra Research Institute (HRI),  Prayagraj (Allahabad) 211019, India}
  	\author{Nick G. Jones}
   \email{nick.jones@maths.ox.ac.uk}
 	\affiliation{St John’s College and Mathematical Institute, University of Oxford, UK}
 		\author{Abhishodh Prakash}
    \email{abhishodhprakash@hri.res.in}
    \altaffiliation{(he/him/his)}
        \affiliation{Harish-Chandra Research Institute (HRI),  Prayagraj (Allahabad) 211019, India}
        \affiliation{Homi Bhabha National Institute (HBNI),  Mumbai 400094, India}
	\begin{abstract}
We study phase diagrams of charge-conserving `class A' non-interacting fermions, focusing on the trivial phase in various dimensions. Such phases are usually termed `featureless' to distinguish them from those others with either symmetry-broken or topological order. We show that the presence of non-trivial topological families of states, including charge pumps and their generalizations, results in phase diagrams being endowed with non-trivial topological textures that can be visualized through Berry phases and their higher-dimensional generalizations. We show that for non-interacting fermion systems with translation invariance, these `higher' Berry phases can be computed using integrals of non-abelian Chern-Simons forms of the Berry-Bloch connection over momentum and parameter spaces. Singularities in these textures correspond to gap-closing loci of `diabolical points', which represent the obstruction to contracting topologically non-trivial families of states, and bulk-boundary correspondence results in a locus of robust boundary modes that terminate at the bulk diabolical points. In the presence of finite chemical potential, we argue that the edge modes are generically robust without any need for fine-tuning for two and higher dimensions, whereas in one dimension they are `estranged' in the phase diagram, i.e. appearing at different parameter values for different edges. We demonstrate our results by constructing several microscopic models of non-interacting fermions. We argue stability to interactions and explore proximate phase diagrams by mapping to continuum field theories. 
\end{abstract}

	\maketitle
\makeatletter
\begingroup
\let\@thefnmark\@empty
\setlength{\footnotemargin}{0pt}
\@footnotetext{The published version of this article is \href{https://doi.org/10.1103/5wf2-j1lm}{Phys.\ Rev.\ B \textbf{114}, 305107 (2026)}.}
\endgroup
\makeatother
    \tableofcontents

\section{Introduction and overview of results}
Over the past decades, topological states of matter~\cite{Haldane_Nobel_RevModPhys.89.040502,Thouless_Nobel_RevModPhys.89.040501} have been subjects of great interest that have been investigated by a vibrant collaboration of condensed matter theorists, high energy theorists, and mathematicians. Much of the interest and focus has been on \emph{non-trivial} phases of matter, which include topological insulators and superconductors~\cite{QiZhang_TITScReview_RevModPhys.83.1057,HasanKane_TI_RevModPhys.82.3045}, spin liquids~\cite{Savary_2017} and the like. Central to the program of understanding non-trivial topological phases is to contrast them to the notion of the trivial phase. In the trivial phase, any ground state can be smoothly connected to either a trivial atomic insulator (in the case of fermions) or a product state (in the case of bosons). Any system with a ground state not satisfying this property is, by definition, non-trivial. Trivial phases are often also called `featureless' due to the absence of long-range order of the kind resulting when symmetries are spontaneously broken and the absence of topological order. Generically, they do not have ground state degeneracy~\cite{WenNiu_GSD_PhysRevB.41.9377},  topological response to gauge fields~\cite{QiZhang_topologicalresponse_PhysRevB.78.195424}, non-trivial local and nonlocal order parameters~\cite{StringOrderPar_PerezGarcia_PhysRevLett.100.167202} or boundary modes~\cite{ChiuTeoSchnyderRyu_SPTReview_RevModPhys.88.035005}. In this work, we demonstrate the presence of rich topological structures that are present even within the featureless trivial phase, which take the form of topological textures in the phase diagram. These occur due to the generic presence of non-trivial \emph{families} of states~\cite{kitaev2019,Wenetal_topologicalfamilies,HsinKapustinThorngren_PhysRevB.102.245113,Shiozaki_SPTPUmp_PhysRevB.106.125108,JonesThorngrenAP_SPTPump,APNickPaul_ChiralClock} and are accompanied by loci of spectral gap closure in the bulk and at the boundary.  

Concretely, we study various phase diagrams of fermionic systems in different dimensions that conserve electric charge and thus belong to class A~\cite{ChiuTeoSchnyderRyu_SPTReview_RevModPhys.88.035005} of the tenfold way classification \cite{AltlandZirnbauer_PhysRevB.55.1142}. We show that the trivial band insulating phase in each spatial dimension contains topological textures in parameter space, reminiscent of superfluid vortices~\cite{Babev_svistunov2015superfluid} and their generalizations~\cite{Mermin_RevModPhys.51.591}. These can be detected by quantized topological invariants and visualized in phase diagrams using geometric invariants.  Non-trivial textures are accompanied by singular hypersurfaces reminiscent of vortex cores~\cite{HsinKapustinThorngren_PhysRevB.102.245113,manjunath2026searchdiabolicalcriticalpoints}, where the system exits the insulating phase through closure of the spectral gap, as well as robust edge modes through bulk-boundary correspondence~\cite{HsinKapustinThorngren_PhysRevB.102.245113,Seiberg_AnomaliesCouplingOne10.21468/SciPostPhys.8.1.001}. We show that in the presence of chemical potentials, edge modes take on surprising forms: in 1d, they become `estranged' where the edge modes on each end appear at different parameter values. For higher dimensions, edge modes become robust, exhibiting the same degree of stability as those protected by non-trivial topological insulators~\cite{ChiuTeoSchnyderRyu_SPTReview_RevModPhys.88.035005,KANE_topbandtheory,QiZhang_TITScReview_RevModPhys.83.1057}.  We argue that topological structures in each spatial dimension can have multiple components, each corresponding to distinct series that are classified by the mathematical framework of generalized cohomology~\cite{kitaev1,kitaev2,kitaev3}. For symmetry class A and low spatial dimensions ($d\le3$), to which we restrict our focus, there are three such relevant series that we refer to as the Rice-Mele, Berry and Qi-Wu-Zhang series. Using a `suspension recipe' \cite{kitaev1,kitaev2,kitaev3,Wenetal_topologicalfamilies}, we construct representative models of non-interacting fermion Hamiltonians for each of these series in each dimension and analyze them using band theory. We show that in every case topological invariants take the form of Chern numbers, and that geometric `higher Berry' invariants can be expressed as integrals of Chern-Simons forms. Both are defined for the Berry-Bloch connection of filled bands and integrated over appropriate manifolds composed of the momentum space (Brillouin zone) and a space of external parameters spanning the phase diagram. By mapping to continuum field theories we give a complementary analysis, showing that these structures are stable to perturbations including interactions, and we explore proximate phase diagrams.

The paper is organized as follows. The three distinct series of textures are studied in \cref{sec:RiceMele,sec:2d_RiceMele,sec:ddim_RiceMele} (Rice-Mele series) \cref{sec:Berry} (Berry series) and \cref{sec:QWZ} (Qi-Wu-Zhang series). For each, we first review the specific namesake model, and then apply the suspension construction, reviewed in \cref{sec:suspension}, to find its higher dimensional `ascendants'. To keep the discussion self-contained, we use the Rice-Mele series to introduce various concepts and constructions that are echoed in the analysis of other series. These discussions are relatively more detailed, and, consequently, the Rice-Mele series is split into three sections (\ref{sec:RiceMele},\ref{sec:2d_RiceMele},\ref{sec:ddim_RiceMele}). For each series, the first non-trivial ascendant is discussed in substantial detail, including its phase diagrams, the nature of bulk and boundary gap closing points, edge modes, field theory, and proximate phase diagrams. Ascendants in arbitrary dimensions are explicitly constructed, and general expectations  stated but not discussed in full detail to restrict the discussions to a reasonable length. 

Let us comment on connection to relevant past work. The study of topological families of quantum states has been the subject of several recent works~\cite{kitaev2019,Wenetal_topologicalfamilies,HsinKapustinThorngren_PhysRevB.102.245113,Shiozaki_SPTPUmp_PhysRevB.106.125108,JonesThorngrenAP_SPTPump,qi2025chartingspacegroundstates,SommerWenVishwanath_HigherMPS1_PhysRevLett.134.146601,SommerWenVishwanath_HigherMPS2_PhysRevB.111.155110,RyuHigherPhysRevB.109.115152,OhyamaRyu_HigherMPS1_PhysRevB.111.035121,OhyamaRyu_HigherMPS1_PhysRevB.111.035121,OhyamaRyu_HigherMPS2_PhysRevB.111.045112,AP_UC_Pump,manjunath2026searchdiabolicalcriticalpoints,Shiozaki_Ohyama_MPS_HigherBerry}  where the focus was primarily on interacting systems and field theories. Our work focuses on non-interacting fermions, where the analysis is simplified considerably. In \cref{sec:Berry}, we discuss the connection between interacting invariants, particularly the Dixmier-Douady invariant, and our non-interacting Chern invariants. Another important class of related works is on the connection between free-fermion systems, Clifford algebras, and K theory that has been uncovered in a variety of contexts~\cite{Horava_PhysRevLett.95.016405,Kitaev_PeriodicTable,teokane}. Refs.~\cite{teokane,Kitaev_PeriodicTable} classified diabolical loci (referred to as `defects') in free-fermion systems using higher Chern numbers. Parts of the field theory analysis presented in this work is also found in Ref~\cite{HsinKapustinThorngren_PhysRevB.102.245113}. Our analysis, motivated by microscopic models, is complementary and agrees where there is overlap. The identification and characterization of topological textures in the phase diagram using higher Berry phases, edge mode structure in the presence of chemical potentials, models constructed using suspension isomorphism, and the topological characterization of their phase diagrams are, to the best of our knowledge, novel to this work.

\section{The Rice-Mele model and Thouless pump} 
\label{sec:RiceMele}
\begin{figure}[!h]
\centering
    \makebox[242.39297pt][c]{\includegraphics[width=\textwidth]{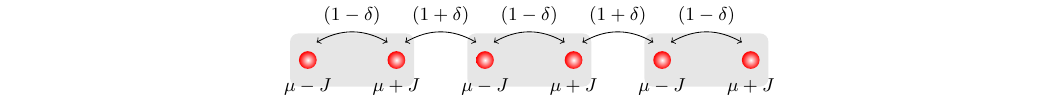}}
    \caption{Schematic representation of the Rice-Mele model. $(1\pm\delta)$ represents staggered hopping amplitudes between even and odd bonds. $\mu \pm J$ represents the staggered chemical potential on even and odd sites. The shaded boxes denote a choice of fiducial unit cell. \label{fig:1d_lattice}}
\end{figure}
In this section, we revisit the phase diagram of the Rice-Mele model~\cite{RiceMele_PhysRevLett.49.1455} and demonstrate the topological structures present in its phase diagram.

\subsection{Model and phase diagram}
\begin{figure}[!h]
    \centering
    \makebox[0.49\linewidth][c]{\includegraphics[width=\textwidth]{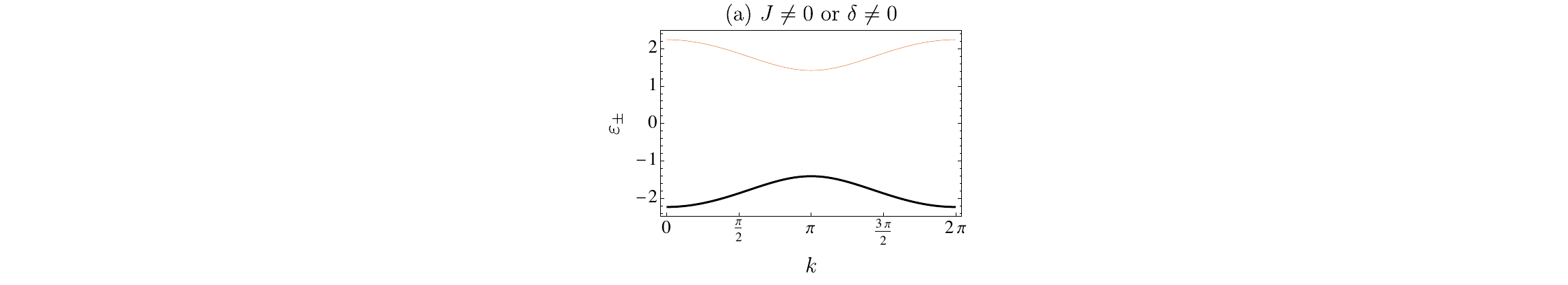}}
    \makebox[0.49\linewidth][c]{\includegraphics[width=\textwidth]{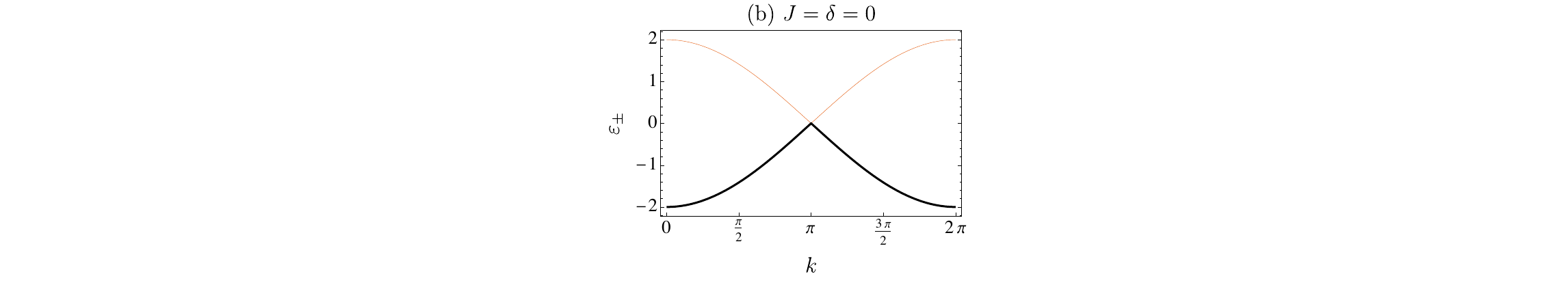}}
    \caption{Single-particle energy dispersion of the Hamiltonian \eqref{eq:H_1d_RiceMele} (a) away from and  (b) at the origin of the $J$-$\delta$ plane. }
    \label{fig:dispersion1d}
\end{figure}
The Rice-Mele model~\cite{RiceMele_PhysRevLett.49.1455,asboth2016short} is a one-dimensional system of non-interacting spinless fermions governed by the following Hamiltonian: 

\begin{align}
    \Ham_{\text{RM}}=\sum_{x\in\mathbb{Z}}t_x \left(c^\dagger(x) c({x+1})+h.c\right) 
    + \sum_{x\in\mathbb{Z}} \mu_x  n(x) \ ,
  \label{eq:H_1d_RiceMele}
\end{align}
where $t_x$ and $\mu_x$ are staggered hopping amplitudes and chemical potentials, defined by
\begin{align}
    t_x = \frac{(1+(-1)^x\delta)}{2}, ~\mu_x = \mu + (-1)^x J \ .  \label{eq:txmux_RiceMele}
\end{align}
We restrict the parameters to the range $J \in (-\infty,\infty),~\delta\in[-1,1]$. \cref{eq:H_1d_RiceMele} has a two-site translation symmetry, $\T_x$, and a $\uone$ symmetry generated by the total charge. These symmetries act as
\begin{align}
    \T_x: c(x) \mapsto c({x+2}),~\uone: c(x) \mapsto e^{i \vartheta} c(x) \label{eq:symmetries}\ .
\end{align}
 All other symmetries can be considered as accidental and can be safely eliminated without changing the results to follow~\footnote{Translations are also needed to allow the use of band theory and are not strictly essential.}. This places this free-fermion model in class A of the Altland-Zirnbauer classification~\cite{AltlandZirnbauer_PhysRevB.55.1142,Ryu2010}. In 1d, this symmetry class has no non-trivial topological phases~\cite{Kitaev_PeriodicTable} and so band insulators belong to the trivial phase.  We consider a two-site unit cell as shown in \cref{fig:1d_lattice} and Fourier transform the fermion operators to get the first-quantized two-band Bloch Hamiltonian
\begin{align}
    \h(k) = \mu \mathbb{1} +   \Vec{g}(k)\cdot\Vec{\sigma}\ , \label{eq:h_Bloch}
\end{align}
where $\Vec{\sigma} = \{\sigma^1, \sigma^2,\sigma^3 \}$ are standard Pauli matrices and 
\begin{align}
g_1(k) &= \frac{(1-\delta)}{2} + \frac{(1+\delta)}{2} \cos k,~\nonumber\\
g_2(k) &= \frac{(1+\delta)}{2} \sin k,~ g_3(k) = -J\ . \label{eq:g_RM}
\end{align}
The two energy bands have dispersion 
\begin{align}
    \varepsilon_\pm(k)=\mu \pm |\vec{g}(k)|,\qquad~|\vec{g}(k)| = \sqrt{\sum_{a=1}^3g_a^2(k)} \ .\label{eq:Epm}
\end{align}

\begin{figure}[!h]
    \centering
       \subfloat[$\mu=0$]{ \makebox[.24\textwidth][c]{\includegraphics[width=\textwidth]{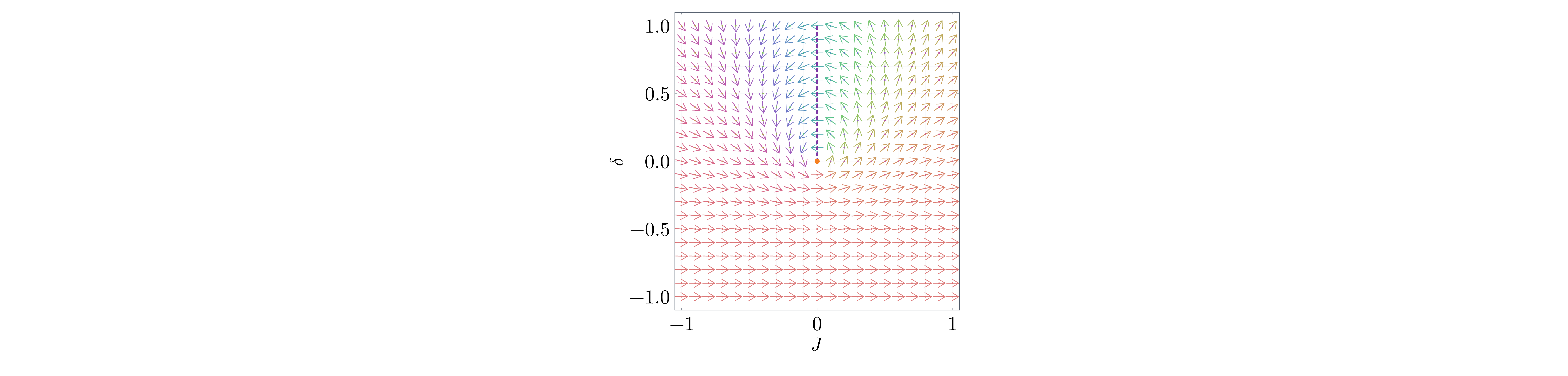}}}
        \subfloat[$\mu=0.4$]{\makebox[.24\textwidth][c]{\includegraphics[width=\textwidth]{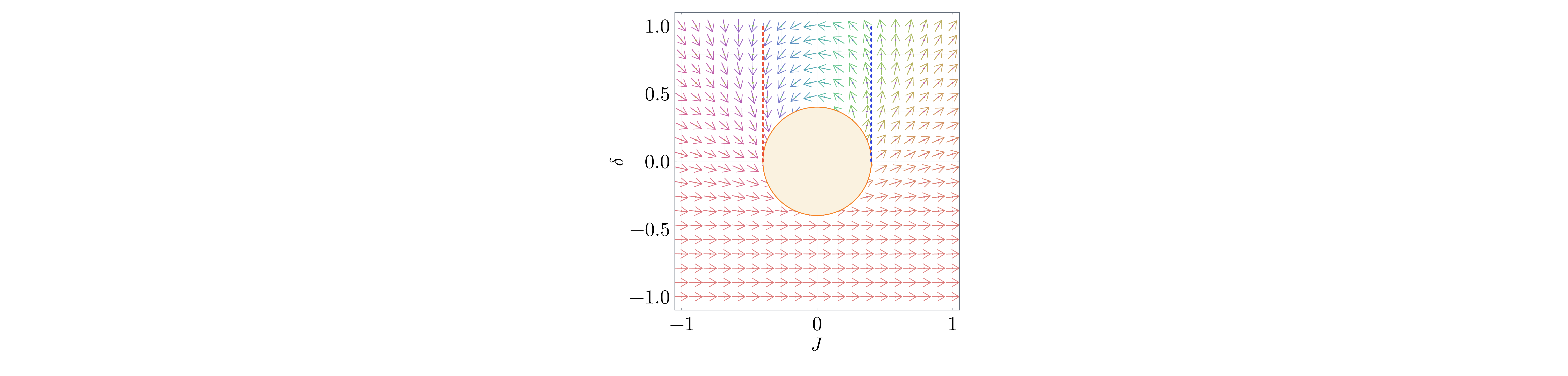}}}\\
     \subfloat[$\mu= \sqrt{2}$]{ \makebox[.45\textwidth][c]{\includegraphics[width=\textwidth]{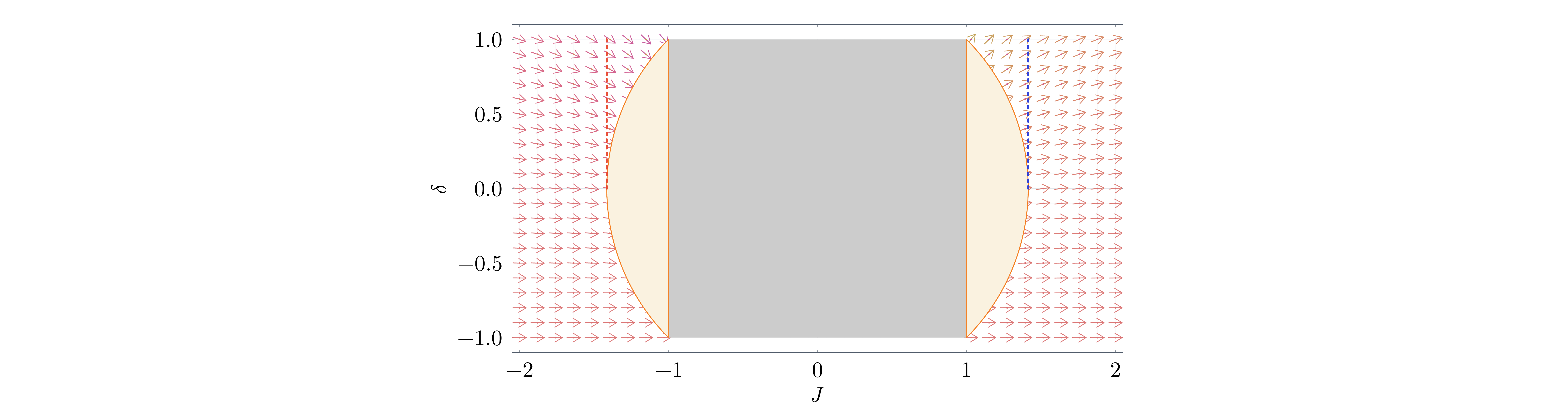}}}
    \caption{The textured phase diagrams of the Rice-Mele model \eqref{eq:H_1d_RiceMele}. The insulating phase with a single filled band has a topological texture similar to that of a vortex, exposed by plotting the Berry phase $\gamma$ (numerically evaluated). The insulator with both bands filled or empty has no texture. The broken line indicates the presence of edge modes. For $\mu \neq 0$, the left and right edge modes are estranged and occur on different parts of the phase diagram as can be seen in $(b),(c)$.}
    \label{fig:phase_diagram_1d}
\end{figure}

Let us now study the phase diagram of the Hamiltonian \eqref{eq:H_1d_RiceMele}. We begin with $\mu=0$ when the system is at half-filling. From \cref{eq:Epm}, we see that away from $J=\delta=0$, the many-body ground state is a band insulator with a filled lower band and a finite gap, as shown in \cref{fig:dispersion1d}(a).  At the origin $J=\delta=0$, the spectral gap closes as shown in \cref{fig:dispersion1d}(b), resulting in a gapless state whose low-energy theory is the relativistic 1+1d Dirac fermion. Thus, the two-dimensional $J$-$\delta$ phase diagram contains a single, trivially gapped phase whose ground states are all connected to the atomic insulator as shown in \cref{fig:phase_diagram_1d}(a). For a finite chemical potential $\mu  \neq 0$, the gapless point at the origin opens up into a gapless phase with a partially filled band. For $|\mu| \le 1$, as shown in \cref{fig:phase_diagram_1d}(b), the gapless phase is a circular island of radius $\mu$. For $|\mu| > 1$, as shown in \cref{fig:phase_diagram_1d}(c) the gapless phase splits into two disks and opens up an insulating phase in between, characterized by both bands being either fully filled or empty depending on the sign of $\mu$. Although the various insulating regions are topologically trivial as phases, we will argue that the region with a single filled band has hidden topological features. 
\subsection{Texture on the phase diagram}
\begin{figure}[!h]
    \centering
    \makebox[0.5\linewidth][c]{\includegraphics[width=\textwidth]{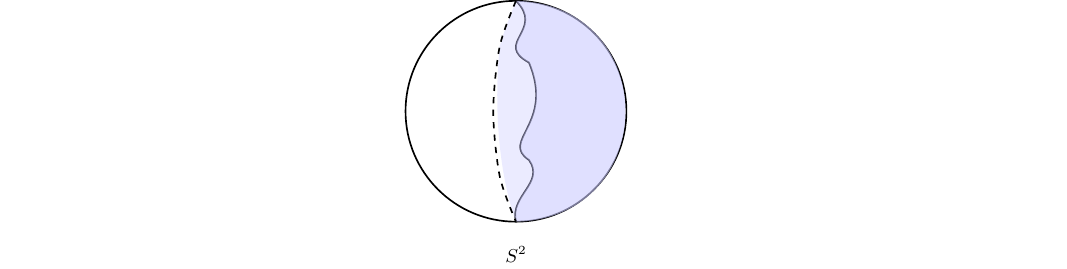}}
    \caption{The image of $\hat{g}$ mapping the BZ at an illustrative parametric point $(J,\delta)$ to a closed curve on $S^2$. The colored cap denotes $\hat{g}(\bar{\cM_2})$ whose boundary gives us $\hat{g}(\partial\bar{\cM_2})=\hat{g}(S^1_{BZ})$ which is the domain of integration, by Stokes' theorem, for computing the Berry phase.}
    \label{fig:stokes}
\end{figure}
Let us focus on the single-band insulator, whose filled band is denoted $\ket{\varepsilon^-(k,J,\delta)}$. For each pair of parameters $(J,~\delta)$, we can define the Berry-Bloch connection as~\cite{KANE_topbandtheory}
\begin{align}
    A_k(J,\delta) = - i \innerproduct{\varepsilon^-(k,J,\delta)|\frac{\partial}{\partial k}}{\varepsilon^-(k,J,\delta)}\ .
\end{align}
Integrating $A_k(J,\delta)$ over the Brillouin zone (BZ) circle $S^1_{\text{BZ}}$, we define the Berry phase $\gamma$ by
\begin{align}
    \gamma(J,\delta) = \oint_{\sonebz} A_k(J,\delta) \rmd k \label{eq:Berry_phase} \ .
\end{align}
This is a geometric invariant that is well-defined as an angle $\gamma \in \mathbb{R}/2\pi\bZ$. It is useful to consider the unit vector $\hat{g} = \vec{g}/|\vec{g}|=\{\hat{g}_1,\hat{g}_2,\hat{g}_3\}$ which defines a map  $\hat{g}:\sonebz \rightarrow S^2$. The expression in \cref{eq:Berry_phase} can be recast in terms of $\hat{g}$ as the solid angle $\Omega$ subtended over the target space unit sphere as shown in \cref{fig:stokes} 
\begin{align}
    \gamma(J,\delta)=\frac{1}{2}\int_{k\in BZ} \frac{\hat{g}_1\partial_k{\hat{g}_2-\hat{g}_2\partial_k \hat{g}_1}}{1+\hat{g}_3}\rmd k=\frac{\Omega}{2}\label{eq:solid_angle} \ .
\end{align}
We plot the numerically evaluated $\gamma$ at each point of the phase diagram that corresponds to the single-band insulator. We represent $\gamma$ as the angle of orientation of an arrowhead, and see in \cref{fig:phase_diagram_1d} that a topological texture emerges taking the form of a vortex. For $\mu=0$ the core of the vortex is a gapless point, whereas for $0<|\mu|<1$ it becomes an extended gapless region. Note that the precise value of $\gamma$ depends on the choice of unit cell. Different choices however, only result in smooth changes to the texture and preserve its non-trivial topological nature.

\subsection{Thouless charge pump}
\begin{figure}[!h]
\centering
    \makebox[\linewidth][c]{\includegraphics[width=\textwidth]{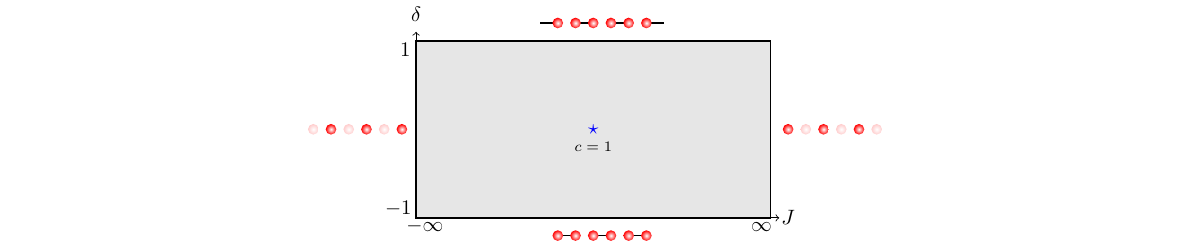}}
    \caption{Phase diagram of \cref{eq:H_1d_RiceMele} the different limits encapsulated; as one takes a trajectory around the origin, we see a transport of charge by one unit cell which is central to the behaviour of the Thouless pump. \label{fig:Periphery_Thouless}}
\end{figure}
The reason for this texture is that the states surrounding the origin form a topologically non-trivial family---a Thouless charge pump~\cite{Thouless83,asboth2016short}. This is easily seen in the periphery of the phase diagram as shown in \cref{fig:Periphery_Thouless}. Tracking the location of one of the charges shows how it is transported by one unit cell as we go around the phase diagram. The charge pump represents a non-contractible loop of states within a single phase and the obstruction to this contraction is manifested by the closure of the spectral gap at `diabolical points' as we approach the core of the vortex \cite{HsinKapustinThorngren_PhysRevB.102.245113,manjunath2026searchdiabolicalcriticalpoints}. 

The Thouless pump can be diagnosed by a topological invariant. For this, we consider the Berry connection for the filled band $\ket{\varepsilon^-(\xi)}$ over the crystal momenta $k$ as well as Hamiltonian parameters,
\begin{align}
    A_\mu(\xi) = -i \innerproduct{\varepsilon^-(\xi)|\frac{\partial}{\partial \xi_\mu}}{\varepsilon^-(\xi)}\ ,
\end{align}
where $\{\xi_\mu\}$ are the collective coordinates that represent all the aforementioned parameters. Now consider any circle $S^1_{\text{par}}$ that surrounds the origin and belongs to the gapped trivial phase (say, $J^2 + \delta^2 =c^2$ for $\mu<c<1$). We parametrize the circle by $\theta$ and compute the first Chern number, defined as the integral of the Berry curvature  $F = \epsilon^{\mu \nu} \partial_{\mu} A_{\nu}$ over the two-dimensional manifold, $\cM_2 = \sonebz  \times \snpar{1}$ 
\begin{align}
    \Ch_1=\frac{1}{2\pi}\oint_{\sonebz} \rmd k \oint_{\snpar{1}}\rmd \theta F(k,\theta) \in \bZ\label{eq:firstchern}\ .
\end{align}
The unit vector $\hat{g} = \vec{g}/|\vec{g}|$ defined earlier now produces a map  $\hat{g}:\cM_2\rightarrow S^2$. Evaluating \cref{eq:firstchern} is equivalent to computing the winding number of $\hat{g}$
\begin{align}
    \Ch_1=\frac{1}{4\pi}\oint_{\sonebz} \rmd k \oint_{\snpar{1}}\rmd \theta~ \hat{g}\cdot(\partial_{k}\hat{g}\times\partial_{\theta}\hat{g})\ . \label{eq:winding_thouless}
\end{align}
 For any parametric circle  surrounding the origin of the phase diagram, \cref{eq:winding_thouless} evaluates to $\Ch_1=1$ indicating the presence of the Thouless pump \cite{RevModPhys.82.1959,KANE_topbandtheory}. Since this is a topological invariant, any small deformation of the loop preserves it. If the parametric circle is subjected to a large deformation such that it no longer encloses the origin, it evaluates to $\Ch_1 =0$. However, at some stage of the deformation when the circle touches the origin, the spectral gap closes and \cref{eq:winding_thouless} becomes ill-defined.

The expression for the Berry phase in \cref{eq:Berry_phase,eq:solid_angle} can be obtained using Stokes' theorem. Let $\bar{\cM}_2$  be any two-dimensional surface in the combined parameter and momentum space where the Bloch Hamiltonian does not touch any gap-closing points, and whose boundary is purely in the momentum direction corresponding to the Brillouin zone at some parameter value $(J,\delta)$. If we evaluate the integrals in \cref{eq:firstchern,eq:winding_thouless} over $\bar{\cM}_2$, after multiplying by a factor of $2 \pi$ we get, by Stokes' theorem, that
\begin{align}
     \gamma(J,\delta) &=  \int_{\bar{\cM}_2} F~\rmd^2\xi   = \oint_{\partial \bar{\cM}_2 = \sonebz} A_k(J,\delta) \rmd k\ . \label{eq:Berry_Stoke}
\end{align}
In terms of $\hat{g}$, $\gamma(J,\delta)$ can be expressed in terms of the integral in \cref{eq:winding_thouless} evaluated over $\bar{\cM}_2$ and multiplied by a factor of $2\pi$
\begin{multline}
    \gamma(J,\delta) =\frac{2\pi}{4\pi}\int_{\bar{\cM}_2} \rmd^2\xi~ \hat{g}\cdot(\partial_{k}\hat{g}\times\partial_{\theta}\hat{g}) \\= \frac{1}{2}\int_{\partial \bar{\cM}_2 = \sonebz} \frac{\hat{g}_1\partial_k{\hat{g}_2-\hat{g}_2\partial_k \hat{g}_1}}{1+\hat{g}_3}\rmd k\ . \label{eq:winding_thouless_gamma}
\end{multline}
This too is a manifestation of Stokes' theorem. Clearly, the choice of $\bar{\cM}_2$ is not unique so long as it does not touch any gap closing points. This is because the quantization condition in \cref{eq:firstchern} guarantees that a different choice, $\bar{\cN_2}$, only shifts $\gamma(J,\delta)$ by integer multiples of $2\pi$. In particular,
\begin{multline}
    \int_{\bar{\cN}_2} \rmd^2\xi F = \int_{\bar{\cM}_2} \rmd^2\xi F - \int_{\bar{\cM}_2 \cup_{\sonebz} \bar{\cN_2}} \rmd^2\xi F \\=  \int_{\bar{\cM}_2} \rmd^2\xi F - 2 \pi \Ch_1(\bar{\cM}_2 \cup_{\sonebz} \bar{\cN_2})
\end{multline}
where $\bar{\cM}_2 \cup_{\sonebz} \bar{\cN_2}$ is the closed manifold obtained by gluing $\bar{\cM}_2$ and $\bar{\cN_2}$ along their common boundary. We see that the integer shift $\Ch_1(\bar{\cM}_2 \cup_{\sonebz} \bar{\cN_2})$ depends on whether $\bar{\cM}_2 \cup_{\sonebz} \bar{\cN_2}$ encloses any gap-closing loci. Thus, $e^{i\gamma}$ is well-defined, irrespective of the choice of $\bar{\cM}_2$. We will use a generalization of \cref{eq:Berry_Stoke} when we study higher dimensional generalizations in subsequent sections.

\subsection{Estranged edge modes}
A tell-tale signature of non-trivial topological phases is the presence of robust edge modes as a result of bulk-boundary correspondence~\cite{HasanKane_TI_RevModPhys.82.3045}. In the trivial phase, such edge modes are not expected to be stable. However, the presence of non-trivial topological families results in a novel variant of bulk-boundary correspondence~\cite{Seiberg_AnomaliesCouplingOne10.21468/SciPostPhys.8.1.001}, and therefore edge modes, even in the trivial phase. Ref~\cite{HsinKapustinThorngren_PhysRevB.102.245113} argued that the boundary gap closes at least once for any parametric circle that represents a Thouless charge pump and predicts a tail of ground states with edge modes that terminate on the diabolical points.

{To study edge modes, let us study the system with open boundaries. For concreteness, we consider our system with an even number of lattice points where the left end belongs to the odd sublattice and rightmost site belongs to the even sublattice.} For $\mu=0$, it is easy to see where and how these states with edge modes arise in our phase diagram. Along the $J=0$ line, \cref{eq:H_1d_RiceMele} has an accidental particle-hole symmetry,
\begin{align}
    \mathscr{P}: c(x) \mapsto c^\dagger(x) \label{eq:particle_hole_symmetry} 
\end{align}
 and we recover the Su-Schreifer-Heeger (SSH) model~\cite{SSH_PhysRevLett.42.1698}.  This hosts a symmetry-protected-topological (SPT)~\cite{ChiuTeoSchnyderRyu_SPTReview_RevModPhys.88.035005} phase for $\delta>0$ that has edge modes. For any $J\neq0$, the accidental particle-hole symmetry $ \mathscr{P}$ is broken and the edge modes are immediately gapped out.  Thus, in the full phase diagram, we have edge modes along the line segment $J=0$, $\delta>0$ as shown in \cref{fig:phase_diagram_1d}(a) with broken lines that terminate on the diabolical points at the core of the vortex. 
\begin{figure}[!h]
    \centering
    \subfloat[left = odd, right = even]{\makebox[0.49\linewidth][c]{\includegraphics[width=\textwidth]{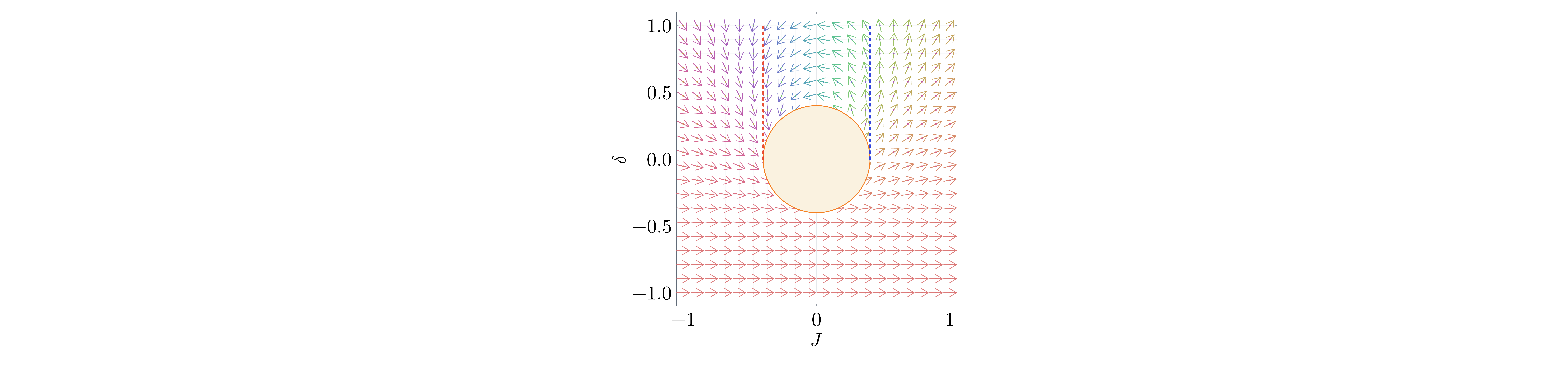}}}
    \subfloat[left = odd, right = odd]{\makebox[0.49\linewidth][c]{\includegraphics[width=\textwidth]{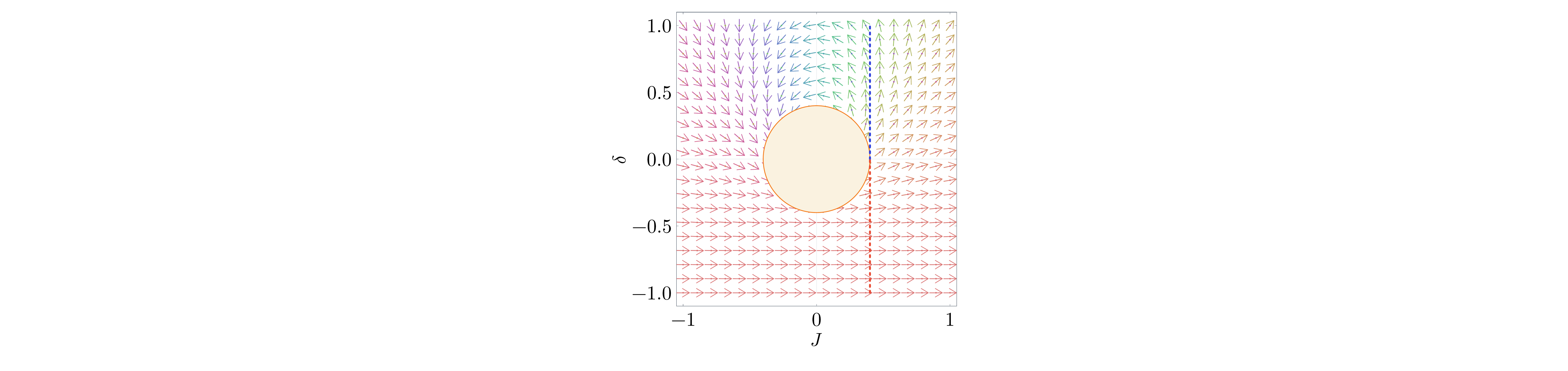}}}\\
    \subfloat[left = even, right = odd]{\makebox[0.49\linewidth][c]{\includegraphics[width=\textwidth]{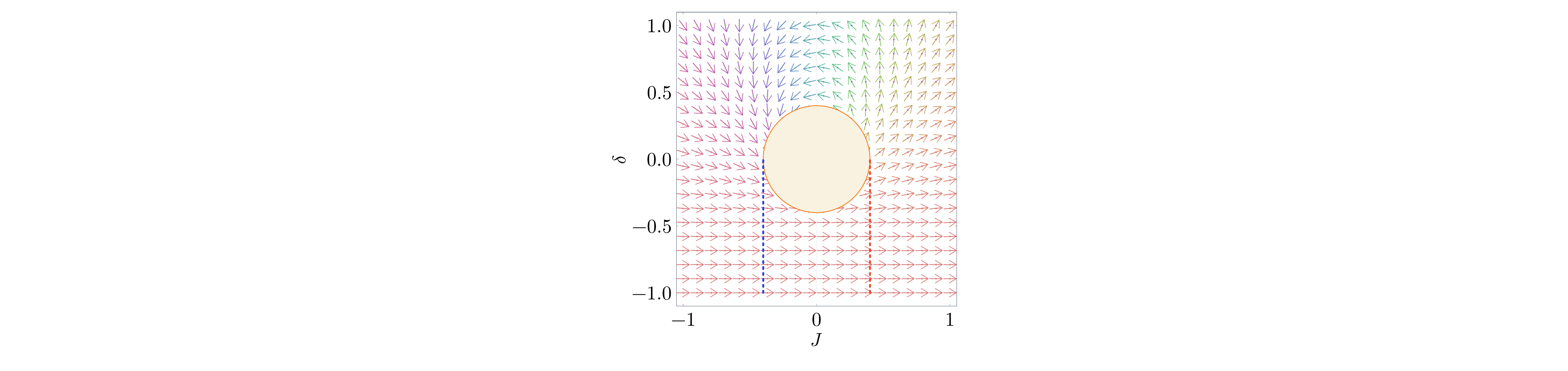}}}
    \subfloat[left = even, right = even]{\makebox[0.49\linewidth][c]{\includegraphics[width=\textwidth]{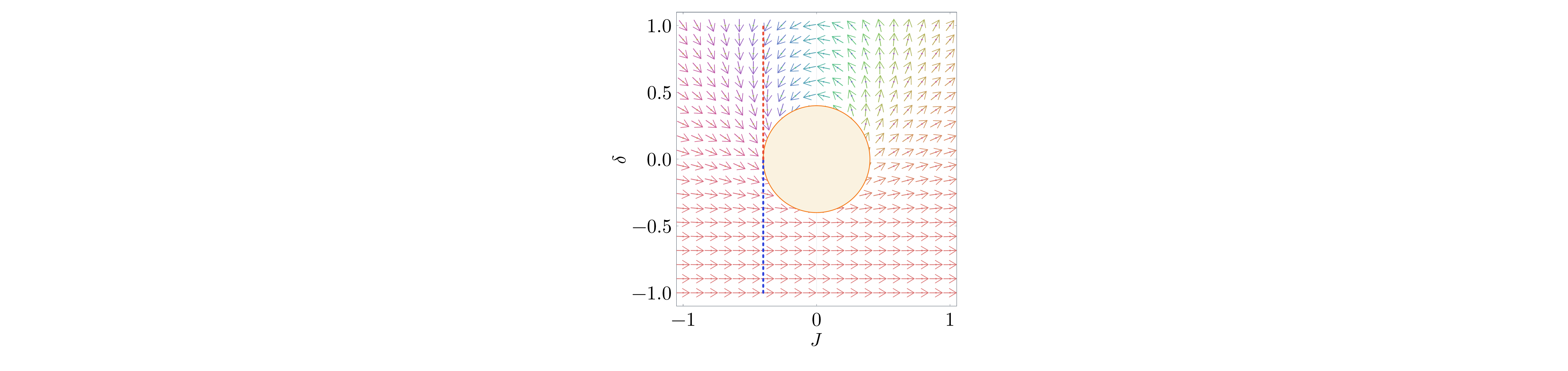}}}
    \caption{ {Estranged edge modes for terminations in which the left and right edges belong to odd or even sublattices. Blue (red) broken lines denote modes localized at the left (right) edge. Different terminations result in the edge modes appearing in different parts of the phase diagram, but the modes on the two ends remain estranged from each other.}}
    \label{fig:estranged}
\end{figure}
However, the accidental symmetry and SPT are \emph{not} necessary to obtain edge modes. To see this, let us study the Hamiltonian \cref{eq:H_1d_RiceMele} at finite chemical potential $\mu \neq 0$. This term eliminates the accidental particle-hole symmetry, the SPT phase and the edge modes along $J=0$. However, the edge modes reappear in a way that cannot be associated with any SPT phases---the left and right edge modes become \emph{estranged} and appear on \emph{different} parts of the phase diagram as shown in \cref{fig:phase_diagram_1d}(b,c). 

This unusual edge mode structure can be easily understood in the limit of $\delta = 1$. For $\mu = 0$, the edge modes appear as decoupled fermions on the left and right edges when $J=0$. For $\mu\neq 0$, the chemical potential increases the energy of these modes, which becomes zero only when compensated by the staggered chemical potential. For the odd left endpoint and even right endpoint chosen here, this compensation occurs at $J=+\mu$ on the left and $J=-\mu$ on the right. To study the edge modes away from the $\delta = +1$ limit,  we consider the first-quantized Hamiltonian of \cref{eq:H_1d_RiceMele} in real space,
\begin{multline}
  \h_{jk} = \delta_{j,k-1} \frac{(1+(-1)^j\delta)}{2} +\delta_{j,k+1}\frac{(1+(-1)^k\delta)}{2}\\ + \delta_{jk}(J(-1)^j+\mu) \ ,   
\end{multline}
and solve for localized zero energy single-particle modes $\alpha$ satisfying $\sum_k\h_{jk} \alpha_k = 0$. If such a solution exists \footnote{Mathematically, so that the edge-mode is normalisable, we are looking for solutions to this equation in the sequence space $l^2$.}, we get an edge mode creation operator $\psi_0^\dagger = \sum_x \alpha_x c^\dagger(x)$ which commutes with the second-quantized Hamiltonian. For a finite chain, the edge modes hybridize and we do not expect to find exact zero energy solutions \cite{Jones23}. However, we can isolate zero modes at the left or right boundaries by considering a half-infinite chain. Let us first consider $0< \delta\leq1$ and $J=+\mu$, then we have a normalisable zero-mode at the left edge. Labeling the lattice sites $j=1,2,\ldots$, the zero-mode wavefunction coefficients are given by
\begin{align}
    \alpha_{j=1,3,5,\ldots} \propto \left(\frac{\delta-1}{\delta+1}\right)^{\left(j-1\right)/2} ,\quad~\alpha_{j=2,4,6,\ldots} = 0  \ . \label{eq:Edge_Zero}
\end{align}
For $J=-\mu$, we have a normalizable zero-mode on the right edge for $0<\delta\leq 1$. Labeling the lattice sites $j=0,-1,-2,\ldots$ from the right, we have the zero-mode wavefunction coefficients, 
\begin{align}
    \alpha_{j=0,-2,-4,\ldots} \propto \left(\frac{\delta-1}{\delta+1}\right)^{-j/2} ,~\alpha_{j=-1,-3,-5,\ldots} = 0  \ . \label{eq:Edge_Zero2}
\end{align} 

{The other terminations can be summarized by assigning to each terminal site the parity $\eta_e=(-1)^{x_e}$, where $e=L,R$ and $\eta_e=-1$ ($+1$) denotes an odd (even) site. The sublattice-polarized edge mode has energy $E_e=\mu+\eta_eJ$, and hence crosses zero at $J=-\eta_e\mu$. Its recurrence relation is normalizable when $-\eta_L\delta>0$ at the left edge and when $\eta_R\delta>0$ at the right edge. These conditions give all four panels of \cref{fig:estranged}; in particular, for an even--odd termination the left and right modes occur at $J=-\mu$ and $J=+\mu$, respectively, both for $\delta<0$, while for an even--even termination they occur at the common value $J=-\mu$ on the $\delta<0$ and $\delta>0$ rays, respectively. For any termination, the localization length in lattice-site units is
\begin{align}
    \xi=\frac{2}{\left|\log\left|\frac{1-\delta}{1+\delta}\right|\right|}\ .
\end{align}
Thus $\xi \rightarrow 0$ as $|\delta|\rightarrow1$, where the mode is localized on a single lattice site, whereas $\xi \rightarrow \infty$ as $\delta\rightarrow0$, where it merges with the bulk states. Finally, as $\mu\rightarrow0$, the two loci merge and we recover the SPT edge modes of the SSH model on a semi-infinite system.}
\subsection{Field theory analysis}
\begin{figure}[!h]
    \centering
    \makebox[0.5\linewidth][c]{\includegraphics[width=\textwidth]{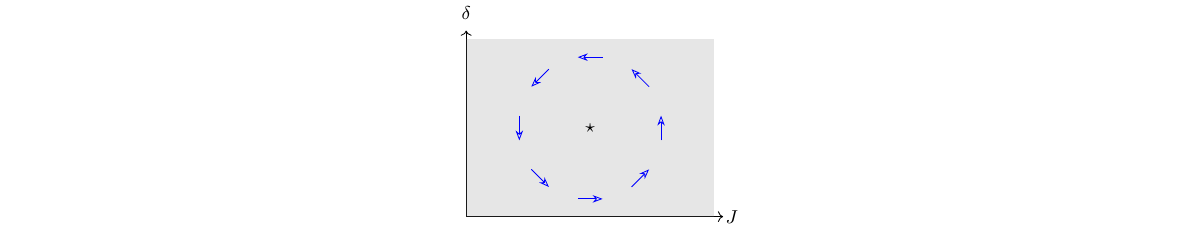}}
    \caption{Topological texture on the parameter space recovered from the field theoretic description.}
    \label{fig:Thouless_vortex_fieldtheory}
\end{figure}
We now reproduce certain results of the previous subsections using an effective field theory. We will also argue that these features (textured phase diagram and edge modes) are stable to interactions. Some parts of the discussion of this section can be found in Ref.~\cite{HsinKapustinThorngren_PhysRevB.102.245113} and are reproduced here for completeness and to set the language for new results.

We begin with $\mu=0$ and focus on the vicinity of the gapless theory at the origin $J=\delta=0$. The low-energy Hamiltonian of \cref{eq:H_1d_RiceMele} can be obtained by expanding around $k=\pi$ to obtain a relativistic two-component Dirac fermion 

\begin{align} 
\Ham \approx \int \rmd x~\psi^\dagger(x) \left( i\sigma^2 \partial_x  - \delta \sigma^1 - J  \sigma^3 \right) \psi(x) \ .
\end{align}
The corresponding Euclidean space-time Lagrangian density can be written as
\begin{align}
    \cL =  -\bar{\psi} \slashed{\partial} \psi - \delta \bar{\psi}  \psi +i J \bar{\psi} \gamma_c  \psi \label{eq:RM_field_fermion}.
\end{align}
We have set the dispersion velocity $v=1$ by rescaling spacetime coordinates and used the standard notation, $\bar{\psi} = \psi^\dagger \gamma_0$ and $\slashed{\partial} = \gamma^\mu\partial_\mu$ with 
\begin{align}
  \gamma_0 = \sigma^1,~ \gamma_1 =-\sigma^3,~\gamma_c = \sigma^2\ .
\end{align}
The microscopic parameters $\delta$ and $J$ couple to precisely the two relevant perturbations for the massless Dirac fermion in   1+1d~\cite{HsinKapustinThorngren_PhysRevB.102.245113,coleman.sine}.  Thus, any quadratic perturbations merely shifts the location of the diabolical point. 

{At $J=\delta=0$, the massless theory also has an emergent axial $\mathrm{U}(1)$ symmetry that rotates the two mass terms into one another~\cite{HsinKapustinThorngren_PhysRevB.102.245113,manjunath2026searchdiabolicalcriticalpoints}, and is distinct from the microscopic charge $\mathrm{U}(1)$. Terms beyond linear order in $q=(k-\pi)$ break the axial rotation, but  are irrelevant at the Dirac fixed point.}

We can also consider adding interactions to the Rice-Mele model of the form
     \begin{align}
            \Ham = \Ham_{\text{RM}} + U \sum_{x \in \bZ} \left(n(x) - \half\right)\left(n({x+1}) - \half \right).
        \end{align}
At the level of field theory, this results in a Thirring-type \cite{Giamarchi} interaction term $(\bar{\psi}\gamma^\mu \psi)(\bar{\psi}\gamma_\mu \psi)$ \cite{Haldane_luttingerliquid}. This is exactly marginal and changes the scaling dimensions of various operators. However, for a finite range of interaction strengths, the two mass terms remain relevant, and the phase diagram remains unchanged. A more detailed analysis can be provided using bosonization \cite{Giamarchi} to obtain an effective continuum description for the interacting system in terms of a compact boson $\phi \equiv \phi + 2\pi$ as
\begin{equation}
    \cL \approx \frac{(\partial_\mu \phi)^2}{8 \pi K }  - \delta \cos \phi +J \sin \phi + \ldots \ . \label{eq:RM_field_boson}
\end{equation}
The operators, $\cos \phi \leftrightarrow \bar{\psi} \psi,~\sin\phi \leftrightarrow i\bar{\psi} \gamma_c \psi$ have scaling dimension equal to $K$, which is set by the interaction strength~\cite{HALDANE1981153,Giamarchi,AP_SEC_PhysRevB.108.245135}. So long as $K < 2$, the operators are relevant and the basic form of the phase diagram is unchanged. For the Rice-Mele model, the relationship between the $U$ and $K$ for $\delta = J = 0$ is known from Bethe ansatz~\cite{HALDANE1981153},
\begin{equation}
    K = \frac{\pi}{2 \arccos{(-U)}}.
\end{equation}
We see that for $U>-1/\sqrt{2}$, $K<2$ and interactions do not introduce any qualitative changes.

We can also obtain the vortex texture using field theory. The two masses in \cref{eq:RM_field_fermion,eq:RM_field_boson} have the same operator scaling dimensions, and it is convenient to reorganize them into the following `complex' form
\begin{multline}
     \cL = - \bar{\psi} \slashed{\partial} \psi -m \left(\bar{\psi} e^{-i \alpha \gamma_c} \psi\right) \leftrightarrow \frac{(\partial_\mu \phi)^2}{8 \pi K } -m \cos (\phi + \alpha), \\
\text{where }        m = \sqrt{\delta^2 + J^2},~\alpha = \arctan\left(\frac{J}{\delta}\right) \ . \label{eq:malpha_dirac1d}
\end{multline}
Under RG, $m$ flows to larger values as $m(\ell)  = m e^{K \ell}$. However, dimensionless $\alpha$ does not flow~\cite{fradkin2013field}. We see that the moduli space of fixed points  has the topology of a circle. This provides a field-theoretic origin for the texture in the phase diagram. As shown in \cref{fig:Thouless_vortex_fieldtheory}, plotting the value of $\alpha$ which represents the fixed point to which the gapped theory away from the origin flows, reproduces the texture of the microscopic phase diagram \cref{fig:phase_diagram_1d} near the origin. Note that the definition of $\alpha$ is not unique and changes with different basis as well as counterterms. However, these only correspond to smooth deformations of $\alpha$ and preserve the topological nature.  

For finite chemical potential, the first-quantized continuum Hamiltonian becomes
\begin{align}
     \h(x) =i\sigma^2 \partial_x  - \delta \sigma^1 - J  \sigma^3 + \mu \mathbb{1}\ .
 \label{eq:RM_firstquant_chemicalpot}
 \end{align}
For $J^2 + \delta^2 <\mu^2$, \cref{eq:RM_firstquant_chemicalpot} favours partially filled Dirac bands resulting in a gapless metal, reproducing the microscopic phase diagram. Finally, let us understand the edge modes using field theory. Open boundaries in the continuum can be modelled by interpolating the dimerization mass $\delta$ to a fiducial vacuum value $\delta^*$ while keeping $J$ fixed~\cite{HsinKapustinThorngren_PhysRevB.102.245113,IanAffleck_1998,AP_UC_Pump,AP_UC_Mutiversality_PhysRevLett.130.256401}, so that \cref{eq:RM_firstquant_chemicalpot} becomes
\begin{align}
      \h(x) =i \sigma^2 \partial_x  - m(x) \sigma^1 - J  \sigma^3 + \mu \mathbb{1}\ .
 \label{eq:RM_firstquant_chemicalpot_edge}
 \end{align}
Here $m(x)$ is a smooth interpolation from $\delta$ to a fiducial vacuum mass $\delta^*$. The precise value of $\delta^*$ and the corresponding boundary condition depend on the microscopic termination. For the natural termination used in the lattice model, we may take $\delta^*<0$ and keep $J$ constant across the interface, since the fiducial vacuum is already gapped by $\delta^*$. Just as in the lattice case, we model the left and right edges using semi-infinite chains extending from $x=0$ to $x = \pm \infty$ with a smooth interpolation at $x=0$ to the fiducial vacuum living at $x<0$ and $x>0$, respectively. Let us begin with $\mu =0$ and solve for zero-energy solutions of \cref{eq:RM_firstquant_chemicalpot_edge}, $\h \cdot \alpha = 0$. Normalizable zero-mode solutions exist for $J=0$, $\delta>0$ on both edges \`{a} la Jackiw and Rebbi \cite{JackiwRebbi_PhysRevD.13.3398} and take the form
\begin{align}
    \alpha_{\text{left}} &\propto \exp\left( - \int_0^x m(x) \rmd x \right) \begin{pmatrix}
        1\\0
    \end{pmatrix}, ~x \in [0,+\infty)\ ,\nonumber\\
    \alpha_{\text{right}} &\propto \exp\left( + \int_0^x m(x) \rmd x \right) \begin{pmatrix}
        0\\1
    \end{pmatrix},~x \in (-\infty,0]\ . \label{eq:continuum_zeromodes}
\end{align}
For $\mu \neq 0$, the wavefunctions in \cref{eq:continuum_zeromodes} remain localized but their energies shift to
\begin{align}
\h \cdot \alpha_{\text{left}} = (\mu - J) \alpha_{\text{left}},~\h \cdot \alpha_{\text{right}} = (\mu + J) \alpha_{\text{right}}\ .   
\end{align}
Thus, the left and right modes cross zero at $J=+\mu$ and $J=-\mu$, respectively, reproducing the estranged edge-mode phenomenon.

{This continuum description also clarifies the even--odd termination effect in \cref{fig:estranged}. The universal field-theory ingredient is a Dirac bound state at an interface where the dimerization mass changes sign. The microscopic termination enters through the boundary condition, or equivalently through the choice of fiducial vacuum, which selects the sublattice polarization of that bound state. Projecting the chemical-potential terms onto the selected spinor gives $E_e=\mu+\eta_eJ$, while normalizability gives $\eta_L\delta<0$ and $\eta_R\delta>0$, in agreement with the lattice analysis above. Thus the rearrangement of the edge-mode rays under a change of termination is captured by the continuum theory once the appropriate boundary condition is specified; what remains microscopic is the identification of a given odd or even termination with that boundary condition, as well as non-universal details of the edge-state profile and the precise location of a crossing in the presence of additional boundary-local perturbations.}

\section{The suspension construction and ascendants}
\label{sec:suspension}
\subsection{The suspension construction: generalities}
\begin{figure}[!h]
    \centering
    \makebox[\linewidth][c]{\includegraphics[width=\textwidth]{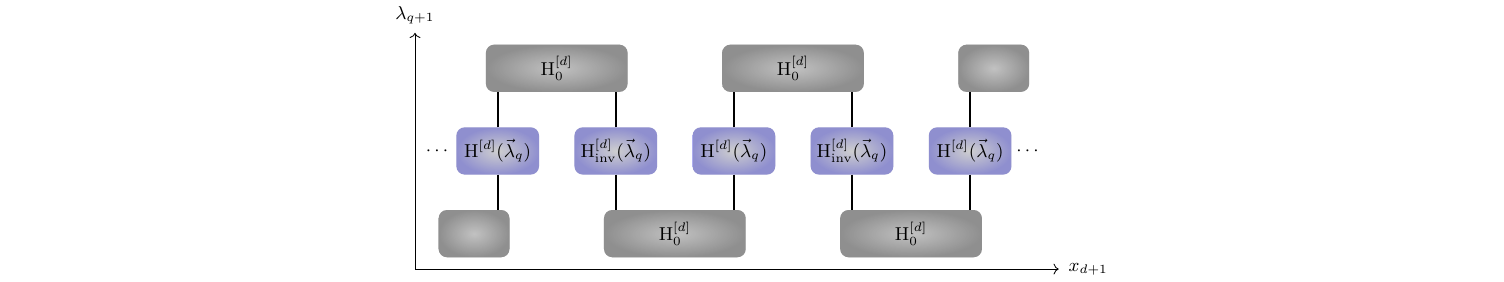}}
    \caption{Schematic representation of Kitaev's pump~\cite{kitaev1,kitaev2} formalized by the suspension construction of Ref~\cite{Wenetal_topologicalfamilies}. }
    \label{fig:suspension}
\end{figure}
We now turn to generalizations of the Rice-Mele model in higher dimensions.  To build such models, a recipe was outlined by Kitaev~\cite{kitaev1,kitaev2} and formalized as the so-called suspension construction in Ref.~\cite{Wenetal_topologicalfamilies}. This recipe takes as input a non-trivial topological Hamiltonian family $\Ham^{[d]}(\vec{\lambda}_q)$ in $d$ spatial dimensions parametrized over a $q$ dimensional parameter space $\cM_q$ with coordinates $\vec{\lambda}_q$. From this, it constructs a Hamiltonian family $\Ham^{[d+1]}(\vec{\lambda}_{q+1})$ in $d+1$ spatial dimensions over a $q+1$ dimensional parameter space $\cM_{q+1} \cong \Sigma\cM_q$, called the \emph{suspension} of $\cM_q$~\cite{fomenko2016homotopical}. We briefly review  the details of this construction here. 

Let $\Ham^{[d]}(\vec{\lambda})$ be a non-trivial topological family with a unique ground state on any closed spatial manifold for all values of $\vec{\lambda}_q$. Such states are termed `invertible'~\cite{Gaiotto_2019} since we can define the inverse family of states, $\Ham^{[d]}_{\text{inv}}(\vec{\lambda}_q)$ such that the stack $\Ham^{[d]}(\vec{\lambda}_q) \oplus \Ham^{[d]}_{\text{inv}}(\vec{\lambda}_q)$ is a \emph{trivial family}; that is, it can be adiabatically deformed to a Hamiltonian $\Ham_0$ 
with a unique ground state that does not vary with $\vec{\lambda}$,
\begin{align}
    \Ham^{[d]}(\vec{\lambda}_q) \oplus \Ham^{[d]}_{\text{inv}}(\vec{\lambda}_q) \sim \Ham_0\ .
\end{align}
Given $\Ham^{[d]}(\vec{\lambda}_q)$, to construct $\Ham^{[d+1]}(\vec{\lambda}_{q+1})$ we alternately stack $\Ham^{[d]}(\vec{\lambda}_q)$ and its inverse along a new spatial direction $x_{d+1} \in \bZ$, as shown in \cref{fig:suspension}, and introduce a new parameter $\lambda_{q+1}$ that couples the neighbouring layers and deforms them into the trivial family in two different ways, as schematically shown in \cref{fig:suspension}.  Given $\Ham^{[d]}(\vec{\lambda}_{q})$, this iterative procedure in principle produces models for us in all higher spatial dimensions $\Ham^{[d+1]}(\vec{\lambda}_{q+1}),\Ham^{[d+2]}(\vec{\lambda}_{q+2}),\dots$. We will call these the \emph{ascendants} of $\Ham^{[d]}(\vec{\lambda}_{q})$. 

\subsection{Constructing the ascendants of the Rice-Mele model}
\begin{figure}[!h]
    \centering
    \makebox[\linewidth][c]{\includegraphics[width=\textwidth]{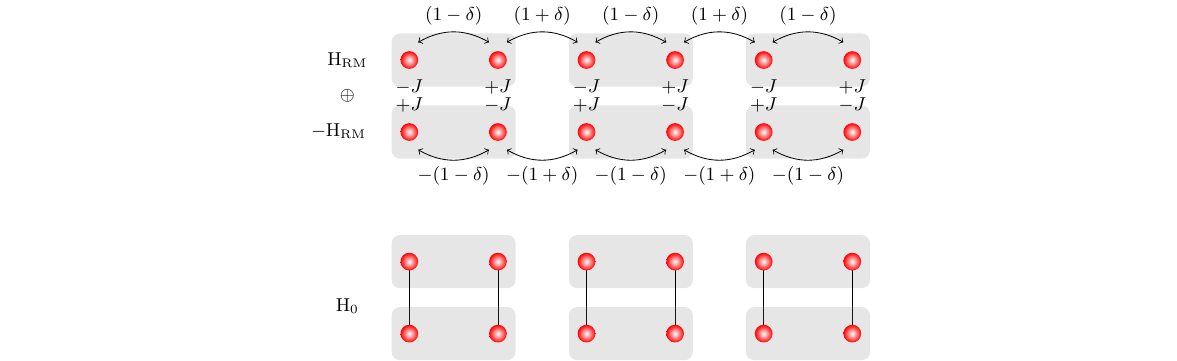}}
    \caption{$\Ham_{\text{RM}}$ for $\delta,J \neq 0,~\mu=0$ stacked with its inverse $-\Ham_{\text{RM}}$  can be smoothly deformed to the trivial Hamiltonian family $\Ham_0$.}
    \label{fig:RM_inversion}
\end{figure}
In this section, we study the ascendants of the Rice-Mele model. To construct these, it is convenient to consider the gapped states of $\Ham_{\text{RM}}$ at $\mu=0$ as an input to the suspension recipe, and then later reintroduce $\mu$ for the higher-dimensional system. First, we need the inverse family for $\Ham_{\text{RM}}$.  We can verify that $\Ham_{\text{RM,inv}} = -\Ham_{\text{RM}}$ by considering the stack of the two systems $\Ham_{\text{RM}} \oplus  -\Ham_{\text{RM}}$.
This can be deformed to the trivial family, 
\begin{align}
    \Ham_0 = \sum_{x \in \bZ} \left(c^\dagger_1(x) c_2^{\vphantom \dagger}(x) + h.c \right) \label{eq:H0_RM}
\end{align}
without any gap closure using the following explicit interpolation,
\begin{align}
    \Ham(\lambda) = (1-\kappa) \left(\Ham_{\text{RM}} \oplus  -\Ham_{\text{RM}} \right)|_{\mu=0} + \kappa \Ham_0, \label{eq:RM_inverse}
\end{align}
see Fig.~\ref{fig:RM_inversion}. Equation~\eqref{eq:RM_inverse} corresponds to the four-band model with the following Bloch Hamiltonian:
\begin{align}
    \h(\kappa) = (1-\kappa) (\vec{g}\cdot \vec{\sigma} \otimes \sigma^z) + \kappa (\mathbb{1} \otimes \sigma^x)
\end{align}
where $\vec{g}$ is as defined in \cref{eq:g_RM} with $\mu=0$. Throughout the interpolation $\kappa \in [0,1]$, the valence and conduction bands do not touch and the many-body ground state remains gapped as long as $J \neq 0 $ or $\delta \neq 0$. 

Having found the inverse family of the Rice-Mele model, let us now apply the suspension construction to construct its ascendant in two spatial dimensions. We first relabel the original lattice coordinate of the one-dimensional model $\Ham_{\text{RM}}$ as $x \mapsto x_1$, and  its hopping anisotropy $\delta \mapsto \delta_1$. We then alternately stack $\Ham_{\text{RM}}$ and its inverse $-\Ham_{\text{RM}}$ in a new direction $x_2$ (the horizontal direction in \cref{fig:suspension}) and couple them to form stacks of $\Ham_0$, defined in \cref{eq:H0_RM}, using a new parameter $\delta_2$ as
\begin{multline}
    \Ham^{[2]}_{\text{RM}}|_{\mu=0} = \sum_{x_2 \in \bZ} (-1)^{x_2} \Ham_{\text{RM},x_2}|_{\mu=0} \\ + \sum_{\vec{x} \in \bZ^2}\left( \frac{1+(-1)^{x_2}\delta_2}{2} \right) (c^\dagger(\vec{x}) c(\vec{x}+\hat{e}_2)+h.c) \ . \label{eq:H_RM_2d_suspension}
\end{multline}
 Generalizing \cref{eq:H_RM_2d_suspension}, we can also schematically write the ascendant for $d$ spatial dimensions $\Ham^{[d]}_{\text{RM}}$. This is constructed from $\Ham^{[d-1]}_{\text{RM}}$ as
\begin{multline}
    \Ham^{[d]}_{\text{RM}} =\mu \sum_{\vec{x} \in \bZ^d} n(\vec{x}) +  \sum_{x_d \in\bZ} (-1)^{x_d}\Ham^{[d-1]}_{\text{RM},x_d}|_{\mu=0} \\+\sum_{\vec{x} \in \bZ^d}\left( \frac{1+(-1)^{x_d}\delta_d}{2} \right) (c^\dagger(\vec{x}) c(\vec{x}+\hat{e}_d)+h.c) \ . \label{eq:Hd_RM_suspension}
\end{multline}
Note that $\vec{\lambda}_1 \leftrightarrow (J,\delta_1)$ and  $\vec{\lambda}_2 \leftrightarrow (J,\delta_1,\delta_2), \ldots$ in the general construction. We also introduced the notation
\begin{align}
    \vec{x} = \sum_{a=1}^d x_a \hat{e}_a,~\text{where }x_a \in \bZ,~\text{and }\hat{e}_a\cdot\hat{e}_b = \delta_{ab}\ . \label{eq:r2_def}
\end{align}

We will study \cref{eq:H_RM_2d_suspension,eq:Hd_RM_suspension} in  detail in the upcoming sections. First, let us make a few comments on the mathematical origins of the suspension construction.

\subsection{Mathematical underpinnings}
\label{sec:Mathematical}
Here we comment on the mathematics that underpins the suspension construction. We will simply state the main results relevant to this work and direct the interested reader to Refs~\cite{kitaev1,kitaev2,kitaev3,kitaev2019,Gaiotto_2019,Xiong_2018} for more details. 

Consider the space of Hamiltonians with fixed attributes (i.e. $d$ spatial dimensions, bosons or fermions, global symmetries) having a unique ground state on any spatial manifold without boundaries. Call $\sI_d$  the space of such \emph{invertible} ground states. $\sI_d$  includes ground states of the trivial phase and also non-trivial invertible phases~\cite{Gaiotto_2019,Xiong_2018} such as topological insulators and the integer quantum Hall states. We are interested in topological properties of $\sI_d$. Its disconnected components, $\pi_0(\sI_d)$, are nothing but the $d$-dimensional invertible phases. In this work, we are interested in $\pi_1(\sI_d), \pi_2(\sI_d),\ldots$.  Kitaev proposed that the full topology of $\sI_d$ is classified by generalized cohomology \cite{kitaev1,kitaev2,kitaev3}. In particular, the collection of spaces, $\{\sI_d\}$, $d=0,1,\ldots$ are conjectured to form a so-called \emph{$\Omega$-spectrum}~\cite{Gaiotto_2019,Wenetal_topologicalfamilies,fomenko2016homotopical,kubota2025stablehomotopytheoryinvertible}. This means that the space of loops of $\sI_d$, denoted $\Omega \sI_d$ has the same topology as that of $\sI_{d-1}$. That is, there exist homotopy equivalences,
\begin{align}
    \sI_{d-1} \cong \Omega \sI_d\ . \label{eq:homotopy_equivalence}
\end{align}
This allows us to determine the homotopy groups of $\sI_d$ as follows. Recall that $\pi_n(\sI_d)$ are homotopy classes of based maps from $S^n$ to $\sI_d$, denoted by $ [S^n,\sI_d]$. From \cref{eq:homotopy_equivalence}, we have that 
\begin{align}
    \pi_n(\sI_d) = [S^n,\sI_d] = [S^n,\Omega \sI_{d+1}]\ .
\end{align}
Using the Freudenthal suspension theorem~\cite{fomenko2016homotopical}, we have
\begin{equation}
     [S^n,\Omega \sI_{d+1}] =  [\Sigma S^n,\sI_{d+1}]\ ,
\end{equation}
where $\Sigma$ is the so-called reduced suspension~\cite{Hatcher:478079,Wenetal_topologicalfamilies} which relates manifolds of different dimensions. In particular, it relates  $n-$spheres as $\Sigma S^n \cong S^{n+1}$. Thus we have
\begin{align}
     \pi_n(\sI_d) = [S^{n+1}, \sI_{d+1}] = \pi_{n+1}(\sI_{d+1})\ .
\end{align}
This tells us that non-trivial families of states in $d$ spatial dimensions over $S^n$ imply the existence of non-trivial families of states over $S^{n+1}$ in $d+1$ spatial dimensions and so on. The suspension construction of Ref.~\cite{Wenetal_topologicalfamilies} reviewed above is a physical way of constructing models for the latter given the former.

\section{The two-dimensional Rice-Mele ascendant}
\label{sec:2d_RiceMele}
\begin{figure}[!h]
    \centering
      \subfloat[$\mu=0$]{\makebox[.79\linewidth][c]{\includegraphics[width=\textwidth,valign=c]{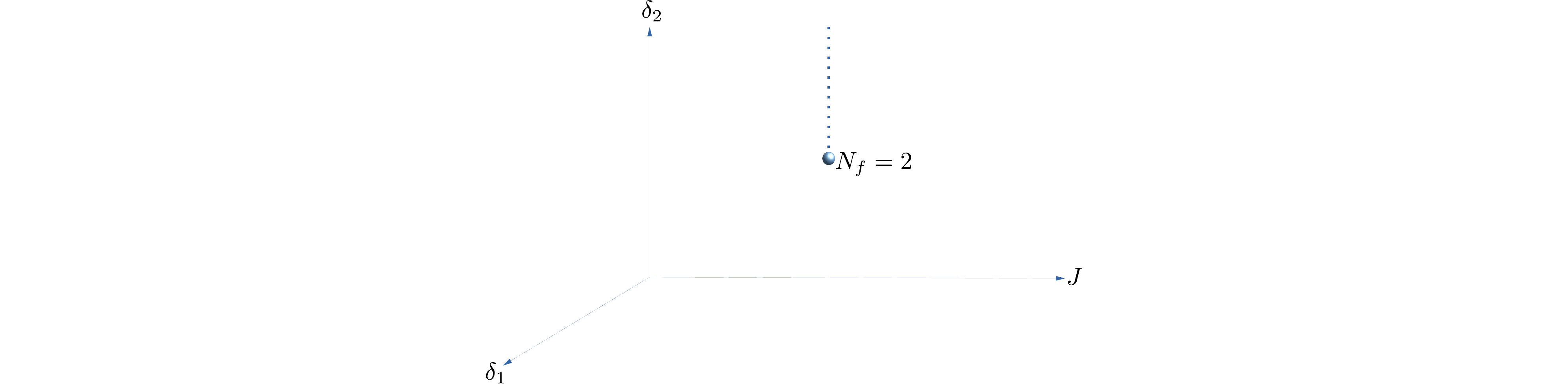}}}\\
  \subfloat[$0<|\mu|<1$]{\makebox[.8\linewidth][c]{\includegraphics[width=\textwidth,valign=c]{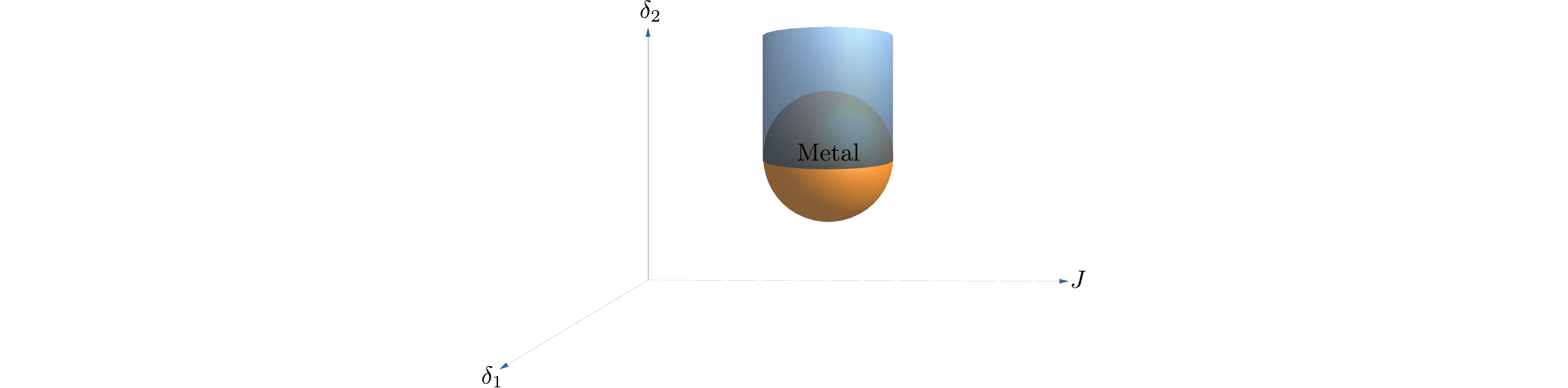}}}
    \caption{The phase diagram of 2d Rice-Mele ascendant for various values of $\mu$ and periodic boundary condition along $x_1$, open boundaries for $x_2$. (a) For $\mu=0$, there exists a single diabolical gapless point at $\delta_1 = \delta_2 = J=0$ described by $N_f=2$ flavours of massless Dirac fermions. Edge modes appear for $J=\delta_1=0,~0<\delta_2 \le1$ and terminate on the diabolical point. (b) For $\mu\neq0$, the diabolical point changes to a metallic phase with partially filled bands. The system transitions to an insulator as the band is emptied or filled for $\delta_1^2 + \delta_2^2 + J^2 = \mu^2$. Edge modes appear for $\delta_1^2+J^2<\mu^2,~0<\delta_2\le1$ and terminate on the metallic phase. 
    } 
    \label{fig:2dRiceMele_mu}
\end{figure}
\subsection{Model and ground states}
In this section we generalize  \Cref{sec:RiceMele}, and identify topological textures in the phase diagram of the $d=2$ Rice-Mele ascendant, $\Ham^{[2]}_{\text{RM}}$ constructed using the suspension recipe in \cref{eq:H_RM_2d_suspension}. Writing out the various terms and reintroducing a chemical potential for the full system, we can express $\Ham^{[2]}_{\text{RM}}$ as 
\begin{align}
   \Ham^{[2]}_{\text{RM}} &= \sum_{\substack{\vec{x} \in \bZ^2\\a=1,2}} t^a_{\vec{x}}~ c^\dagger(\vec{x}) c(\vec{x}+ \hat{e}_a) + h.c. + \sum_{\vec{x} \in \bZ^2}\mu_{\vec{x}} n(\vec{x}), \nonumber\\
    t^1_{\vec{x}} &= (-1)^{x_2} \left(\frac{1+(-1)^{x_1} \delta_1}{2} \right),~t^2_{\vec{x}} =  \left(\frac{1+(-1)^{x_2} \delta_2}{2} \right), \nonumber\\ \mu_{\vec{x}} &= \mu + (-1)^{x_1 + x_2} J \label{eq:H_RiceMele_2d} \ .
\end{align}
This Hamiltonian has a two-site translation symmetry in each direction $a=1,2
$, as well as a $\uone$ symmetry,
\begin{align}
    \T_{a}: c(\vec{x}) \mapsto c(\vec{x}+2\hat{e}_a),~\uone:c(\vec{x}) \mapsto e^{i \vartheta} c(\vec{x}) \ . \label{eq:RiceMele_symmetries}
\end{align}
Choosing a 4-site unit cell, \cref{eq:H_RiceMele_2d} can be written in momentum space as a four-band model with the following Bloch Hamiltonian 
\begin{align}
   \h(\vec{k}) = \mu \mathbb{1} +  \sum_{m=1}^5  g_m(\vec{k}) \Gamma^m\ , \label{eq:h_fq_RiceMele_2d}
\end{align}
 where corresponding to a particular choice of unit cell and arrangement of the modes within the spinor $\psi$, we have
\begin{align}
    g_{1} &= \frac{(1-\delta_1)}{2}+\frac{(1+\delta_{1})}{2}\cos(k_{1}),\nonumber~g_{2} = \frac{(1+\delta_{1})}{2}\sin(k_{1}),\nonumber\\g_{3} &= \frac{(1-\delta_{2})}{2}+\frac{(1+\delta_{2})}{2}\cos(k_{2}),\nonumber ~g_4 = \frac{(1+\delta_2)}{2}\sin(k_{2}),\\
    ~g_5 &= J\ . \label{eq:bloch2d}
    \end{align}
The $\{\Gamma^{\mu}\}$, represent the four-dimensional irreducible representation of the Clifford algebra $\mathrm{Cl}(5)$ , satisfying
\begin{align}
    \{\Gamma^m,\Gamma^n \} = 2 \delta_{mn}\ . \label{eq:Clifford}
\end{align}
The matrix representation of $\{\Gamma^\mu\}$ for the same choice of unit cell is
\begin{align}
    \Gamma_{1} &= \sigma^3 \otimes \sigma^{1},~\Gamma_{2} = \sigma^3 \otimes \sigma^{2},~\Gamma_3 = \sigma^1 \otimes \mathbb{1}, \nonumber\\
    \Gamma_4 &= \sigma^2 \otimes \mathbb{1},\Gamma_{5} = \sigma^3 \otimes \sigma^{3}\ . \label{eq:Gamma_2d_RM}
\end{align}

\begin{figure}
    \centering
    \subfloat[$\delta_1=\delta_2=J=0$]{\makebox[0.5\linewidth][c]{\includegraphics[width=\textwidth]{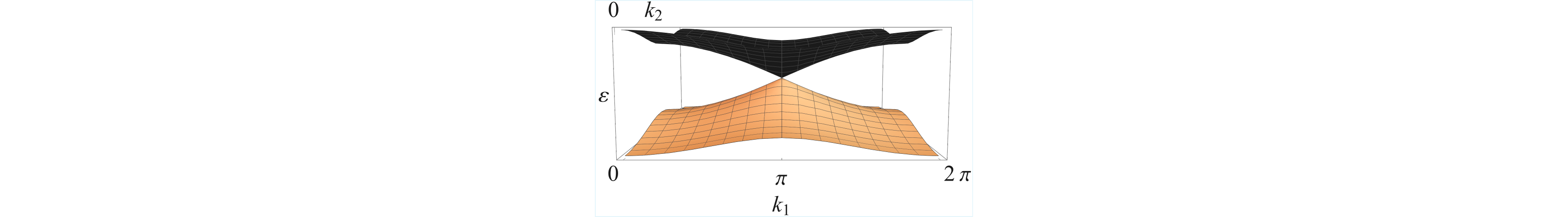}}}
    \subfloat[Away from the origin]{\makebox[0.5\linewidth][c]{\includegraphics[width=\textwidth]{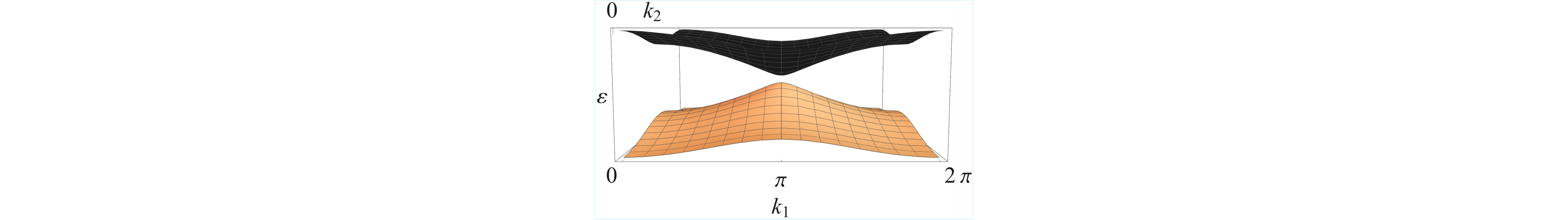}}}
    \caption{Single particle band dispersions for the 2d Rice-Mele ascendant in \cref{eq:H_RiceMele_2d,eq:h_fq_RiceMele_2d}. The bands, each two-fold degenerate, are separated unless $\delta_1=\delta_2=J=0$.}
    \label{fig:2ddirac}
\end{figure}

For $\mu=0$, except at the origin $J=\delta_1 = \delta_2 = 0$, the three-parameter phase diagram obtained by tuning $J \in (-\infty,\infty),~\delta_{1,2} \in[-1,1]$ contains a single trivial gapped phase whose many-body ground state is obtained by filling the two degenerate lower bands of the single-particle Hamiltonian \eqref{eq:h_fq_RiceMele_2d}, see \cref{fig:2ddirac}. 
At $J= \delta_1 = \delta_2=0$, the bands touch to produce a linearly dispersing massless Dirac spectrum. The phase diagram is summarized in \cref{fig:2dRiceMele_mu}(a).
The presence of a chemical potential $\mu \neq 0$ changes the gapless point to a gapless metallic phase for $J^2 + \delta_1^2 + \delta_2^2 < \mu^2$ in \cref{fig:2dRiceMele_mu}(b). We will study the non-trivial textured nature of the gapped states. 

\subsection{Topological invariant}
\begin{figure}[!h]
    \centering
    \makebox[.8 \linewidth][c]{\includegraphics[width=\textwidth]{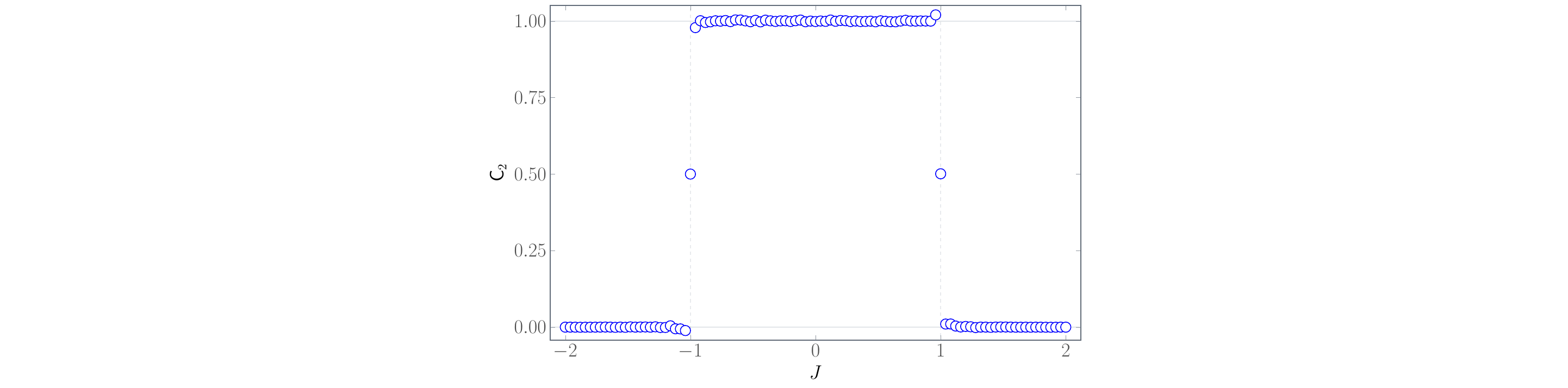}}
    \caption{The second Chern number $\Ch_2$ evaluated by computing the expression in \cref{eq:C2_Pontryagin} numerically. The integral is performed over $\cM_4 = \snpar{2} \times \tnbz{2}$ where $\snpar{2}$ is a parametric unit 2-sphere centered on $J,~\delta_1=\delta_2 = 0$ and $\tnbz{2}$ is the two-dimensional Brillouin zone. For $|J|<1$, $\snpar{2}$ encloses the gap-closing diabolical point, and evaluates to $\Ch_2=1$ whereas for $|J|>1$, $\snpar{2}$ does not enclose the diabolical point and is contractible, therefore evaluating to $\Ch_2  =0$. Finally, at $J = \pm 1$, $\snpar{2}$ contains a gapless point and $\Ch_2$ is formally not well-defined. Numerically we find the average $\Ch_2$ of the neighbouring gapped phases, i.e., $\Ch_2 = 1/2$. This is consistent with the analysis in Ref.~\cite{Verresen20}.}
    \label{fig:C2_numerics}
\end{figure}
Recall that the states surrounding the origin of the two-dimensional phase diagram of the Rice-Mele model form a non-trivial topological family over a circle $S^1$, as determined by the first Chern number of the Berry connection. The suspension construction guarantees that the states surrounding the origin of the three dimensional phase diagram of $\Ham^{[2]}_{\text{RM}}$ forms a non-trivial family over $\Sigma S^1 = S^2$. To detect this non-trivial family, let us consider the non-Abelian Berry connection of the two filled bands $\ket{\varepsilon^-_{1,2}}$, defined over momentum and parameter space represented by collective coordinates $\xi$:
\begin{align}
    [A_\mu(\xi)]_{ab}=-i\langle \varepsilon^-_a(\xi) |\frac{\partial}{\partial \xi_\mu}|\varepsilon^-_b(\xi)\rangle \ .\label{eq:NonAbelian_Berry}
\end{align}
Now, consider the states on some sphere $S^2_{\text{par}}$ surrounding the origin containing the trivial gapped trivial phase, say $J^2 + \delta_1^2 + \delta_2^2 = c^2,~c^2 > \mu^2$. Define $\cM_4 = S^2_{\text{par}} \times T^2_{\text{BZ}}$, where $T^2_{\text{BZ}}$ is the two-dimensional Brillouin zone. The non-trivial nature of the family of states over $S^2_{\text{par}}$ can be determined by computing the second Chern number~\cite{Avron1983,Avron1988,Avron1989,Avron1994,Zhang2001,QiZhang_topologicalresponse_PhysRevB.78.195424,Ryu2010},
\begin{align}
    \Ch_2 =  \frac{1}{32 \pi^2} \int_{\mathcal{M}_4} \rmd^4\xi ~\epsilon^{\mu \nu \alpha \beta}\tr(F_{\mu \nu}  F_{\alpha \beta}) \label{eq:Second_Chern}
\end{align}
where $F_{\mu \nu}$ is the non-abelian Berry curvature,
\begin{align}
    F_{\mu \nu} = \partial_\mu A_\nu -\partial_\nu A_\mu + [A_\mu,A_\nu]\label{eq:F_nonabelian}\ .
\end{align}

For our model, \cref{eq:Second_Chern} can also be given a simple expression in terms of $\hat{g}_m = g_m/|g|$, as 
\begin{align}
    \Ch_2 =\frac{3}{8 \pi^2} \int_{\mathcal{M}_4}\rmd^4\xi\epsilon_{abcde}\epsilon^{\mu \nu \alpha \beta}   \hat{g}^a \partial_\mu \hat{g}^b  \partial_\nu \hat{g}^c \partial_\alpha\hat{g}^d \partial_\beta\hat{g}^e \label{eq:C2_Pontryagin}
\end{align}
where $\hat{g}$ now defines a map $\hat{g}: \cM_4 \rightarrow S^4$ and $\Ch_2$ counts the degree of winding of this map. 

We can verify by numerically evaluating \cref{eq:C2_Pontryagin}  that $\Ch_2 = 1$ as expected. Furthermore, by moving the center of $S^2_{\text{par}}$ away from the origin along the $J$ direction, we can verify that $\Ch_2 = 1$ holds as long as $S^2_{\text{par}}$ wraps around the band-touching point but drops to $\Ch_2= 0$ when it does not as shown in \cref{fig:C2_numerics}. Note that at the transition points, the integration gives $\Ch_2 \simeq 1/2$. This is consistent with previous work on phase transitions in Chern insulators \cite{Verresen20}, where a half-integer invariant is found by excluding an infinitesimal region in the BZ around the gap closing point.

\subsection{Textured phase diagram}
\begin{figure}
    \centering
   \subfloat[]{ \makebox[0.7\linewidth][c]{\includegraphics[width=\textwidth,valign=b]{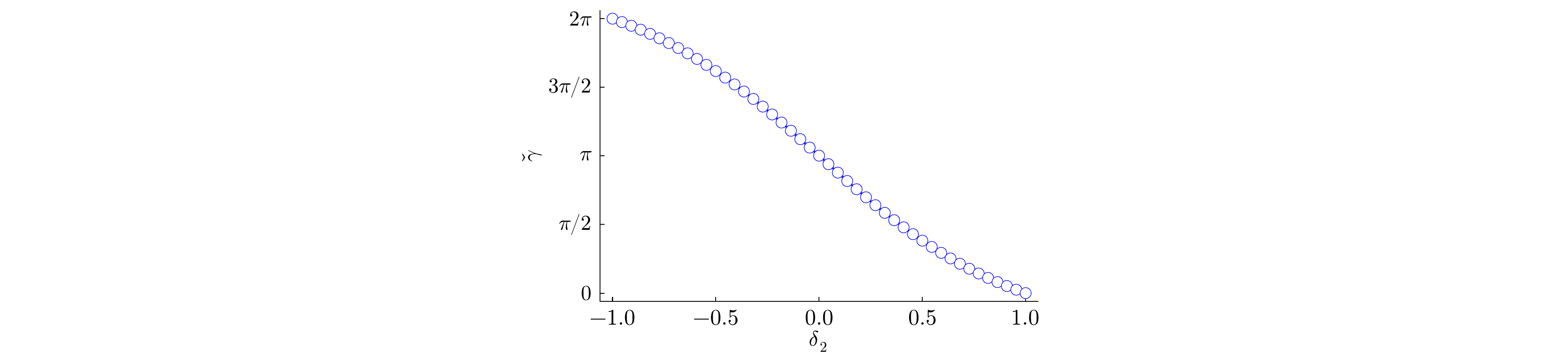}}}\\
     \subfloat[]{\includegraphics[width = 0.7\linewidth,valign = b]{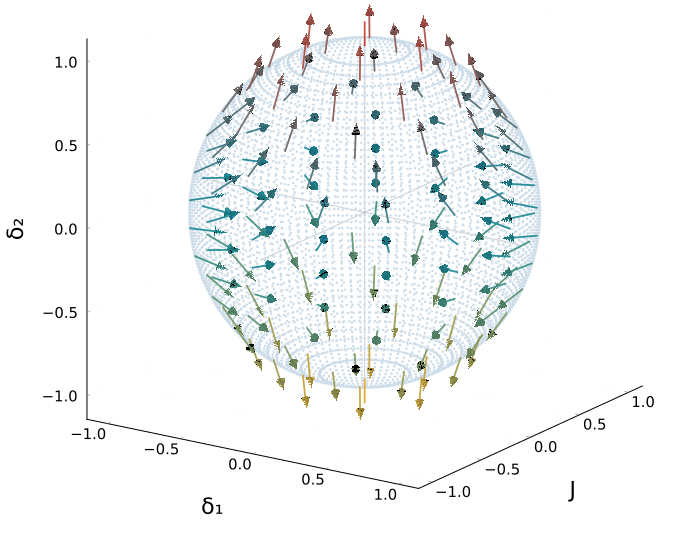}}
     \caption{(a) The higher Berry phase $\check{\gamma}$  computed on a sphere $\delta_1^2 + \delta_2^2 + J^2 =1$ for each latitude, $\delta_2 \in [-1,1]$ by evaluating \cref{eq:C2_Pontryagin_Geometric} numerically.  (b) Texture on a single sphere obtained by plotting $\check{\gamma}$ for each fixed latitude $\delta_2$ using the recipe in \cref{fig:Higher_Berry_visualize}. Repeating this for various spheres foliates the texture on  the full phase diagram.  }
         \label{fig:Rice_Mele_2d_texture}
\end{figure}
In the study of the Rice-Mele model in \cref{sec:RiceMele}, we were able to uncover a topological texture in its two-parameter phase diagram by computing the Berry phase. The ascendant in two spatial dimensions from the suspension construction is expected to have a texture in the three-parameter phase diagram. How can we visualize the texture? What can play the role of a geometric invariant similar to Berry's phase? We proceed by analogy to the Rice-Mele model. Recall that the Berry phase was related to the expression for the first Chern number by Stokes' theorem as shown in \cref{eq:Berry_Stoke}. Let us now evaluate the expression for the second  Chern number, \cref{eq:Second_Chern}, over a 4-manifold with a boundary $\bar{\cM}_4$, $\partial\bar{\cM}_4 = \cM_3= T^2_{\text{BZ}} \times S^1_{\text{par}}$, where $S^1_{\text{par}}$ is some circle in parameter space where there is no gap closure. Multiplying this by a factor of $2 \pi$, we get 
\begin{multline}
    \check{\gamma}(S^1_{\text{par}})   =   \frac{1}{16 \pi} \int_{\bar{\cM}_4} \rmd^4\xi ~\epsilon^{\mu \nu \alpha \beta}\tr(F_{\mu \nu}  F_{\alpha \beta}) \\= \frac{1}{4 \pi} \int_{\cM_3} \rmd^3\xi ~\epsilon^{\mu \nu \alpha}\tr\left(A_\mu \partial_\nu A_\alpha + \frac{2}{3} A_\mu A_\nu A_\alpha \right) . \label{eq:Higher_Berry_ChernSimons}
\end{multline}
\cref{eq:Higher_Berry_ChernSimons} is an integral of the Chern-Simons three form $\CS_3$ which assigns an unquantized $\uone$ geometric invariant to every circle (more generally a closed one-dimensional loop) in  parameter space. Indeed, Berry's phase in \cref{eq:Berry_phase} can be interpreted as an integral over the Chern-Simons form $\CS_1$, making the expression in \cref{eq:Higher_Berry_ChernSimons} a natural generalization. Higher dimensional extensions, to which we will return shortly, also follow from the same line of reasoning. \cref{eq:Higher_Berry_ChernSimons} does not depend on the choice of $\bar{\cM}_4$. To see this, consider two bounding 4-manifolds $\bar{\cM}_4$ and $\bar{\cN}_4$ with the same boundary $\partial\bar{\cM}_4 =\partial \bar{\cN}_4  = \cM_3= T^2_{\text{BZ}} \times S^1_{\text{par}}$. We then have
\begin{multline}
     \int_{\bar{\cN}_4} \frac{\rmd^4\xi}{16 \pi} \epsilon^{\mu \nu \alpha \beta}\tr(F_{\mu \nu}  F_{\alpha \beta}) =  \int_{\bar{\cM}_4} \frac{\rmd^4\xi}{16 \pi}\epsilon^{\mu \nu \alpha \beta}\tr(F_{\mu \nu}  F_{\alpha \beta}) \\- 2\pi  \Ch_2\bigg(\bar{\cM}_4 \cup_{\bar{\cM}_3} \cN_4\bigg).
\end{multline}
$\bar{\cM}_4 \cup_{\cM_3} \bar{\cN}_4$ is a 4-manifold without any boundaries that is obtained by gluing $\bar{\cM}_4$ and $\bar{\cN}_4$ along their common boundary. The multiplicative factor of $2\pi$ compared to \cref{eq:Second_Chern} allows $\check{\gamma}$ to be restricted to the fundamental domain, $\check{\gamma} \in [0,2\pi)$, making $e^{i \check{\gamma}}$ a well-defined phase for any choice of $\bar{\cM}_4$ which depends only on the boundary $\cM_3 = \snpar{1}  \times \tnbz{2}$ and ultimately only on $\snpar{1}$. \cref{eq:Higher_Berry_ChernSimons} can also be evaluated using a modification of the expression in \cref{eq:C2_Pontryagin}:
\begin{align}
    \check{\gamma}(S^1_{\text{par}}) = \frac{3}{4 \pi} \int_{\bar{\cM}_4}\rmd^4\xi\epsilon_{abcde}\epsilon^{\mu \nu \alpha \beta}   \hat{g}^a \partial_\mu \hat{g}^b  \partial_\nu \hat{g}^c \partial_\alpha\hat{g}^d \partial_\beta\hat{g}^e \ .\label{eq:C2_Pontryagin_Geometric}
\end{align}
\cref{fig:Rice_Mele_2d_texture}(a) shows $\check{\gamma}$ computed at each latitude $\delta_2\in[-1,1]$ of a unit sphere $\delta_1^2 +\delta_2^2 + J^2 = 1$.

\begin{figure}[!h]
    \centering
     \makebox[0.6\linewidth][c]{\includegraphics[width=\textwidth,valign=b]{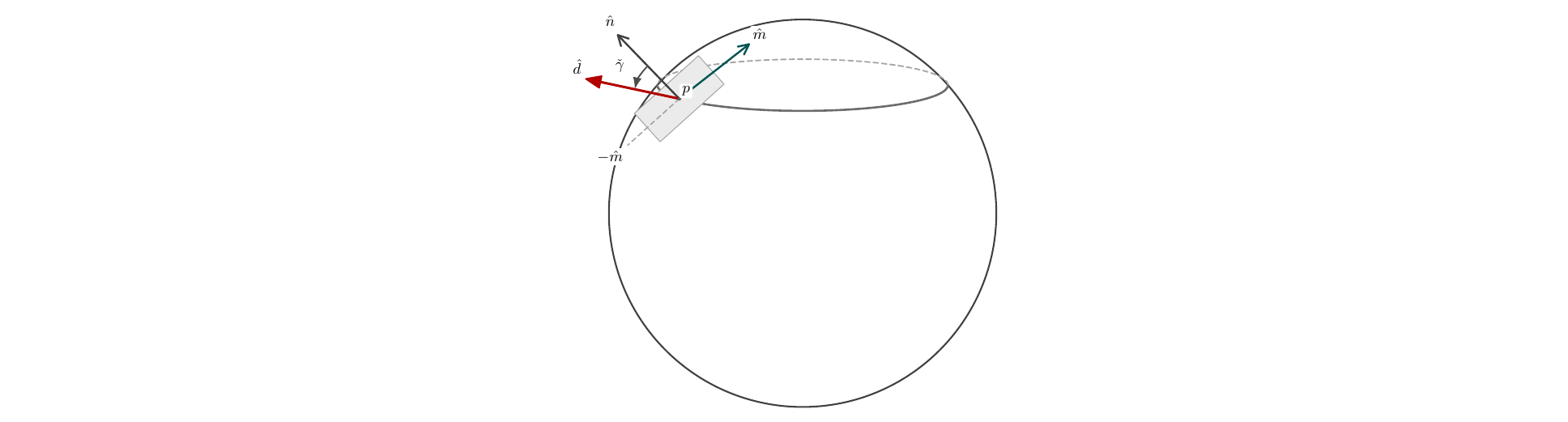}}\\
    \caption{For each fixed latitude $\delta_2$, $\check{\gamma}$ is plotted as an arrowhead, oriented at angle $\check{\gamma}$ to the normal $\hat{n}$ in the plane spanned by  $\hat{n}$ and  $\hat{m} = \hat{n} \times \hat{t}$  where $\hat{t}$ is the tangent along the latitude circle.}
    \label{fig:Higher_Berry_visualize}
\end{figure}

To visualize \cref{eq:Higher_Berry_ChernSimons} on the phase diagram, consider foliating the trivial phase surrounding the origin with a collection of 2-spheres. Let us further break each of these spheres into a series of parallel circles at different latitudes, for which we assign a $\check{\gamma}$ (\cref{fig:Rice_Mele_2d_texture}(a)) evaluating \cref{eq:Higher_Berry_ChernSimons,eq:C2_Pontryagin_Geometric}. We can plot this on the circle as an arrowhead on each point, inclined at an angle $\check{\gamma}$ to the normal of the circle on the sphere at the point, along the plane formed by the normal $\hat{n}$ and $\hat{m} = \hat{n} \times \hat{t}$, where $\hat{t}$ is the tangent along the latitude circle (\cref{fig:Higher_Berry_visualize}). This $\hat{n} - \hat{m}$ plane is ambiguous at the poles when the latitudes reduce to a single point. However, at the poles $\check{\gamma} = 0$ (mod $2\pi$) and the arrowheads point along the normal $\hat{n}$ making the $\hat{n} - \hat{m}$ plane irrelevant.

As shown in \cref{fig:Rice_Mele_2d_texture}(b), this reveals a hedgehog-like texture over the 2-sphere, generalizing the vortex texture of the 1d Rice-Mele model. This recipe is certainly not unique, and different ways of foliating the phase diagram result in different values of $\check{\gamma}$, generalizing the ambiguity of the Berry phase resulting from the choice of unit cell in the Rice-Mele model. Various choices, however, are related by smooth deformations of the texture, preserving the non-trivial topological nature.

\subsection{Stable edge modes} \label{subsec:edg2d}
\begin{figure}[!h]
    \centering
 {\makebox[0.3\linewidth][c]{\includegraphics[width=\textwidth,valign=c]{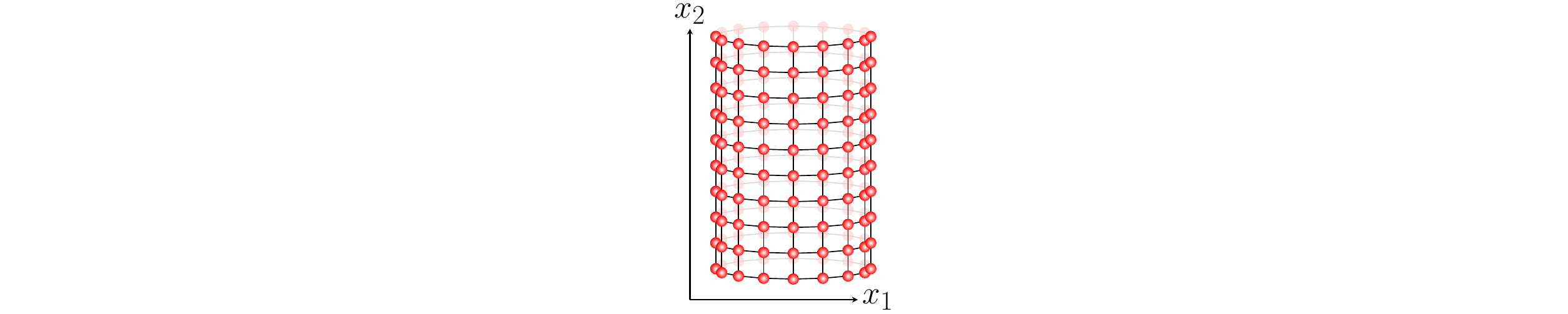}}}
    \caption{Boundary conditions used to study edge modes}
    \label{fig:obc_2dRiceMele}
\end{figure}
For the Rice-Mele model, we were able to show that the topological texture in the phase diagram was associated, through a bulk-boundary correspondence with edge modes of an unusual variety, which we termed `estranged'. We now study the edge modes present in the two-dimensional ascendant with open boundaries. We will consider the system on a finite cylinder, retaining periodic boundary conditions in one direction, $x_1$, but terminating in the $x_2$ direction, as shown in \cref{fig:obc_2dRiceMele}, and we assume an even number of lattice sites in each direction. This introduces two disjoint 1d edges, one at each end.  To understand the structure of the edge modes, let us consider the limit of $\delta_2 = 1$. This decouples the Hilbert space that lives at the circular ends of the cylinder with a width of one lattice point. The effective Hamiltonian of the boundary Hilbert space is nothing but the one-dimensional Rice-Mele model $\Ham_{\text{RM}}$, with parameters $J,~\delta = \delta_1,~\mu$. 

Let us begin with $\mu=0$. We know that $\Ham_{\text{RM}}$ is gapless when $J=\delta_1=0$, which tells us that its higher dimensional ascendant $\Ham^{[2]}_{\text{RM}}$ has gapless boundary modes for $J=\delta_1=0$ and $\delta_2=1$, described by a free massless Dirac fermion. Even as we move away from $\delta_2 = 1$, along $0<\delta_2<1$, we expect these boundary modes to persist on the line $J = \delta_1 =0$ over a Hilbert space that is localized on the edges of the cylinder, but with a larger width. The edge modes terminate on the bulk gapless point on the origin $J=\delta_1 = \delta_2 = 0$. The resulting phase diagram is shown in \cref{fig:2dRiceMele_mu}(a). For the 1d Rice-Mele model, the edge modes were obtained by tuning one parameter, whereas for the 2d model, at $\mu=0$, we see that we need to fine-tune two parameters to access the edge modes. This picture changes significantly for $\mu \neq 0$. 

Indeed for non-zero $\mu$, the chemical potential is raised for both the bulk \emph{and boundary}. To see the consequence of this, let us again begin at the $\delta_2=1$ parametric plane, where we have the Rice-Mele model at finite chemical potential. This has an extended region of gapless states in the circular disc $\delta_1^2 + J^2 < \mu^2$ described by a massless Dirac fermion. This disc is expected to extend along $0<\delta_2\leq 1$ and terminates on the equator of the bulk metallic phase at the origin, as shown in \cref{fig:2dRiceMele_mu}(b).  We see that unlike the Rice-Mele model in 1d, the edge modes of the 2d ascendant are parametrically {stable} for generic $\mu \neq 0$, requiring no fine-tuning whatsoever. This is reminiscent of strong topological free-fermion phases, such as the quantum spin Hall insulator \cite{Kane_Mele} which are characterized by similarly stable edge modes. In fact, the edge modes of 2d QSH systems have the same boundary degrees of freedom i.e. a massless Dirac fermion! Unlike the QSH phase, whose \emph{topologically protected} edge modes can only be eliminated by a bulk gap closure, the edge modes of the 2d Rice-Mele ascendant can disappear either by a bulk transition (when it meets the bulk metal near the origin of the phase diagram) or a boundary Lifshitz transition (which completely empties or fills up the boundary bands) as seen in \cref{fig:2dRiceMele_mu}(a).

\subsection{Stability to band deformations}
\begin{figure}[!h]
    \centering
    \makebox[\linewidth][c]{\includegraphics[width=\textwidth]{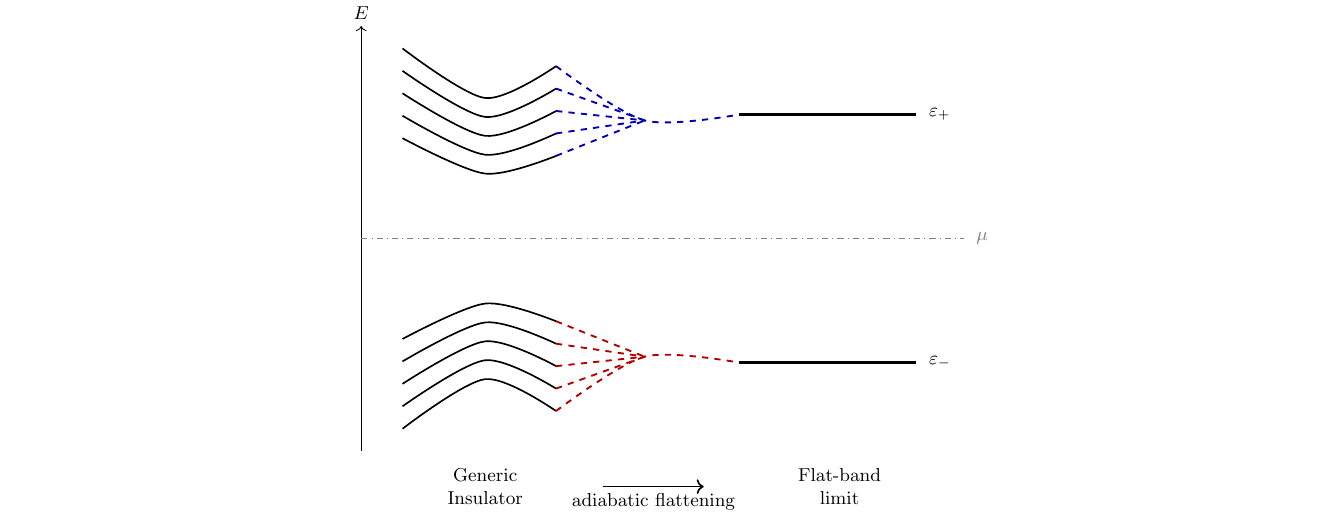}}
    \caption{Adiabatic deformation to a flat-band limit.}
    \label{fig:flatband}
\end{figure}
We have shown that the free-fermion Hamiltonian family in \cref{eq:H_RiceMele_2d,eq:h_fq_RiceMele_2d} belongs to a non-trivial  class quantified by the second Chern number over the {unified} parameter and momentum space. However, the model in \cref{eq:H_RiceMele_2d,eq:h_fq_RiceMele_2d} has an additional structure---the filled bands are exactly degenerate. One may wonder if this degeneracy, and the accidental symmetries protecting it, are necessary to preserve the integer invariant and texture in the phase diagram. The answer is no. The only symmetry that needs to be preserved is the $\uone$ particle conservation since this symmetry gives meaning to the charge from which all the textures ascend.  All other symmetries can be safely broken~\footnote{Lattice translations allow us to use band theory but even this can be broken weakly. An analysis of this case would require more sophisticated tools~\protect\cite{KITAEV20062,RealSpaceSecondChern}}. In particular, we can introduce the following perturbation on the single particle Hamiltonian, \cref{eq:h_fq_RiceMele_2d}, which can split the degeneracy of the filled bands,
\begin{align}
    \delta h = m_s \left(\sigma^1 \otimes \sigma^2 \right). \label{eq:singlet_mass}
\end{align}
The phase diagram of this perturbed system can again be analyzed easily by studying its bands. We set $\mu=0$ for simplicity.  So long as we are sufficiently far from the origin, $J^2 + \delta_1^2 + \delta_2^2 > m_s^2$, the many-body ground states are adiabatically connected to the original $m_s = 0$ system, except that the filled bands characterizing them are no longer degenerate. How then do we infer the topological nature of the ground state family? As explained in \cite{QiZhang_topologicalresponse_PhysRevB.78.195424}, as we continuously deform away from the limit of exact degeneracy, as long as there is no level crossing at the Fermi level, the topological nature of the ground states remains the same. In particular, to compute and analyze the non-abelian Berry connection in \cref{eq:NonAbelian_Berry}, we do not need the filled levels to be degenerate.

Conversely, consider a generic Bloch Hamiltonian family $\h(\vec{\lambda},k)$ with eigenstates and eigenvalues $\{ \ket{\varepsilon_a(\vec{\lambda},\vec{k})}, \varepsilon_a(\vec{\lambda},\vec{k}) \}$. So long as this represents a family of gapped states with no partially filled bands, we can always deform this system to one described by a spectrally flattened Bloch Hamiltonian,
\begin{align}
    \q(\vec{\lambda},\vec{k}) = \sum_{a } \frac{\varepsilon_a(\vec{\lambda},\vec{k})}{|\varepsilon_a(\vec{\lambda},\vec{k})|}  \outerproduct{\varepsilon_a(\vec{\lambda},\vec{k})}{\varepsilon_a(\vec{\lambda},\vec{k})}\ ,
\end{align}
and perform the analysis as in the unperturbed, degenerate case. This is schematically shown in Fig.~\ref{fig:flatband}. 

\subsection{Field theory and proximate phase diagrams}
\begin{figure}[!h]
    \centering
    \subfloat[Phase diagram of the 2d Rice-Mele ascendant with singlet mass]{\makebox[.8\linewidth][c]{\includegraphics[width=\textwidth]{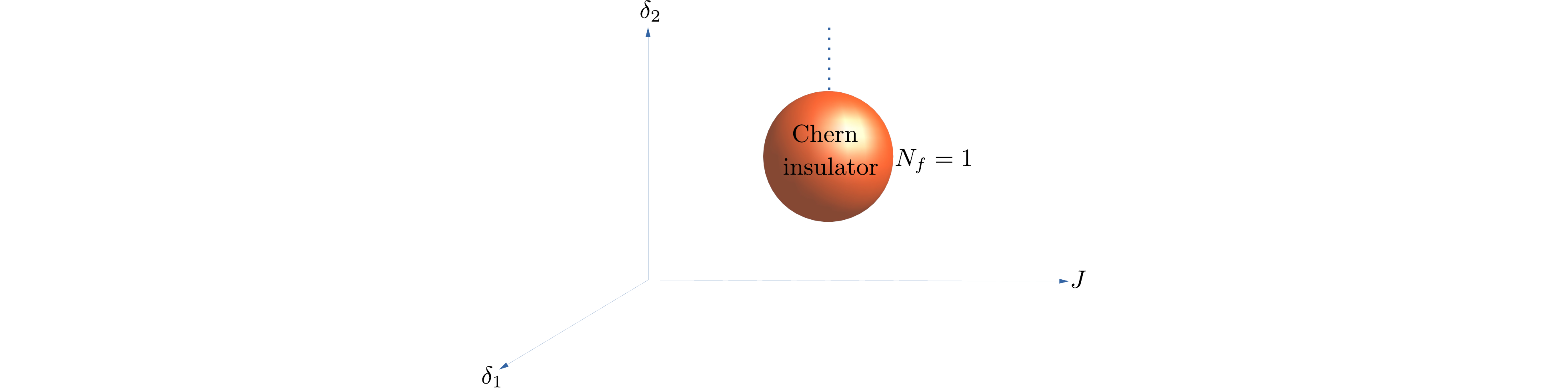}}}\\
    \subfloat[$J=\delta_1=\delta_2=0$]{\makebox[0.49\linewidth][c]{\includegraphics[width=\textwidth]{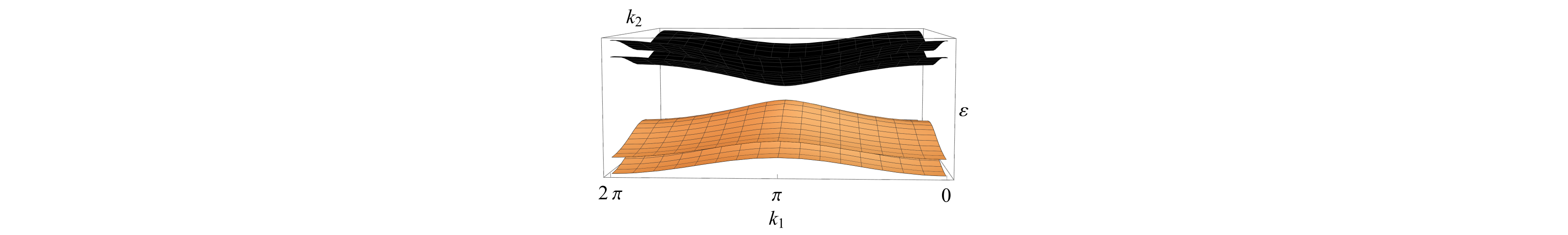}}}
    \subfloat[$J^2+\delta_1^2+\delta_2^2=m_s^2$]{\makebox[0.49\linewidth][c]{\includegraphics[width=\textwidth]{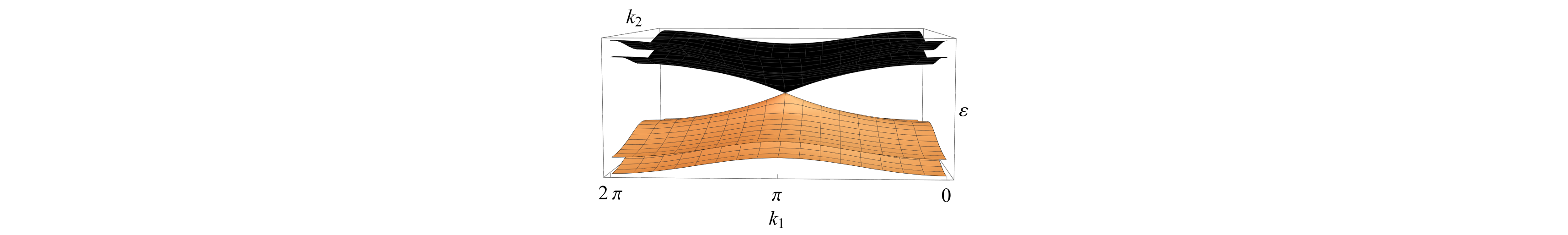}}}
    \caption{(a) The phase diagram of the 2d Rice-Mele ascendant upon the addition of the singlet mass term $m_s$ shown in \cref{eq:singlet_mass}. This gaps out the $N_f=2$ diabolical point and produces a Chern insulator (a,b) [$N_f$ counts the number of flavours of fermion at the gapless point]. As we go radially outward in the phase diagram, the system transitions back to a trivial insulator with two filled bands, through an $N_f=1$ massless Dirac fermion along $\delta_1^2 + \delta_2^2 + J^2 = m_s^2$ (c). }
    \label{fig:2dRiceMele_singletmass}
\end{figure}
We now analyze the Rice-Mele ascendant using a continuum field theory formulation to show that the non-trivial phase diagram topology is stable. We will only focus on bulk properties. Consider the Hamiltonian in \cref{eq:H_RiceMele_2d,eq:h_fq_RiceMele_2d} with $\mu=0$ taken for simplicity (the extension to $\mu\neq 0$ is straightforward). By performing a small-wavelength expansion about $(k_1,k_2)=(\pi,\pi)$, we can approximate the nearby Hamiltonian using a continuum field theory formulation. i.e.,
\begin{align}
    \Ham &\approx \int \rmd^2x~ \Psi^\dagger(\vec{x}) \h(\vec{x}) \Psi(\vec{x}), \label{eq:2d_HRM_Continuum} \\
    \h(\vec{x}) &= \left(-i \Gamma_2 \partial_{x_1} - i \Gamma_4 \partial_{x_2} - \delta_1 \Gamma_1 - \delta_2 \Gamma_3 + J \Gamma_5 \right), \nonumber
\end{align}
where  $\Psi$ are 4-component spinors and the $\Gamma_k$ are as in \cref{eq:Gamma_2d_RM}. It is illuminating to write the Lagrangian density of \cref{eq:2d_HRM_Continuum} in relativistic notation~\cite{Abanov_2000,HsinKapustinThorngren_PhysRevB.102.245113} 
\begin{align}
    \mathcal{L} = - \sum_{a=1}^2\bar{\psi}_a \slashed{\partial} \psi_a -  \sum_{a,b=1}^2 \bar{\psi}_a \left(\vec{m}\cdot\vec{\tau}_{ab} \right) \psi_b\ , \label{eq:Nf2_Dirac}
\end{align}
where $\psi_{a}$ are each two-component irreducible Dirac spinors for $a=1,2$. We have defined
\begin{align}
    \vec{m} = \left(\delta_1,\delta_2,J \right)
\end{align}
and chosen a basis such that
\begin{align}
    &\gamma_0 = \sigma^1 \otimes \sigma^2,\gamma_1= \sigma^2 \otimes \mathbb{1},\gamma_2 =-\sigma^3 \otimes \sigma^2, \nonumber\\
    & \tau^1  = -\sigma^2 \otimes \sigma^3,\tau^2  = \mathbb{1} \otimes \sigma^2,\tau^3  = -\sigma^2 \otimes \sigma^1\ .
\end{align}

\cref{eq:Nf2_Dirac} represents $N_f = 2$ flavours of 2+1 dimensional Dirac fermions perturbed by the so-called triplet mass term \cite{fradkin2013field,HsinKapustinThorngren_PhysRevB.102.245113}.  At first glance, \cref{eq:Nf2_Dirac} appears fine-tuned---it is known that an additional relevant singlet mass of the form $m_s \sum_a \bar{\psi}_a \psi_a$, with the same scaling dimensions as the triplet mass could be added~\cite{HsinKapustinThorngren_PhysRevB.102.245113}, resulting in a four-dimensional space of relevant coupling constants that produces a unique vacuum. Prima facie, this seems to suggest that the gapless diabolical point can be gapped out and the non-trivial family, if any, is spanned by four relevant parameters, contradicting the suspension construction. However, this is not the case as the triplet and singlet masses are not independent. At the single-particle level,  the singlet mass commutes with all triplet masses. Thus, the space of \emph{independent} maximally anticommuting relevant masses is three-dimensional, resulting in a three-parameter texture. {At $\vec{m}=0$, an emergent $\mathrm{SU}(2)$ flavor symmetry rotates the three triplet masses $(\delta_1,\delta_2,J)$~\cite{manjunath2026searchdiabolicalcriticalpoints}. The singlet mass is invariant under this symmetry. }

But what is the effect of the singlet mass on the phase diagram? Clearly, the gapless point at the origin would be eliminated. To see what state results in its place, and how the surrounding gapped states are affected, it is easiest to study a lattice version of the singlet mass. This, in fact, was already presented in \cref{eq:singlet_mass}. In addition to lifting the degeneracy of bands as explained above, this term also gaps out the theory at the origin $\delta_1=\delta_2=J=0$  shown in \cref{eq:Nf2_Dirac} . For $|m_s|<1$ and $\mu=0$, this results in two non-degenerate valence bands but one of them, closest to the Fermi level, {carries non-zero Chern number}, resulting in a bubble of Chern insulating state as shown in \cref{fig:2dRiceMele_singletmass}(a). The opposite Chern number is carried by the lowest conduction band.  As we increase $r = \sqrt{\delta_1^2 + \delta_2^2 + J^2}$, the two Chern bands again meet at $r = |m_s|$, resulting in a gapless single-flavor $N_f=1$ massless Dirac theory, and open into two trivial bands, reproducing the trivial phase with the texture as shown in \cref{fig:2dRiceMele_singletmass}(b,c). Introducing chemical potential $\mu \neq 0$ can lead to the partial population of various bands, introducing various metallic phases into \cref{fig:2dRiceMele_singletmass}. We do not discuss this here. Finally, interactions are irrelevant and do not change the nature of the field theory. Thus, \emph{any} symmetric microscopic perturbations merely changes the bare parameters of the field theory preserving the phase diagram in \cref{fig:2dRiceMele_singletmass}.

We end with comments on related work that offer complementary insight into the discussion of this section. Ref~\cite{KapustinSpodyneiko2020higherdimensionalgeneralizationsthoulesscharge} discusses higher dimensional generalizations of the Thouless pump using quantized non-linear response analogous to the spirit of the original Thouless pump~\cite{Thouless83}. They also present an alternative microscopic model to \cref{eq:H_RiceMele_2d}. Ref~\cite{HsinKapustinThorngren_PhysRevB.102.245113}  provides a field theory analysis of \cref{eq:Nf2_Dirac} perturbed by the singlet mass that is complementary to our discussion.

\section{Rice-Mele ascendants in arbitrary dimensions}
\label{sec:ddim_RiceMele}
We now summarize the essential features of of the Rice-Mele ascendants in arbitrary dimensions. The general Hamiltonians are shown in \cref{eq:Hd_RM}. 
Similar to \cref{eq:H_RiceMele_2d}, we can write
\begin{align}
    \Ham^{[d]}_{\text{RM}} = \sum_{\substack{\vec{x} \in \bZ^d\\a=1\ldots d }} t^a_{\vec{x}} ~c^\dagger(\vec{x}) c(\vec{x}+ \hat{e}_a) + h.c. + \sum_{\vec{x} \in \bZ^d}\mu_{\vec{x}} n(\vec{x}) \ .\label{eq:Hd_RM}
\end{align}
Here, the staggered hopping amplitudes $t^{a}_{\vec{x}}$ for $a=1,\ldots,d$ and chemical potential $\mu_{\vec{x}}$ are defined as
\begin{align}
    t^{a}_{\vec{x}} &= (-1)^{\sum_{k=a+1}^d x_k} \left(\frac{1+(-1)^{x_a}\delta_a}{2}\right),~a=1,\ldots,d-1 \nonumber\\
        t^{d}_{\vec{x}} &= \left(\frac{1+(-1)^{x_d}\delta_d}{2}\right),~
    \mu_{\vec{x}} = \mu + (-1)^{\sum_{k=1}^d x_k} J\ .
\end{align}
The model has a two-site translation invariance in each spatial direction as well as a $\uone$ symmetry as shown in \cref{eq:RiceMele_symmetries}. Choosing a suitable $2^d$ site unit cell, we get a $2^d$ band model with Bloch Hamiltonian 
\begin{align}
    \h(\vec{k}) = \mu \mathbb{1} + \sum_{m=1}^{2d+1} g_m(\vec{k}) \Gamma^m \ ,\label{eq:Hd_RM_firstquant}
\end{align}
where for a particular choice of unit cell, we have
\begin{align}
    g_{2m-1}(\vec{k}) &= \frac{(1-\delta_m)}{2} + \frac{(1+\delta_m)}{2} \cos(k_m), \nonumber\\
    g_{2m}(\vec{k}) &=  \frac{(1+\delta_m)}{2} \sin(k_m),~m=1,\ldots,d, \nonumber\\
    g_{2d+1}(\vec{k}) &= (-1)^d J \ .\label{eq:Hd_ddim_firstquant_gvec}
\end{align}
The $\{ \Gamma^m\}$ represent the $d$ dimensional irrep of the Clifford algebra $\mathrm{Cl}(2d+1)$, satisfying~\cref{eq:Clifford}, whose matrix representation for the same unit cell choice is given by
\begin{align}
    \Gamma^{2m+1} &= \underbrace{\sigma^3 \otimes \sigma^3 \cdots \otimes \sigma^3}_{\text{$d-m$ terms}} \otimes \sigma^1 \otimes \underbrace{\mathbb{1} \otimes \mathbb{1} \otimes \cdots \otimes \mathbb{1}}_{\text{$m-1$ terms}} \nonumber\\
        \Gamma^{2m} &= \underbrace{\sigma^3 \otimes \sigma^3 \cdots \otimes \sigma^3}_{\text{$d-m$ terms}} \otimes \sigma^2 \otimes \underbrace{\mathbb{1} \otimes \mathbb{1} \otimes \cdots \otimes \mathbb{1}}_{\text{$m-1$ terms}} \nonumber\\
        \Gamma^{2d+1} &= \underbrace{\sigma^3 \otimes \sigma^3 \cdots \otimes \sigma^3}_{\text{$d$ terms}},\qquad ~m=1,\ldots,d\ .
\end{align}
The model has two bands each of which is $2^{d-1}$ fold degenerate with eigenvalues
\begin{eqnarray}
    \varepsilon_{\pm}(\vec{k}) = \mu \pm |g(\vec
    k)|,~|g(\vec{k})| = \sqrt{\sum_{m=1}^{2d+1} |g_m(\vec{k})|^2} \ .\label{eq:Hd_RM_singleparticle_energies}
\end{eqnarray}

For simplicity, we consider only $\mu=0$. It can be easily checked from \cref{eq:Hd_RM_singleparticle_energies} that for any of $J,\delta_1,\delta_2,\ldots,\delta_d$ being non-zero, the system is a gapped insulator with $2^d$ filled bands. Any family of states parametrized by a sphere surrounding the origin $J= \delta_1 = \ldots=\delta_d =0$ is topologically non-trivial. 

To study this, note that the $2^d$ filled bands can be used to construct a $U(2^d)$ non-abelian Berry connection of the form shown in \cref{eq:NonAbelian_Berry}, where  $a=1,\ldots,2^d$ labels the filled bands and $\xi_a$ represents the collective coordinates for $\cM_{2d} = S_{\text{par}}^d \times T_{\text{BZ}}^d$, where $S_{\text{par}}^d$ is the parametric $d$-sphere surrounding the origin and $T^d$ is the Brillouin zone. We can define the Chern number $\Ch_d$ by
\begin{align}
    \Ch_{d} = \frac{1}{d!}  \int_{\cM_{2d}} \tr \left( \overbrace{\frac{F}{2\pi} \wedge \frac{F}{2\pi} \wedge \cdots \wedge \frac{F}{2\pi}}^\text{$d$ terms} \right)\ . \label{eq:Chern number ddim}
\end{align} 
We expect this to evaluate to $\Ch_d = 1$ for the band structure of \cref{eq:Hd_RM_firstquant} signaling the non-trivial nature of the family through a quantized invariant. In \cref{eq:Chern number ddim} and henceforth, we use the notation of differential forms~\cite{nakahara2018geometry} to simplify expressions,
\begin{align}
    A &= A_\mu(\xi)\, d\xi^\mu,  \text{  }\mu=1,\dots,2d\ , \nonumber\\
      G \wedge H &=   \epsilon^{\mu_1 \ldots \mu_k  \ldots \mu_D} G_{\mu_1 \ldots \mu_k} H_{\mu_{k+1} \ldots \mu_D} \ ,\nonumber\\  
    \rmd K &= \epsilon^{\mu_1  \ldots  \mu_{D}} \partial_{\mu_1} K_{\mu_2 \ldots \mu_D}, \nonumber \\
     F &= \rmd A + A \wedge A 
    = \tfrac{1}{2} F_{\mu\nu}\, \rmd x^\mu \wedge \rmd x^\nu\ .
\end{align}

 Analogous to \cref{eq:C2_Pontryagin,eq:C2_Pontryagin_Geometric}, for the first quantized Hamiltonian shown in \cref{eq:Hd_RM_firstquant,eq:Hd_ddim_firstquant_gvec},  we can define the $2d+1$ component unit vector  $\hat{g}_m  = g_m / |g| $ which defines a map
\begin{align}
    \hat{g}: \cM_{2d}   \rightarrow S^{2d} \label{eq:ghat_map}\ .
\end{align}
The expression for \cref{eq:Chern number ddim} reduces to the calculation of the winding of the map in \cref{eq:ghat_map} 

\begin{align}
\Ch_{d} = \frac{\epsilon^{a_1 a_2 \cdots a_{2d+1}}}{\mathcal{V}(S^{2d})} \int_{\cM_{2d}}  \hat{g}_{a_1} \wedge \rmd\hat{g}_{a_2}\cdots \wedge     \rmd\hat{g}_{a_{2d+1}}\ , \label{eq:Pontryagin ddim}
\end{align}
where $\mathcal{V}(S^k)$ denotes the volume of the $k$ unit sphere,
\begin{align}
    \mathcal{V}(S^k) = \frac{2 \pi^{k/2}}{\Gamma\left(k/2\right)}\ .
\end{align}
The non-trivial nature of the family of states can also be determined through a topological texture visible by studying the higher Berry geometric invariant, $\check{\gamma}$. This is defined for any closed $d-1$ dimensional surface, such as $S^{d-1}$, on the parameter space. Let $\cM_{2d-1} = S^{d-1} \times T^d_{\text{BZ}}$ and let $\bar{\cM}_{2d}$ be a manifold such that its boundary contains $\cM_{2d-1}$ i.e., $\partial \bar{\cM}_{2d} = \cM_{2d-1}$. $\check{\gamma}$ is defined using the integral in \cref{eq:Chern number ddim} evaluated on $\bar{\cM}_{2d}$ as 
\begin{multline}
  \check{\gamma} =  \frac{2\pi}{d!}  \int_{\bar{\cM}_{2d}} \tr \left( \overbrace{\frac{F}{2\pi} \wedge \frac{F}{2\pi} \wedge \cdots \wedge \frac{F}{2\pi}}^\text{$d$ terms} \right) \\= \int_{\cM_{2d-1}} \CS_{2d-1}\ , \label{eq:Higher_Berry_CS2d-1}
\end{multline}
where we have used Stokes' theorem to rewrite $\check{\gamma}$ as an integral over the Chern-Simons form $\CS_{2d-1}$. The exact expressions for $\CS_{2d-1}$ varies with $d$. We have already encountered the form for $d=1,2$. For $d=3$, we have
\begin{multline}
    \CS_5 = \frac{1}{3!} \frac{1}{(2\pi)^3} \tr\biggl( A \wedge F \wedge F + \frac{3}{2} A \wedge A \wedge A \wedge F 
    \\+ \frac{3}{5} A \wedge A \wedge A \wedge A \wedge A  \biggr).
\end{multline}
Written in terms of $\hat{g}$, the expression for $\check{\gamma}$ reduces to 
\begin{align}
    \check{\gamma} = \frac{2\pi}{\mathcal{V}(S^{2d})} \int_{\bar{\cM}_{2d}} \epsilon^{a_1 a_2 \cdots a_{2d+1}}  \hat{g}_{a_1} \wedge \rmd\hat{g}_{a_2}\cdots \wedge     \rmd\hat{g}_{a_{2d+1}}\ . \label{eq:Higher_Berry_ddim_ghat}
\end{align}

Although direct visualization in higher dimensions is not possible, the higher Berry phases evaluated using \cref{eq:Higher_Berry_CS2d-1,eq:Higher_Berry_ddim_ghat} can be used to reveal the topological textures present in the phase diagrams following strategies analogous to those used in one and two spatial dimensions. 

In \cref{sec:RiceMele,sec:2d_RiceMele,sec:ddim_RiceMele}, we have considered a series of models hosting topological families of quantum states and textured phase diagrams in various spatial dimensions that ascend from the one-dimensional Rice-Mele model. In the upcoming sections, we consider two different series of models with textured phase diagrams for Class A systems. 

\section{Berry's model and its ascendants} 
\label{sec:Berry}
\begin{figure}[!h]
    \centering
    \includegraphics[width=\linewidth]{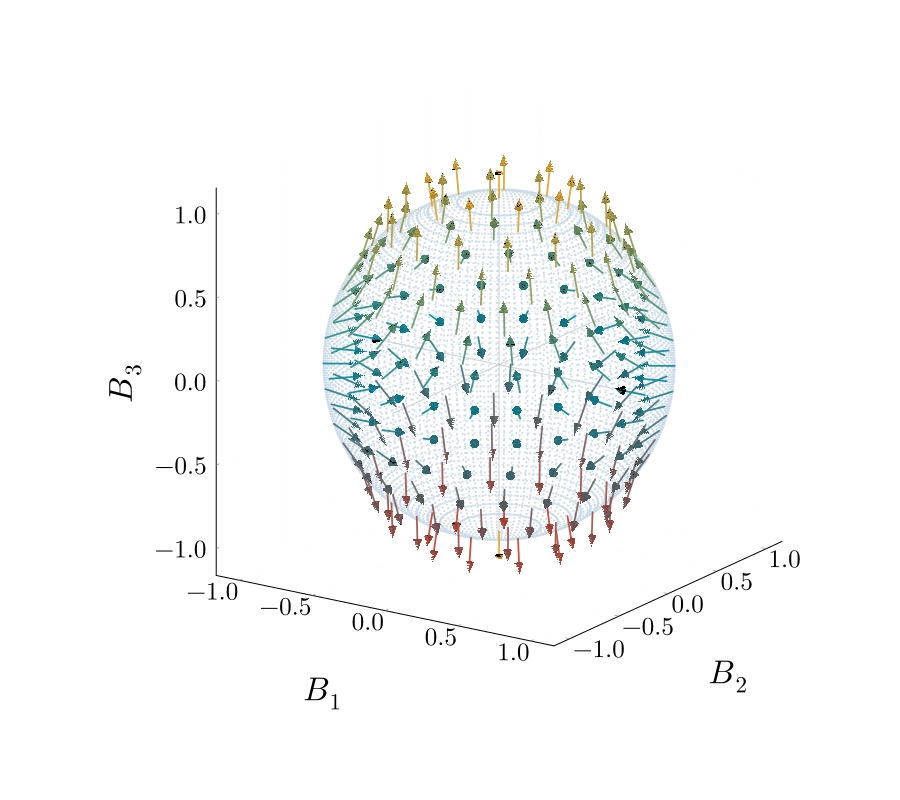}
    \caption{Texture on $S^2$ for Berry's model of a spin in a magnetic field which is plotted by computing the Berry phase for each latitude.}
    \label{fig:berrytexture}
\end{figure}
In this section, we study a  class of topological textures of charge conserving fermions that ascend from Berry's problem~\cite{BerryOG} of a quantum spin in magnetic field. We review this model in the context of our previous sections---textured phase diagrams---and proceed to produce higher dimensional examples of free-fermion models with a texture using the suspension construction. 

\subsection{Quantum spin in a magnetic field}
We begin with a review of the classic example considered by Berry~\cite{BerryOG}: a quantum spin half moment in a magnetic field with the following Hamiltonian
\begin{align}
	\h_{\text{Berry}} =  \vec{B}\cdot \vec{\sigma} \label{eq:H_Berry_QM} \ .
\end{align}
$\vec{B} = (B_1, B_2, B_3)$ is a magnetic field with three components that produces a three-dimensional phase diagram for \cref{eq:H_Berry_QM}, characterized by a unique ground state away from the origin $\vec{B} \neq (0,0,0)$. It is well known that the family of states living on any 2-sphere surrounding the origin is topologically non-trivial. We show that this also results in a textured phase diagram. To see this, we again consider foliating the three-dimensional phase diagram spanned by $\vec{B}$ by spheres. To study the non-trivial topological nature of states living on a single unit 2-sphere surrounding the origin $\snpar{2}$, we construct the Berry connection and Chern number,
\begin{align}
    A^\mu(\xi)  = -i \innerproduct{\varepsilon(\xi)|\frac{\partial}{\partial \xi_\mu}}{\varepsilon(\xi)}, \qquad  \Ch_1 = \frac{1}{2\pi} \int_{S^2} F\ ,
\end{align}
where $\xi_\mu$ are coordinates representing $\snpar{2}$, and verify that $\Ch_1 = 1$. We can also determine the non-trivial texture over the sphere by tracking Berry's geometric phase. To do this, we use the strategy of \cref{sec:2d_RiceMele} and split the 2-sphere into circles at different latitudes, labeled by $B_3$. For a circle $\snpar{1}$ at a fixed $B_3$, we can easily compute Berry's geometric phase $\gamma$ exactly as half the solid angle subtended by it, i.e.
\begin{align}
    \gamma(B_3) = \int_{\snpar{1}} A = \pi (1-B_3) \ ,
\end{align}
and use this to extract a texture over $\snpar{2}$ using the recipe shown in \cref{fig:Higher_Berry_visualize}. This results in \cref{fig:berrytexture}, giving an alternative view of the non-trivial nature of the quantum spin in magnetic field.

\subsection{1d Berry ascendant and multi-texture phase diagrams}
\begin{figure}[!h]
    \centering
    \makebox[0.49\linewidth][c]{\includegraphics[width=\textwidth]{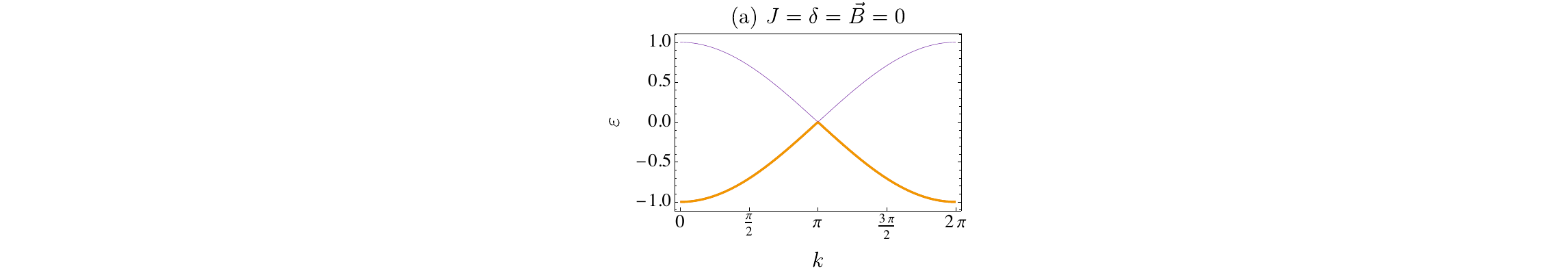}}
    \makebox[0.49\linewidth][c]{\includegraphics[width=\textwidth]{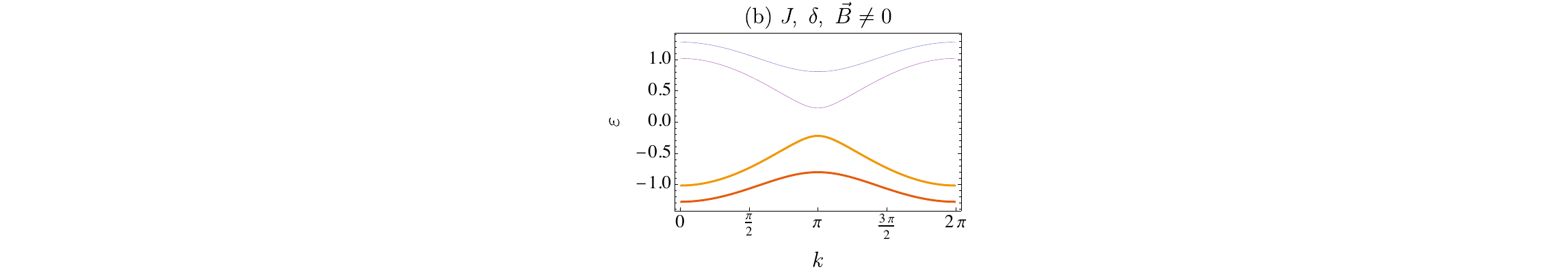}}
    \makebox[0.49\linewidth][c]{\includegraphics[width=\textwidth]{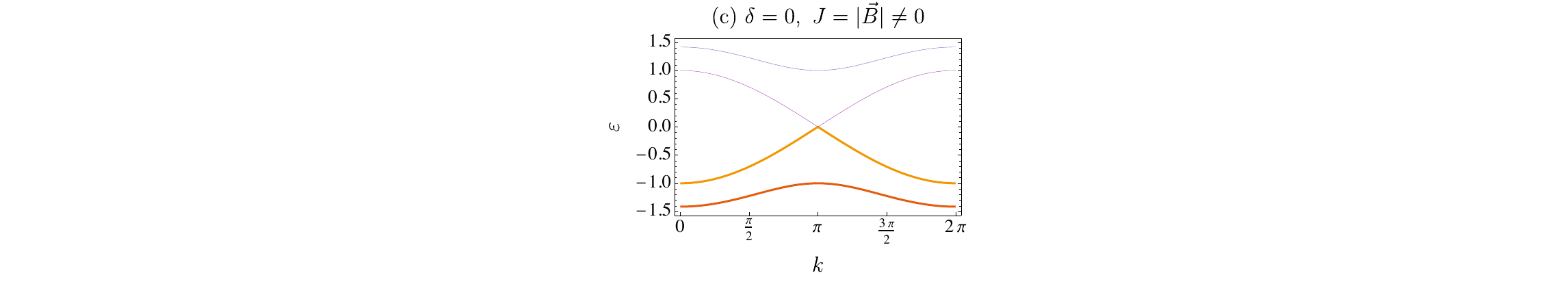}}
    \makebox[0.49\linewidth][c]{\includegraphics[width=\textwidth]{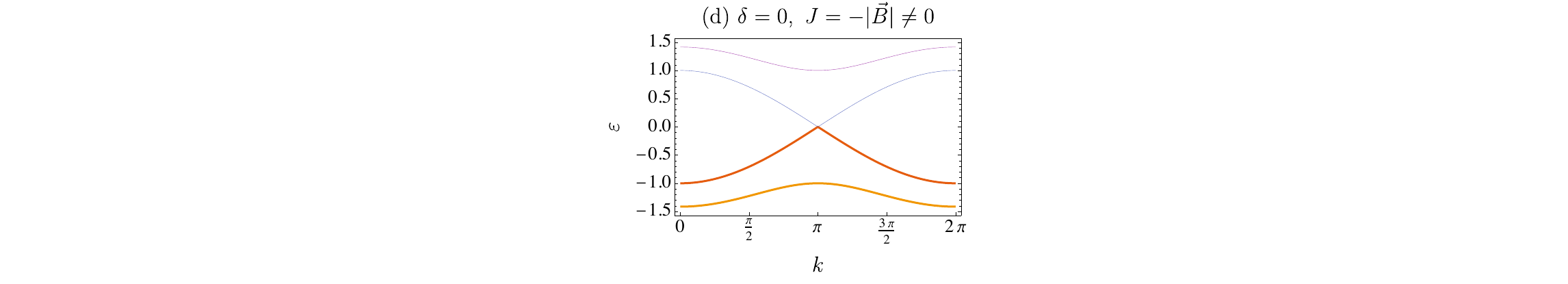}}
    \caption{Single-particle Bloch energies of the 1d Berry ascendant \eqref{eq:H_Berry_1d} with $\mu=0$ and perturbed by a staggered chemical potential \eqref{eq:H_Berry_1dplusRiceMele}. For no staggered potential $J=0$, the diabolical point at $\vec{B}=\delta=0$ has two degenerate bands touching to produce an $N_f=2$ 1d massless Dirac fermion (a). For generic values of $\delta,\vec{B},J$, we have four distinct bands, with two filled (b). For $J\neq 0$, the diabolical point becomes a codimension-2 surface $\delta = 0, |\vec{B}| = |J|$ where we have a single band touching resulting in an $N_f=1$ massless Dirac fermion (c,d).}
    \label{fig:Berry_1d_bands}
\end{figure}

We now want to construct ascendants of Berry's model in higher dimensions using the suspension construction. However, a direct application of suspension to the spin model in \cref{eq:H_Berry_QM} results in higher dimensional spin models as well. These have already been studied in previous work~\cite{Wenetal_topologicalfamilies,qi2025chartingspacegroundstates,SommerWenVishwanath_HigherMPS1_PhysRevLett.134.146601}. The difficulty with such spin models is that they are naturally interacting and require more sophisticated tools to analyze. We will seek a more tractable route in this work, and construct non-interacting fermion ascendants of Berry's model in higher dimensions. We will comment on the connection to interacting models later. 

We begin by fermionizing Berry's model through second-quantization as
\begin{align}
	\Ham_{\text{Berry}} = \sum_{\alpha, \beta \in \{\uparrow,\downarrow\}} c^\dagger_\alpha (\vec{B}\cdot \vec\sigma_{\alpha \beta}) c_\beta + \mu n\ , \label{eq:Berry_fermion} 
\end{align}
where we have defined the number operator in the usual way, $ n = \sum_{\alpha \in \{\uparrow,\downarrow\}}c^\dagger_\alpha c_\alpha$. The original spin Hamiltonian in \cref{eq:H_Berry_QM} is now the first-quantized Hamiltonian for \cref{eq:Berry_fermion}. \cref{eq:Berry_fermion} introduces some additional complexity in the phase diagram in comparison to \cref{eq:H_Berry_QM}. For $|\vec{B}|>|\mu|$, the fermionic ground state of \cref{eq:Berry_fermion} corresponds to occupying the ground state of \cref{eq:H_Berry_QM}. Since the latter forms a topologically non-trivial family, so does the former. For $|\vec{B}|<|\mu|$, depending on the sign of $\mu,$ either both or no eigenstates of \cref{eq:H_Berry_QM} are occupied and we have a trivial family. For $|\vec{B}|=|\mu|$, the system has a zero-mode degeneracy, representing a diabolical locus. 

 Let us now construct ascendants of \cref{eq:Berry_fermion}. First, it is easy to verify that $\Ham_{\text{Berry},\text{inv}} = -\Ham_{\text{Berry}}$. Using the suspension construction, we have
 \begin{multline}
     \HamdBerry{1} = \sum_{x \in \bZ} (-1)^x \Ham_{\text{Berry,x}}|_{\mu=0} + \mu \sum_x n(x) \\
     + \sum_{x \in \bZ} \sum_{\alpha \in \{\uparrow,\downarrow\}} \left(\frac{1+(-1)^x \delta}{2} \right) c^\dagger_\alpha (x) c_\alpha(x)\ .
 \end{multline}
 {A closely related one-dimensional free-fermion representative, with a different parametrization was also recently studied in Ref.\cite{LoWenBoundaryScattering}.} Expanding out $\Ham_{\text{Berry}}$, we get the following Hamiltonian form for the $d=1$ ascendant,
\begin{multline}
    \HamdBerry{1} =  \sum_{\substack{x \in \bZ}} \biggl[  \frac{(1+ (-1)^x \delta )}{2}  \sum_{\alpha \in \{\uparrow,\downarrow\}}c_\alpha^\dagger(x) c_\alpha(x+1) + h.c. \\
    +(-1)^x\sum_{\alpha,\beta \in \{\uparrow,\downarrow\}} c^\dagger_\alpha({x}) (\vec{B}\cdot\vec{\sigma}_{\alpha \beta}) c_\beta({x}) + \mu~  n(x) \biggr] \ .\label{eq:H_Berry_1d}
\end{multline}
This can be represented by a 4-band Bloch Hamiltonian,
\begin{align}
  \h({k}) = \mu \mathbb{1} + \sum_{m=1}^{5} g_m({k}) \Gamma^m \ ,\label{eq:H1_Berry_firstquant}
\end{align}
where for a particular choice of unit cell, 
\begin{multline}
    g_{a} = -B_a,~a=1,2,3\\
    g_{4} = \frac{(1-\delta )}{2} + \frac{(1+\delta)}{2} \cos k, ~
    g_{5} = \frac{(1+\delta)}{2} \sin k\ ,
    \end{multline}
    and $\Gamma^a$ form the $4$ dimensional irreducible representations of the Clifford algebra $\mathrm{Cl}(5)$ with matrix representation,
\begin{multline}
    \Gamma_{1,2,3} = {\sigma^3 } \otimes \sigma^{1,2,3},~
    \Gamma_{4} =   \sigma^1 \otimes \mathbb{1},
    \Gamma_{5} = \sigma^2 \otimes \mathbb{1}\ . \label{eq:1d_Berry_Gammas}
\end{multline}
The two single-particle energy bands, each two-fold degenerate, are
\begin{eqnarray}
    \varepsilon({k}) = \mu \pm |g(
    k)|,~|g({k})| = \sqrt{\sum_{m=1}^{5} |g_m({k})|^2} \ .\label{eq:H1_Berry_singleparticle_energies}
\end{eqnarray}

\begin{figure}[!h]
     \centering
\subfloat[$J=0$]{    \makebox[0.3\linewidth][c]{\includegraphics[width=\textwidth,valign=c]{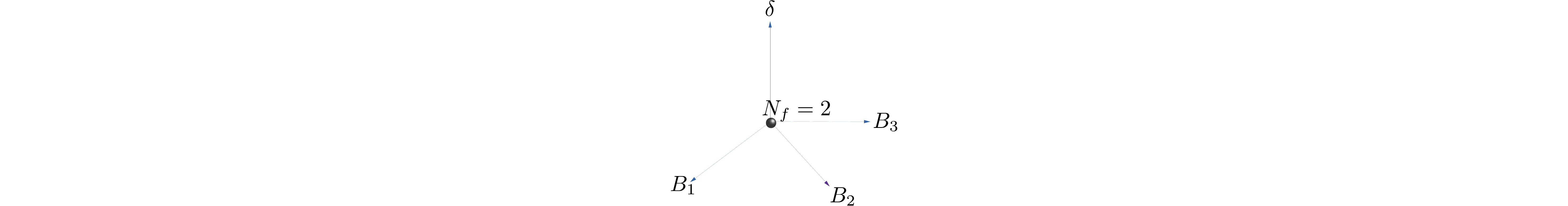}}}
     \subfloat[$J\neq 0$]{ \makebox[0.3\linewidth][c]{\includegraphics[width=\textwidth,valign=c]{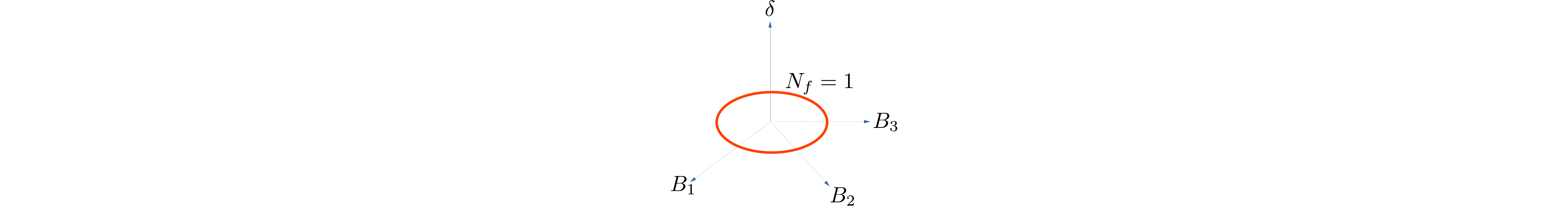}}}
     \subfloat[]{ \makebox[0.3\linewidth][c]{\includegraphics[width=\textwidth,valign=c]{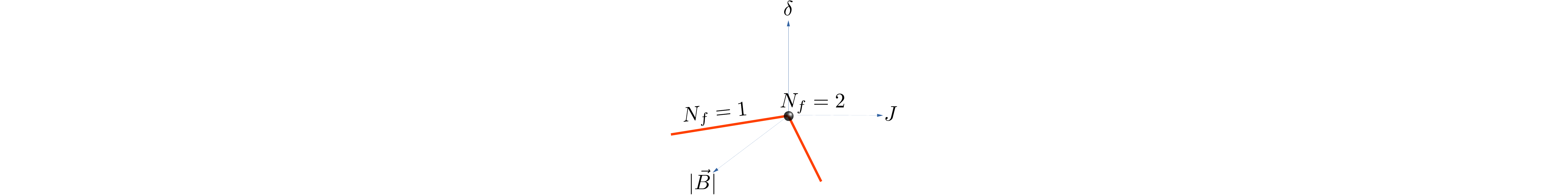}}}
     \caption{Evolution of the diabolical point of the $\mu=0$ 1d Berry ascendant \eqref{eq:H_Berry_1d} with the inclusion of a staggered chemical potential \eqref{eq:H_Berry_1dplusRiceMele} of strength $J$. For $J=0$, the phase diagram contains a codimension-4 diabolical point with $N_f=2$ massless Dirac fermions (a) which transforms into a codimension-2 $N_f=1$ massless Dirac fermion along $|\vec{B}|= |J|,~\delta=0$ (b). In the $J-\delta$ plane, $\vec{B}=0$ represents two copies of the Rice-Mele model with a single $N_f=2$ diabolical point. For $\vec{B}\neq 0$, this splits into two $N_f=1$ gapless points along $J = \pm |\vec
     B|$ (c).}
     \label{fig:Berry_1d}
 \end{figure}
Consistent with the general picture, we see from \cref{fig:Berry_1d_bands} that the system is gapped for all $|\mu|<r<1$, where $r = \sqrt{\vec{B}\cdot\vec{B}+ \delta^2}$, with the degenerate valence bands fully filled. For simplicity, we will assume $\mu=0$ for the rest of this subsection unless stated otherwise. \cref{fig:Berry_1d}(a) shows the phase diagram. For $\delta=\vec{J}=0$, the degenerate bands touch as shown in \cref{fig:Berry_1d_bands}(a) resulting in a diabolical point described by an $N_f=2$ flavour massless 1+1d Dirac fermion. Any parametric $S^3_{\text{par}}$ surrounding the origin for $|\mu|<r<1$ hosts a non-trivial family. This can be verified by taking the non-abelian Berry-Bloch connection, $A$, defined for the two filled valence bands, and evaluating the second Chern number $\Ch_2$ over $\cM_4 = S^3_{\text{par}} \times \sonebz$,
\begin{align}
    \Ch_2 = \frac{1}{8\pi^2} \int_{S^3_{\text{par}} \times \sonebz} \tr\left( F \wedge F \right)\ . \label{eq:C_2_Berry1d}
\end{align}

A texture in the four-dimensional  $\{ \vec{B},\delta\}$ phase diagram can be exposed by foliating it with 3-spheres, splitting each of those into various 2-spheres $S^2_{\text{par}}$ at fixed latitudes and computing a geometric invariant by integrating the Chern-Simons 3-form $\CS_3$ over $\cM_3 = S^2_{\text{par}} \times \sonebz$,
\begin{align}
    \check{\gamma} = \frac{1}{4\pi} \int_{S^2_{\text{par}} \times \sonebz} \tr\left(A \wedge dA +\frac{2}{3} A\wedge A \wedge A \right)\ . \label{eq:CS_Berry_1d}
\end{align}

\begin{figure}[!h]
     \centering
         \subfloat[$\vec{B}=0,~\mu=0$]{ \makebox[.5\linewidth][c]{\includegraphics[width=\textwidth,valign=c]{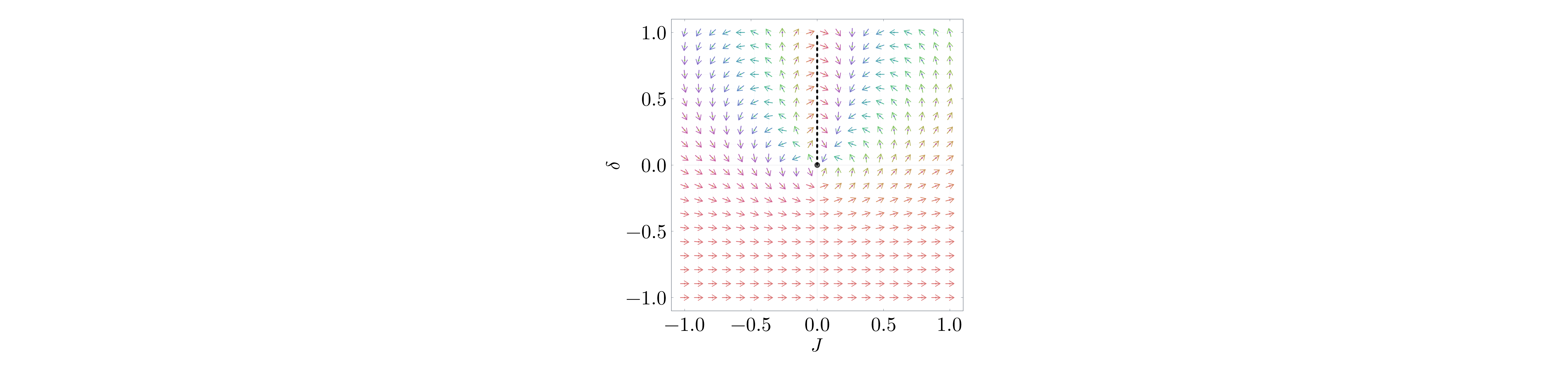}}}
     \subfloat[$\vec{B}\neq 0,~\mu=0$]{ \makebox[0.5\linewidth][c]{\includegraphics[width=\textwidth,valign=c]{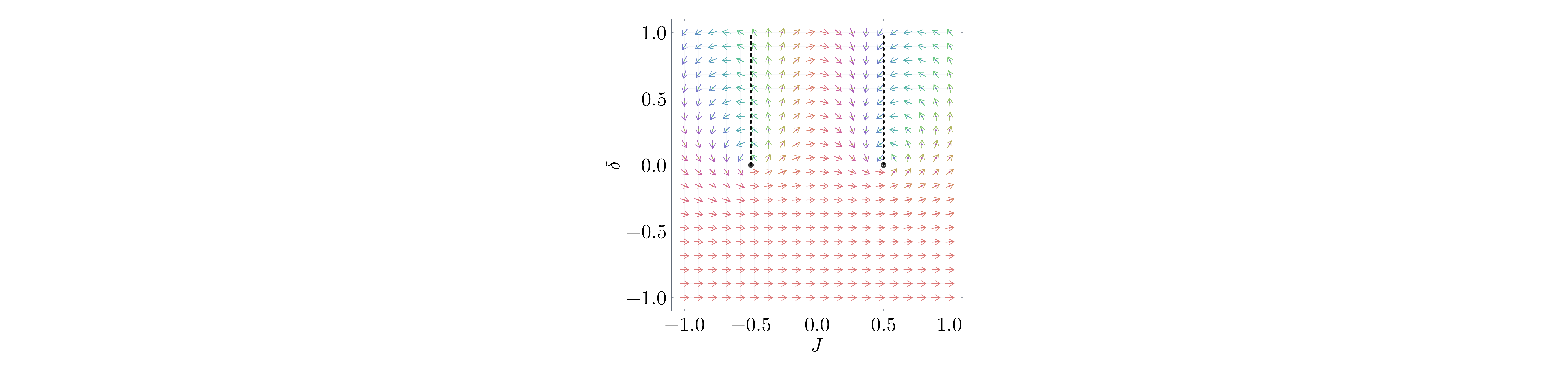}}}\\
      \subfloat[$\vec{B}=0,~\mu \neq 0$]{ \makebox[.5\linewidth][c]{\includegraphics[width=\textwidth,valign=c]{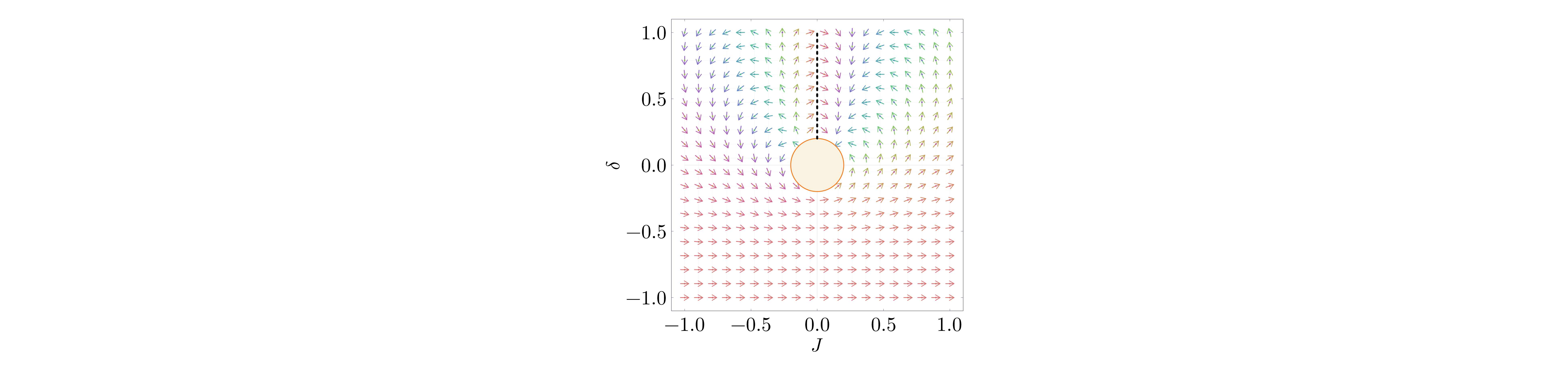}}}
     \subfloat[$\vec{B}\neq 0,~\mu \neq 0$]{ \makebox[0.5\linewidth][c]{\includegraphics[width=\textwidth,valign=c]{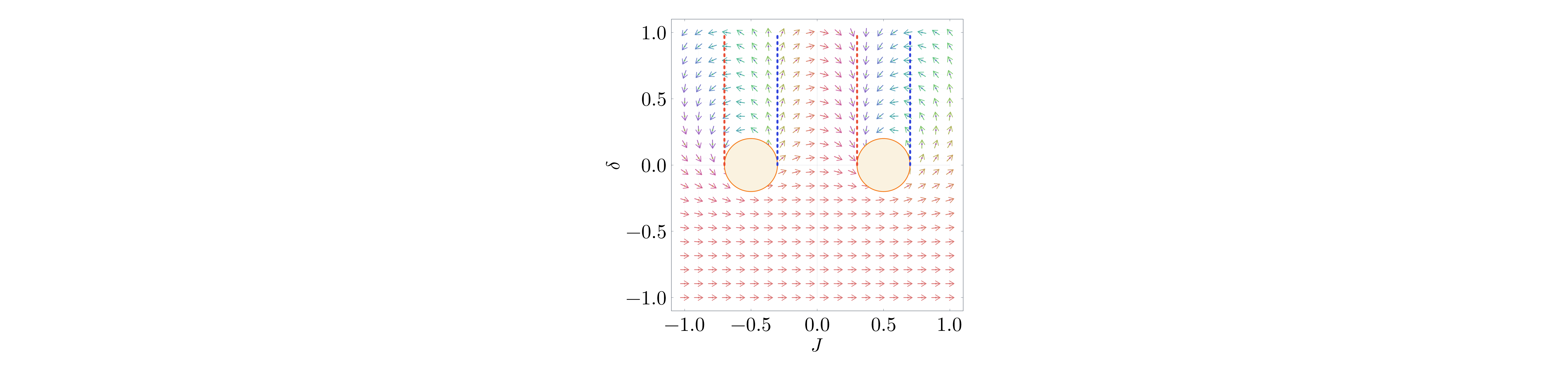}}}
     \caption{Rice-Mele textures in the Berry ascendant perturbed by staggered chemical potential \eqref{eq:H_Berry_1dplusRiceMele}. For $\vec{B}=0$ (a,c), the texture along the $J-\delta$ phase diagram emerges from two disconnected copies of the Rice-Mele model resulting in a charge 2 vortex. (b,d): For $\vec{B}\neq 0$, the charge-2 vortex core splits into charge-1 vortices located at $\delta=0,~J = \pm |\vec{B}|$. For $\mu = 0,$ the core of the vortex is point-like, described by relativistic $N_f = 2$ (a) and $N_f=1$ (b)  Dirac fermions. For small but finite $\mu$, the vortex cores are replaced by metals, with partially filled bands (c,d). Edge modes are represented by broken lines and terminate at the vortex core. For $\mu \neq 0, J\neq 0$ (d), the edge modes are estranged similar to the Rice-Mele model.}
     \label{fig:Berry_1d_texture}
 \end{figure}
We have so far seen two textures in 1d systems: (1) the two parameter phase diagram of the Rice-Mele model, exposed by computing the Berry phase $\gamma$ and (2) the four parameter phase diagram of the Berry ascendant, exposed by the higher Berry phase $\check{\gamma}$ shown in \cref{eq:CS_Berry_1d}. Can the two co-exist? We will now show that they can indeed. This phase diagram can be obtained by perturbing the 1d Berry ascendant, \cref{eq:H_Berry_1d} with a staggered chemical potential.
\begin{align}
    \Ham_\text{multi} &=  \HamdBerry{1} + J \sum_x  (-1)^x n(x)\ , \nonumber\\
    \h_{\text{multi}}(k) &=  \mu \mathbb{1} + \sum_{m=1}^{5} g_m({k}) \Gamma^m - J (\sigma^3 \otimes \mathbb{1})\ .
    \label{eq:H_Berry_1dplusRiceMele}
\end{align}

 To see that this model contains the Rice-Mele texture, let us begin by setting $\vec{B} = 0$. The Hamiltonian reduces to two disconnected copies of the Rice-Mele model
\begin{align}
    \Ham_\text{multi} \xrightarrow[]{\vec{B}=0} \sum_{\alpha \in \{\uparrow,\downarrow\}} \Ham_{\text{RM},\alpha} \ .
\end{align}
The texture resulting in the $J-\delta$ plane can be exposed by studying the sum of Berry phases for each band, for fixed $J,~\delta$.  This can be expressed in terms of the non-abelian Berry connection for the two filled bands used in \cref{eq:CS_Berry_1d} as
\begin{align}
    \gamma = \int_{\sonebz} \tr(A)\ . \label{eq:Berry_Berry1d}
\end{align}
As shown in \cref{fig:Berry_1d_texture}(a,c), this results in a double vortex with a single core. \cref{eq:Berry_Berry1d} applies even away from the $\vec{B}=0$ limit, and the same $\un{2}$ non-abelian Berry connection can be used to expose both classes of textures arising from the Rice-Mele and Berry ascendants.  For $\vec{B} \neq 0$, the band degeneracy is lifted, as shown in \cref{fig:Berry_1d_bands}(b,c,d), with eigenvalues
\begin{align}
    \varepsilon(k)  =\mu \pm \sqrt{\cos^2\left(\frac{k}{2}\right) + \delta^2 \sin^2\left(\frac{k}{2}\right)+(|\vec{B}| + \xi  |J|)^2 }\ ,\nonumber
    \\ \xi \in \pm 1\ .
\end{align}
The four-band gap closure is replaced by two-band closure at $|\vec{B}| = \pm J,~\delta=0$ as shown in \cref{fig:Berry_1d_bands}(c,d). This results in a gapless $N_f=1$ flavour massless Dirac fermion. The codimension-4 $N_f=2$ gap closing point at $\vec{B}=0,~\delta=0$ now changes to a codimension-2 surface, $|\vec{B}| = |J|,~\delta=0$, as shown in \cref{fig:Berry_1d}(b). In the $J-\delta$ planes, the gap-closing point at the origin, $J=\delta = 0$ splits into two, at $J = \pm |\vec{B}|$ as shown in \cref{fig:Berry_1d}(c). Consequently, the core of the Thouless-pump vortex texture also splits into two, resulting in two separate vortices as shown in \cref{fig:Berry_1d_texture}(b,d). For $\mu=0$, the vortex core is described by $N_f = 2$ ($\vec{B}=0$) and $N_f = 1$ ($\vec{B} \neq 0$) massless relativistic Dirac fermions. Let us now introduce a small but finite chemical potential $0<|\mu|\ll1$. This partially populates the Dirac fermion bands and changes the nature of the diabolical points at the vortex core to metallic phases. For larger $|\mu|$, these metallic phases can merge as various bands get populated leading to a complex picture. We will not consider those here. 

Finally, a comment on edge modes. For $\mu = J= 0$, let us consider open boundary conditions where the system is labeled $x = 1, \ldots, L$ with $L$ being even so that the system begins on an odd site and ends on an even site. For $\delta = 1,~\vec{B}=0$, this exposes decoupled edge modes on both ends which are expected to survive in the range $\delta \in (0,1],~\vec{B}=0$ and terminate on the bulk diabolical point at $\delta = \vec{B}=0$ as  shown in \cref{fig:Berry_1d_texture}(a).  Let us now consider the more general situtaion of $J,\mu \neq 0$. For $\delta = 1$, the effective boundary energy levels on the left and right edges are, respectively,
\begin{align}
    \varepsilon_{\text{left}} = \mu -J \pm |\vec{B}|,~\varepsilon_{\text{right}} = \mu +J \pm |\vec{B}|\ .
\end{align}
We see that zero-modes on both edges occur simultaneously for $|\vec{B}| = |J|$ when $\mu = 0$ (\cref{fig:Berry_1d_texture}(b,c)) and  $|\vec{B}| = |\mu|$ when $J = 0$. When both $\mu \neq 0$ and $J \neq 0$ simultaneously, the edge modes are estranged and occur on  $|\vec{B}| = |\mu \pm J|$ on the right and left edges respectively (\cref{fig:Berry_1d_texture}(d)). We expect this picture to hold more generally for $1>\delta>0$ and the edge modes to terminate on the metallic phase near the origin of the parameter space. In summary, for $J = \mu =0$, the edge modes form a codimension 3 hypersurface. When either $\mu \neq 0$ or $J \neq 0$, they generically form a codimension 1 hypersurface and are estranged when   both $\mu \neq 0$ and $J \neq 0$. These edge modes terminate on the metallic states at the vortex cores,  and can also undergo complex restructuring as the metallic states merge with increase in chemical potential. 

The entire discussion can be reproduced using effective field theory. Let us comment on this briefly. A suitable starting point is the limit of small $|\vec{B}|,|\delta|,|J|$ where the system can be described by $N_f=2$ Dirac fermions with various mass terms
\begin{align}
    \Ham &\approx \int \rmd x ~\Psi^\dagger(x) \h(x) \Psi(x)\ , \nonumber\\
    \h(x) &= -i \Gamma_5 \partial_x  - \sum_{a=1}^3B_a \Gamma_a 
    -\delta~ \Gamma_4 +J (i\Gamma_4 \Gamma_5)\  , \label{eq:H_Berry_1d_continuum}
\end{align}
where $\Psi$ are 4-component spinors and the $\Gamma_j$ are as defined in \cref{eq:1d_Berry_Gammas}. The Lagrangian density can be written as two irreducible Dirac fermions as follows
\begin{multline}
    \cL = -\sum_{a=1}^2 \bar{\psi}_a \slashed{\partial} \psi_a  -\delta \sum_{a=1}^2 \bar{\psi}_a  \psi_a + i J \sum_{a=1}^2\bar{\psi}_a \gamma_c   \psi_a \\- i \sum_{a,b=1}^2 (\vec{B}\cdot\vec{\tau}_{ab}) \left(\bar{\psi}_a  \gamma_c  \psi_b\right)\ , \label{eq:H_Berry_1d_continuum_Lagrangian}
\end{multline}
where $\gamma_0 = \sigma^1,~\gamma_1 = \sigma^3,~\gamma_c = \sigma^2.$ 
There are precisely eight relevant quadratic perturbations to the $N_f = 2$ massless Dirac fermion theory, of which five are present in \cref{eq:H_Berry_1d_continuum_Lagrangian}. Of these, four are maximally anticommuting, coupled to $\delta, \vec{B}$ and appear in the suspension construction, spanning the space of independent relevant operators and generating the non-trivial texture. {At $J=0$, these four mass operators are related by an emergent symmetry at the gapless point. To see this explicitly, define six generators,} 
\begin{align}
    L_{ab}=\frac{i}{2}\Gamma_a \Gamma_b\ ,\qquad 1\leq a<b\leq4\ .
    \label{eq:Berry_Spin4_generators}
\end{align}
{These commute with the massless kinetic Hamiltonian and satisfy}
\begin{align}
    [L_{ab},\Gamma_c]=i\left(\delta_{bc}\Gamma_a-\delta_{ac}\Gamma_b\right)\ .
\end{align}
{They therefore generate a $\mathrm{Spin}(4)$ symmetry of the massless continuum theory and induce $\mathrm{SO}(4)$ rotations of the mass vector $(B_1,B_2,B_3,\delta)$ which realizes the mass-rotation symmetry discussed in Ref.\cite{manjunath2026searchdiabolicalcriticalpoints}. In the representation \cref{eq:1d_Berry_Gammas},}
\begin{align}
    L_{ij}&=-\frac{1}{2}\epsilon_{ijk}(\mathbb{1}\otimes\sigma^k)\ ,
    &&1\leq i<j\leq3\ ,\nonumber\\
    L_{i4}&=-\frac{1}{2}(\sigma^2\otimes\sigma^i)\ ,
    &&i=1,2,3\ .
\end{align}
{The $L_{ij}$ generate the physical spin $\mathrm{SU}(2)$, which is an exact lattice symmetry at $\vec{B}=0$, whereas the $L_{i4}$ mix $\delta$ with the components of $\vec{B}$. The latter rotations are emergent: Writing $q=k-\pi$, the leading omitted term is proportional to $q^2\Gamma_4$; it breaks the $L_{i4}$ rotations but is irrelevant at the Dirac fixed point. Equivalently, $\mathrm{Spin}(4)\simeq \mathrm{SU}(2)_L\times\mathrm{SU}(2)_R$, with the physical spin rotation as its diagonal subgroup.}

The term coupled to $J$ commutes with the ones coupled to $\vec{B}$ and changes the nature of the diabolical points, as we have seen. An analysis of \cref{eq:H_Berry_1d_continuum_Lagrangian} for $J=0$ can be found in Ref~\cite{HsinKapustinThorngren_PhysRevB.102.245113}. We can also include three more relevant terms, 
\begin{align}
    \delta \cL =\sum_{a,b=1} \bar{\psi}_a (\vec{m}\cdot\vec{\tau}_{ab}) \psi_b\ ,
\end{align}
which, on the lattice, corresponds to the following change to the Bloch Hamiltonian 
\begin{align}
    \delta \h = \vec{m}\cdot(\sigma^1 \otimes \vec{\sigma} )\ .
\end{align}
This is likely to lead to interesting modifications of the diabolical locus.  We will not explore these modifications further here.

\subsection{Ascendants of Berry's model in arbitrary dimensions}

 Using the suspension construction of \cref{sec:suspension}, we find Hamiltonians of Berry's ascendants in arbitrary dimensions. These are defined recursively as
 \begin{multline}
     \Ham^{[d]}_{\text{Berry}} =\mu \sum_{\vec{x} \in \bZ^d} n(\vec{x}) +  \sum_{x_d \in\bZ} (-1)^{x_d}\Ham^{[d-1]}_{\text{Berry},x_d}|_{\mu=0} \\+\sum_{\vec{x} \in \bZ^d} \sum_{\alpha \in \{\uparrow,\downarrow\}} \left( \frac{1+(-1)^{x_d}\delta_d}{2} \right) (c^\dagger_\alpha(\vec{x}) c_\alpha(\vec{x}+\hat{e}_d)+h.c)\ ,\nonumber
 \end{multline}
 similar to \cref{eq:Hd_RM}, we can expand out $\Ham^{[d-1]}_{\text{Berry},x_d}$ and write the Hamiltonians as follows
\begin{multline}
    \HamdBerry{d} = \sum_{\substack{\vec{x} \in \bZ^d,\\a=1\ldots d}}  t^a_{\vec{x}} \sum_{\alpha \in \{\uparrow,\downarrow\}}c_\alpha^\dagger(\vec{x}) c_\alpha(\vec{x}+ \hat{e}_a) + h.c. \\
    + \sum_{\vec{x} \in \bZ^d}\sum_{\alpha,\beta \in \{\uparrow,\downarrow\}} c^\dagger_\alpha(\vec{x}) \mu^{\alpha \beta}_{\vec{x}} c_\beta(\vec{x})\ ,  \label{eq:Hd_Berry}
\end{multline}
where 
\begin{align}
    &t^{a}_{\vec{x}} = (-1)^{\sum_{k=a+1}^d x_k} \left(\frac{1+(-1)^{x_a}\delta_a}{2}\right), a = 1,\ldots,d-1,\nonumber\\
    &t^{d}_{\vec{x}} = \left(\frac{1+(-1)^{x_d}\delta_d}{2}\right),\nonumber \\
    &\mu^{\alpha \beta}_{\vec{x}} = \mu \delta_{\alpha \beta} + (-1)^{\sum_{k=1}^d x_k} \vec{B}\cdot \vec\sigma_{\alpha \beta},~ \alpha,\beta \in \{\uparrow,\downarrow\}\ .    
\end{align}
By passing to momentum space, using a unit cell containing $2^{d+1}$ fermions (including spin) we get the $2^{d+1}$ band Bloch Hamiltonian,
\begin{align}
  \h(\vec{k}) = \mu \mathbb{1} + \sum_{m=1}^{2d+3} g_m(\vec{k}) \Gamma^m \ ,\label{eq:Hd_Berry_firstquant}
\end{align}
where for a particular choice of unit cell, 
\begin{align}
    g_{a} &= (-1)^d B_a,~a=1,2,3,\nonumber\\
    g_{2i+2} &= \frac{(1-\delta_{i} )}{2} + \frac{(1+\delta_{i} )}{2} \cos k_i, \nonumber\\
    g_{2i+3} &= \frac{(1+\delta_{i})}{2} \sin k_i, \text{ for } i=1,\ldots,d \ .
    \end{align}
$\Gamma^a$ form the $2^{d+1}$ dimensional irreducible representations of the Clifford algebra $\mathrm{Cl}(2d+3)$ with matrix representation, corresponding to the particular choice used above
\begin{align}
      \Gamma_{a} &=  \underbrace{\sigma^3 \otimes \cdots \otimes\sigma^3}_{\text{$d$ terms}} \otimes \sigma^{a},\nonumber\\
    \Gamma_{2m+2} &= \underbrace{\sigma^3 \otimes \sigma^3 \cdots\otimes \sigma^3}_{\text{$d-m$ terms}} \otimes  \sigma^1 \otimes \underbrace{\mathbb{1}\otimes \mathbb{1} \cdots \otimes \mathbb{1}}_{\text{$m$ terms}}, \nonumber\\
    \Gamma_{2m+3} &= \underbrace{\sigma^3 \otimes \sigma^3 \cdots\otimes \sigma^3}_{\text{$d-m$ terms}} \otimes  \sigma^2 \otimes \underbrace{\mathbb{1}\otimes \mathbb{1} \cdots \otimes \mathbb{1}}_{\text{$m$ terms}} \nonumber\\
    &\text{ for }a=1,2,3, ~m = 1,\ldots,d\ .
\end{align}
We have two single-particle energy bands 
\begin{eqnarray}
    \varepsilon(\vec{k}) = \mu \pm |g(\vec
    k)|,~|g(\vec{k})| = \sqrt{\sum_{m=1}^{2d+3} |g_m(\vec{k})|^2} \ ,\label{eq:Hd_Berry_singleparticle_energies}
\end{eqnarray}
each with degeneracy $2^d$. For fixed $\mu$, the Hamiltonians in \cref{eq:Hd_Berry} produce a $(d+3)$-dimensional phase diagram spanned by $\{B_a,\delta_m\}$ for $a=1,2,3$ and $m = 1,\ldots,d$. Let $r$ denote the radius from the origin of this phase diagram
\begin{align}
r = \left(\sqrt{\sum_{a=1}^{3} B_a^2 + \sum_{m=1}^{d} \delta_a^2}\right)\ . \label{eq:mu_condition_Berry}    
\end{align}
For $|\mu|<r<1$, the many-body ground state is an insulator that corresponds to filling exactly half the bands.  For $r<|\mu|$, we have a metal with partially filled bands.

From the suspension construction, the half-filled insulator generates a non-trivial topological family over any parametric $d+2$ sphere, $|\mu| < r < 1$ surrounding the origin. The non-triviality is inherited from the topological nature of Berry's model. As in the case of the Rice-Mele series, this can be diagnosed using topological invariants as well as phase-diagram textures. More precisely, we can define a non-Abelian $\un{2^d}$ Berry connection of the filled $2^d$ bands over parameter and momentum space. Consider the $2d+2$ dimensional manifold $\cM_{2d+2} = S_{\text{par}}^{d+2} \times \tnbz{d}$ composed of the parametric  $d+2$ sphere and $d$-dimensional Brillouin zone. The non-trivial family over $S_{\text{par}}^{d+2}$ can be verified by computing the Chern number $\Ch_{d+1}$, defined in \cref{eq:Chern number ddim} over $\cM_{2d+2}$, and verifying that $\Ch_{d+1} \neq 0$. Similarly, the phase-diagram topological texture can also be extracted for any $d+1$ sphere, $\snpar{d+1}$, living in a parameter space containing the half-filled insulating phase $|\mu| < r<1$. This is achieved by evaluating the geometric invariant $\check{\gamma}$ through integrating the Chern-Simons $2d+1$ form, $\CS_{2d+1}$ defined in \cref{eq:Higher_Berry_CS2d-1} over $\cM_{2d+1} =  S_{\text{par}}^{d+1} \times T^d$. Finally, expressions for both topological and geometric invariants can also be given using $\hat{g}_a = {g_a}/{|g|}$ as in \cref{eq:Pontryagin ddim,eq:Higher_Berry_ddim_ghat}. 

\subsection{Connection with Dixmier-Douady and other interacting invariants} 
Let us return to Berry's original spin model~\cref{eq:H_Berry_QM}. Instead of second-quantizing it as in \cref{eq:Berry_fermion}, we could have applied the suspension construction to it directly, and considered its ascendants in spin form. In 1d, the ascendant would take the form of a spin chain, with a Hamiltonian form  
\begin{align}
    \h^{[1]} =  \sum_{x \in \bZ} \left[\frac{(1+ (-1)^x \delta )}{2} \vec{\sigma}_x\cdot\vec{\sigma}_{x+1} +  (-1)^x \vec{B}\cdot\vec{\sigma}_x\right].  \label{eq:H_Berry_DD}
\end{align}
Such models have been constructed and studied in recent work~\cite{Wenetal_topologicalfamilies,qi2025chartingspacegroundstates,SommerWenVishwanath_HigherMPS1_PhysRevLett.134.146601}. Since \cref{eq:H_Berry_DD} is an interacting spin-chain model, characterizing its ground state family requires sophisticated physical and mathematical tools such as matrix product states (MPS)~\cite{qi2025chartingspacegroundstates,RyuHigherPhysRevB.109.115152,OhyamaRyu_HigherMPS1_PhysRevB.111.035121,OhyamaRyu_HigherMPS1_PhysRevB.111.035121,OhyamaRyu_HigherMPS2_PhysRevB.111.045112,SommerWenVishwanath_HigherMPS1_PhysRevLett.134.146601,SommerWenVishwanath_HigherMPS2_PhysRevB.111.155110,Shiozaki_Ohyama_MPS_HigherBerry},  $\mathrm{C}^*$ algebras~\cite{KapustinSpodyneikoo_HigherBerry_PhysRevB.101.235130,Wenetal_topologicalfamilies,SommerWenVishwanath_HigherMPS1_PhysRevLett.134.146601,SommerWenVishwanath_HigherMPS2_PhysRevB.111.155110} and higher gerbes~\cite{qi2025chartingspacegroundstates,RyuHigherPhysRevB.109.115152,OhyamaRyu_HigherMPS1_PhysRevB.111.035121,OhyamaRyu_HigherMPS1_PhysRevB.111.035121,OhyamaRyu_HigherMPS2_PhysRevB.111.045112}. In particular, the topological invariant that characterizes the non-trivial nature of the ground-state family of \cref{eq:H_Berry_DD} over $S^3_{\text{par}}$ is the so-called Dixmier-Douady invariant $\mathrm{DD}(S^3_{\text{par}}) \in \bZ$. 

{ This should be compared with the one-dimensional fermionic ascendant obtained in \cref{eq:H_Berry_1d}, whose ground state can be analyzed using band theory and fiber bundles. Its non-trivial family over $S^3_{\text{par}}$ is characterized by the second Chern invariant $\Ch_2(S^3_{\text{par}} \times \sonebz) \in \bZ$. The two constructions provide, respectively, interacting-spin and free-fermion representatives of the same suspended Berry series. We now explain the directly established relation between the band invariant of the free-fermion family and the many-body higher-Berry invariant of that same family.

We first note the relevant cohomology groups. Since the integral cohomology of $S^1$ and $T^d$ is torsion-free, the K\"{u}nneth formula for the products used below takes the form~\cite{Hatcher:478079}
\begin{multline}
     \Ham^\ell(\cM \times \cN,\bZ) \cong \bigoplus_{p+q=\ell}
     \Ham^p(\cM,\bZ) \otimes_{\bZ} \Ham^q(\cN,\bZ)\ .
     \label{eq:Kunneth}
\end{multline}

We will also use the following standard identities~\cite{Hatcher:478079}:
\begin{align}
  \Ham^1(S^1,\bZ) \cong \bZ,\qquad
  \mathcal{A}\otimes_{\bZ}\bZ \cong \mathcal{A}\ .
  \label{eq:tensor_Z}
\end{align}
Consequently, for a three-dimensional parameter manifold $\cM_3$,
\begin{align}
     \Ham^4(\cM_3 \times S^1,\bZ) \cong \Ham^3(\cM_3,\bZ)\ .
     \label{eq:Ch4_DD3}
\end{align}
Equation~\eqref{eq:Ch4_DD3} shows that the second Chern invariant and the Dixmier--Douady invariant take values in isomorphic cohomology groups. It does not, by itself, show that the two invariants are equal.

The classes can instead be compared directly using a calculation of Kapustin and Spodyneiko~\cite{KapustinSpodyneikoo_HigherBerry_PhysRevB.101.235130}. Let $E$ be the occupied Bloch bundle over $\cM_3\times S^1_{\mathrm{BZ}}$, and let $\pi:\cM_3\times S^1_{\mathrm{BZ}}\rightarrow\cM_3$ be the projection. For a translationally invariant one-dimensional free-fermion family, Appendix B of Ref.~\cite{KapustinSpodyneikoo_HigherBerry_PhysRevB.101.235130} shows that the many-body higher Berry curvature $\Omega^{(3)}$ obeys
\begin{align}
 \left[\frac{\Omega^{(3)}}{2\pi}\right]
 = \left[\pi_*\operatorname{ch}_2(E)\right]
 \in \Ham^3(\cM_3,\mathbb{R})\ ,
\end{align}
where $\pi_*$ denotes integration over the Brillouin zone and 
\begin{equation}
\operatorname{ch}_2(E)    = \frac{1}{8 \pi^2} \tr\left( F \wedge F \right)
\end{equation}
is the degree-four Chern character whose integral over a four manifold gives the second Chern number. Thus, using the same orientation and curvature conventions on both sides, for $\cM_3=S^3_{\mathrm{par}}$,
\begin{multline}
 \mathrm{DD}(S^3_{\text{par}})=\frac{1}{2\pi}\int_{S^3_{\mathrm{par}}}\Omega^{(3)}
 \\= \int_{S^3_{\mathrm{par}}\times S^1_{\mathrm{BZ}}}\operatorname{ch}_2(E)
 = \Ch_2(S^3_{\mathrm{par}}\times S^1_{\mathrm{BZ}})\ .
\end{multline}
With the conventions of \cref{eq:C_2_Berry1d}, the last integral is precisely the second Chern number used above. Since $\Ham^3(S^3,\bZ)\cong\bZ$ has no torsion, this integral completely determines the Dixmier--Douady class. The free-fermion second Chern number therefore equals the many-body higher-Berry, or Dixmier--Douady, invariant of the same family after its free-fermion structure is forgotten.

A similar comparison can be made in higher dimensions. In $d$ spatial dimensions, our free-fermion ascendant is characterized by the Chern number $\Ch_{d+1}$, evaluated over the $d$-dimensional Brillouin zone and a $(d+2)$-dimensional parameter space. Applying the K\"{u}nneth formula as above gives
\begin{align}
    \Ham^{2d+2}(\cM_{d+2} \times T^d,\bZ)
    \cong \Ham^{d+2}(\cM_{d+2},\bZ)\ .
\end{align}
The corresponding many-body higher-Berry invariant for a ground-state family over $\cM_{d+2}$, described by a higher gerbe, belongs to $\Ham^{d+2}(\cM_{d+2},\bZ)$~\cite{qi2025chartingspacegroundstates}. Thus, the K\"{u}nneth formula explains why the free-fermion Chern number and the many-body invariant can be related, but it does not prove that they are equal. Kapustin and Spodyneiko conjectured that, in arbitrary dimensions, the many-body invariant is obtained by integrating the degree-$(2d+2)$ Chern character over the Brillouin zone~\cite{KapustinSpodyneikoo_HigherBerry_PhysRevB.101.235130}. Our ascendants provide explicit lattice realizations of the free-fermion side of this conjecture: their Chern number $\Ch_{d+1}$ can be computed directly from the occupied bands. 
}

We conclude this section with a few general comments on the overall picture. Recall that the conjectured $\Omega$ spectrum nature of invertible states discussed in \cref{sec:Mathematical} tells us that 
\begin{equation}
    \pi_{0}(\sI_{d}) = \pi_{q}(\sI_{d+q})\ .
\end{equation}
In other words, non-trivial invertible phases in $d$ spatial dimensions result in non-trivial families over $S^q$ in $d+q$ spatial dimensions. The Rice-Mele series ascended from $d=0$ dimensional phases, corresponding to different charges. In the next section, we will consider a series ascending from the $d=2$ Chern insulator phase. These examples suggest that topological families can be labeled by invertible phases of matter in some lower dimensions. However, the Berry series ascended, not from any invertible phase, but from a non-trivial $S^2$ family for $d=0$. This gives rise to textures over $d+2$ spheres in parameter space, in $d$ spatial dimensions since 
\begin{equation}
    \pi_2(\sI_0) = \pi_{d+2}(\sI_d).
\end{equation}
This tells us that non-trivial families need not be associated with any phases. Together, the conjectural classification of topological families of invertible states can be completely determined from the knowledge of (i) invertible phases in all spatial dimensions $\pi_0(\sI_d)$ and (ii) all topological families  for $d=0$ spatial dimensions $\pi_q(\sI_0)$.

\section{The Qi-Wu-Zhang model, Chern insulator pump and ascendants}
\label{sec:QWZ}
\subsection{The Qi-Wu-Zhang model}
\begin{figure}[!h]
    \centering
    \makebox[.9\linewidth][c]{\includegraphics[width=\textwidth]{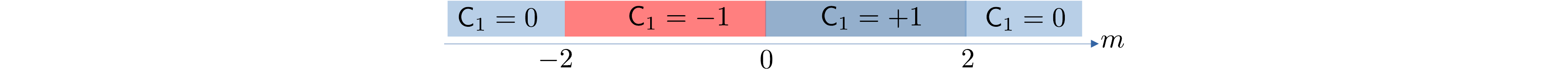}}
    \caption{Phase diagram of the Qi-Wu-Zhang model \eqref{eq:H_QWZ}. This exhibits three phases with the filled band exhibiting Chern number $\Ch_1=0,\pm 1$ in its occupied band. }
    \label{fig:QWZ_phasediagram}
\end{figure}

The final series of textured phase diagrams we will consider in this work ascends from the Chern insulator phase in $d=2$. Let us begin with a review of the Qi-Wu-Zhang model~\cite{QiWuZhang_PhysRevB.74.085308} which hosts various Chern insulator phases. The Hamiltonian is as follows:
\begin{multline}
    \Ham_{\text{QWZ}}=\sum_{\vec{x} \in \bZ^2} \sum_{\alpha,\beta \in \{\uparrow,\downarrow\} }\biggl[  m c_\alpha^\dagger(\vec{x})\sigma^3_{\alpha \beta} c_\beta(\vec{x})\\ - \frac{1}{2}\sum_{a=1,2}c^\dagger(\vec{x})\left({\sigma^3_{\alpha \beta} + i\sigma^a_{\alpha \beta}}\right)c(\vec{x}+\hat{e}_a)+h.c \biggr]\ . \label{eq:H_QWZ}
\end{multline}
The Bloch Hamiltonian corresponds to 
\begin{multline}
    \h(\vec{k}) = \vec{g}(\vec{k})\cdot\vec{\sigma},~\text{where }
    g_1= \sin k_1,\ g_2=\sin k_2, \\ 
    g_3=(m+\cos k_1 +\cos k_2 )\nonumber\ .~~~~
\end{multline}
The system consists of two energy bands with dispersion $\varepsilon_\pm(\vec{k}) = \pm |\vec{g}(\vec{k})|$ which are separated for all values of $m$ except at $m =0, \pm 2$. The phase corresponding to the many-body ground state can be determined by computing the Chern number for the abelian $\uone$ connection, $A$, constructed for the single filled band and evaluated over the Brillouin zone $\tnbz{2}$
\begin{align}
    \Ch_1 = \int_{\tnbz{2}} \frac{F}{2\pi} \ .
\end{align}
The resulting phase diagram, shown in \cref{fig:QWZ_phasediagram} consists of two Chern insulators with $\Ch_1 = \pm 1$ (for $m \in (0,\pm 2)$ respectively) and  a trivial insulator with $\Ch_1 = 0$ (for $|m|>2$).

\subsection{The 3d QWZ ascendant}
\begin{figure}[!h]
    \centering
    \makebox[0.49\linewidth][c]{\includegraphics[width=\textwidth]{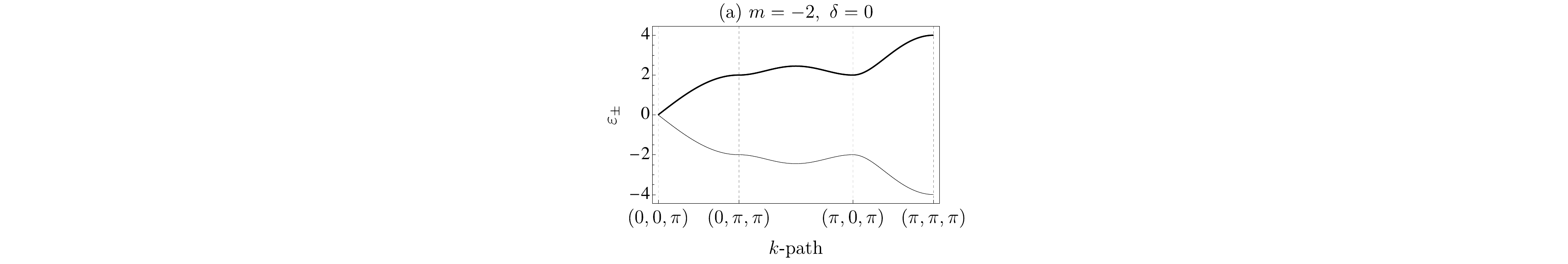}}
    \makebox[0.49\linewidth][c]{\includegraphics[width=\textwidth]{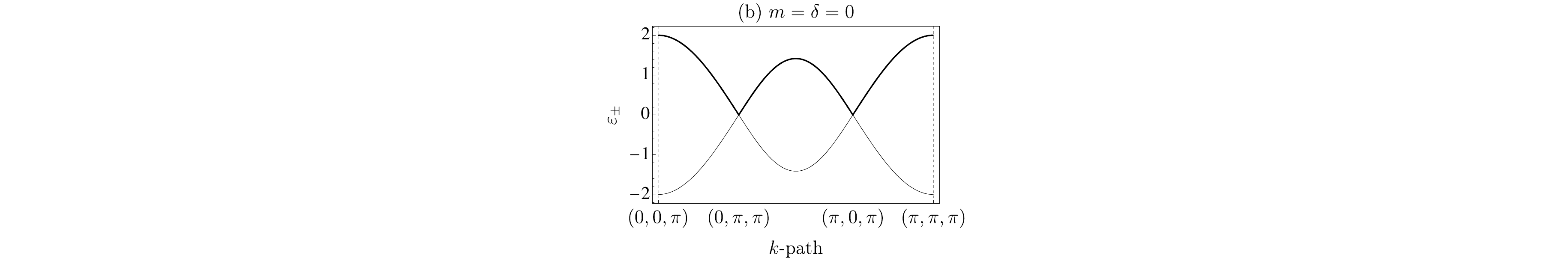}}
    \makebox[0.49\linewidth][c]{\includegraphics[width=\textwidth]{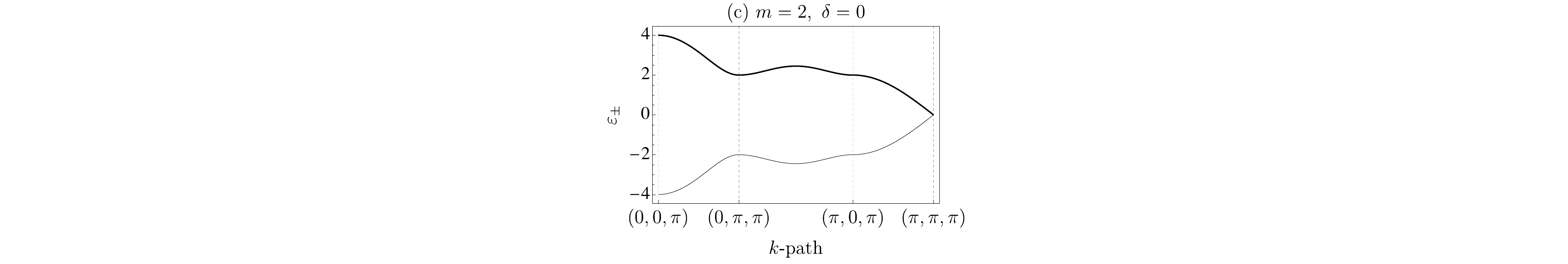}}
    \makebox[0.49\linewidth][c]{\includegraphics[width=\textwidth]{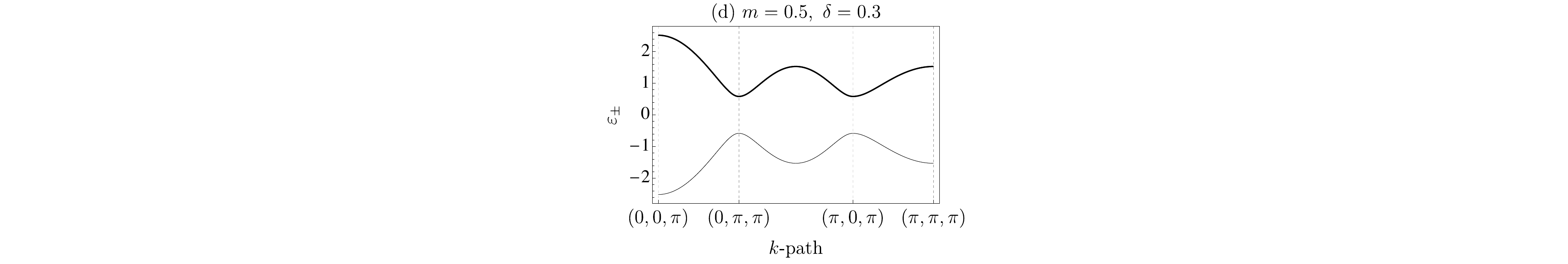}}
    \caption{Single particle band dispersions for the $d=3$ QWZ ascendant interpolating high symmetry points in the Brillouin zone. The system remains gapped for all values of $m,\delta,m$ except $m=0,\pm2,~\delta=0$. Each band is two-fold degenerate. }
    \label{fig:QWZ_3d_bands}
\end{figure}
We will now study the ascendants of the QWZ model using the suspension construction. First, we need to identify the inverse of $\Ham_{\text{QWZ}}$ for a fixed value of $m$. This turns out to be $-\Ham_{\text{QWZ}}$. Using the same arguments as in the Rice-Mele and Berry series, the $d=3$ dimensional ascendant of the QWZ model can be written as
\begin{multline}
\Ham^{[3]}_{\text{QWZ}}=\sum_{x_3\in\mathbb{Z}}(-1)^{x_3}\Ham_{\text{QWZ},x_3} + \mu \sum_{\vec{x} \in \bZ^3} n(x)\\+\sum_{\vec{x}\in \mathbb{Z}^3}\sum_{\alpha \in \{\uparrow,\downarrow \}}\left(\frac{1+(-1)^{x_3}\delta}{2}\right)(c^\dagger_\alpha(\vec{x})c_\alpha(\vec{x}+\hat{e}_3)+h.c) \ . \nonumber
\end{multline}
 Expanding $\Ham_{\text{QWZ}}$, we get,
\begin{multline}
   \Ham^{[3]}_{\text{QWZ}} = \sum_{\vec{x} \in \bZ^3} \sum_{\alpha, \beta \in \{\uparrow,\downarrow \}}  \sum_{a=1}^3 t^{a}_{\alpha \beta}(\vec{x}) (c^\dagger_\alpha(\vec{x})c_\beta(\vec{x}+\hat{e}_a)+h.c)\\+\sum_{\vec{x} \in \bZ^3} \sum_{\alpha, \beta \in \{\uparrow,\downarrow \}} \mu_{\alpha \beta}(\vec{x}) c^\dagger_\alpha(\vec{x})c_\beta(\vec{x})\ , \label{eq:H_QWZ_3d}
\end{multline}
where 
\begin{align}
    t^{1,2}_{\alpha \beta} &=  (-1)^{1+x_3} \left(\frac{\sigma^3_{\alpha \beta} + i\sigma^{1,2}_{\alpha \beta}}{2}\right),\nonumber\\t^{3}_{\alpha \beta} &= \delta_{\alpha \beta}\left(\frac{1+(-1)^{x_3}\delta}{2} \right),\nonumber\\ 
    \mu_{\alpha \beta} &=     \mu \delta_{\alpha \beta} + (-1)^{x_3} m ~\sigma^3_{\alpha \beta}\ .
\end{align}
{A closely related formulation of the Hall-conductivity pump and its higher-dimensional generalizations was also studied by Spodyneiko~\cite{SpodyneikoHallPump}}. This has a two-site translation symmetry in the $x_3$ direction and a single site translation symmetry in $x_1,x_2$, as well as $\uone$ particle conservation symmetry. Choosing a two-site fiducial unit cell, \cref{eq:H_QWZ_3d} can be written in momentum space as a four-band model (including the spin indices) with the following Bloch Hamiltonian
\begin{align}
    \h(\vec{k})=\mu \mathbb{1}+\sum_{m=1}^5 g_m(\vec{k})\Gamma^m\ , \label{eq:h_QWZ_3d}
\end{align}
where, corresponding to a particular choice of unit cell, we have,
\begin{multline}
    g_{1,2}= \sin k_{1,2}, \  g_3=\left(\frac{1-\delta}{2}\right)+\left(\frac{1+\delta}{2}\right)\cos{k_3}  \\ g_4=\left(\frac{1+\delta}{2}\right)\sin k_3, \ g_5=m+\cos k_1+\cos k_2 \ .
\end{multline}
The $\{\Gamma_m\}$ are elements of the Clifford algebra $\mathrm{Cl}(5)$ with matrix representation,
\begin{align}
    \Gamma_1&=\sigma_3\otimes\sigma_1,\ \Gamma_2=\sigma_3\otimes \sigma_2, \nonumber\\    \Gamma_3&=\sigma_1\otimes \mathbb{1}, ~~\Gamma_4=\sigma_2\otimes \mathbb{1},  \ \Gamma_5=\sigma_3\otimes \sigma_3\ .
\end{align}
\begin{figure}
    \centering
    \makebox[\linewidth][c]{\includegraphics[width=\textwidth,valign=c]{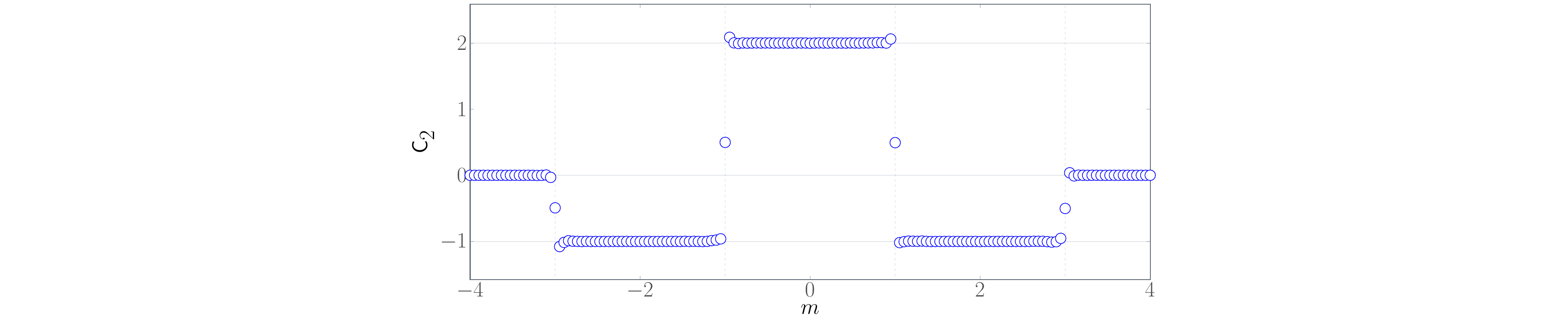}}
    \caption{Chern number for the filled bands of the 3d ascendant of the QWZ model,~\cref{eq:H_QWZ_3d,eq:h_QWZ_3d} evaluated by computing \cref{eq:Chern_3dQWZ} numerically over $\cM_4 = \snpar{1} \times \tnbz{3}$. $\snpar{1}$ is a parametric circle of unit radius centered at $\{m,\delta=0\}$. As we vary $m$, $\snpar{1}$ wraps various diabolical points leading to different values of $\Ch_2 = 0, -1,2$. When $m=\pm1,\pm3$, $\snpar{1}$ touches the gap closing point and represents transitions between the Chern numbers. As expected from Ref.\cite{Verresen20}, we numerically obtain half-integer values for $\Ch_2$ equal to the mean of the straddling values of $\Ch_2$.}
    \label{fig:Chern_QWZ_MC}
\end{figure}
\begin{figure}
    \centering
   \subfloat[$\mu=0$]{\makebox[\linewidth][c]{\includegraphics[width=\textwidth,valign=c]{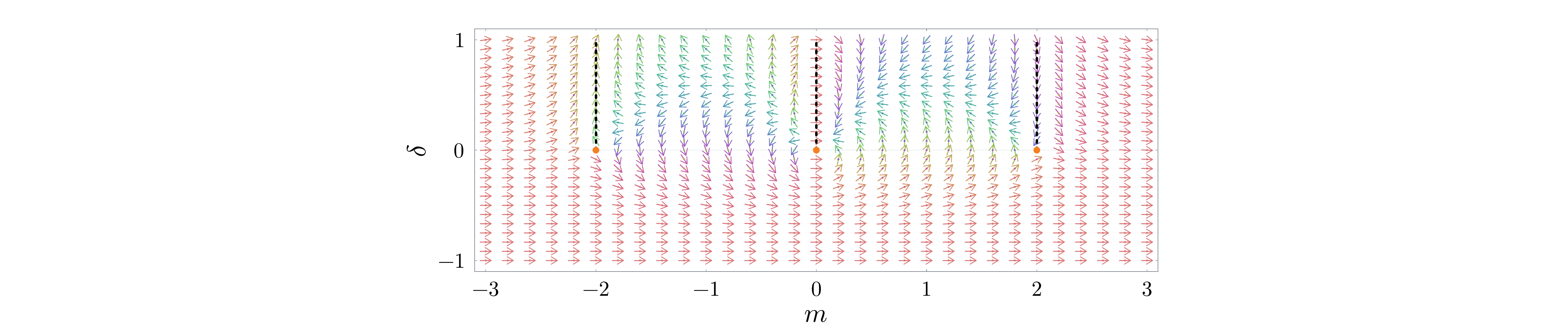}}} \\
    \subfloat[$\mu \neq 0$]{\makebox[\linewidth][c]{\includegraphics[width=\textwidth,valign=c]{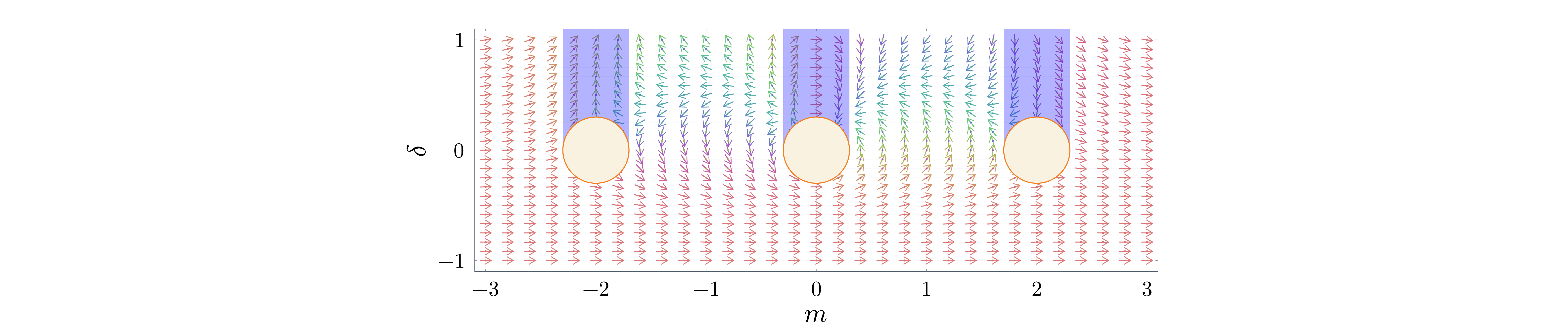}}}
    \caption{Textured phase diagrams of the 3d QWZ ascendant in \cref{eq:H_QWZ_3d} for $\mu=0$ (a) and   small but finite $\mu \neq 0$ (b). We see three vortices of strength $2,-1$ located at $m=0,\pm2,~\delta=0$ respectively. For $\mu=0$ (a), the core of the vortex corresponds to a gapless diabolical point described by massless Dirac fermions of flavours $N_f=2$  for $m=0$ and $N_f =1$ for $m=\pm 2$. These expand to stable gapless phases for small $\mu \neq 0$. For $\mu=0$, edge modes form codimension 1 surfaces terminating on the diabolical points, represented by broken lines in (a). For $\mu \neq 0$, these expand to stable boundary gapless metallic phases, represented by shaded regions crowning the metallic circles in (b). }
    \label{fig:ChernPump}
\end{figure}
As shown in \cref{fig:QWZ_3d_bands}, the system has two bands, each two-fold degenerate, that are separated for all values of parameters except $\delta = 0$, $m=0,\pm 2$, when the bands touch at two ($m=0$) and one ($m=\pm 2$) point(s) in the three-dimensional Brillouin zone. The topological nature of the phase diagram can be studied by considering the non-abelian Berry-Bloch connection and evaluating the second Chern number $\Ch_2$ over $\cM_4 = \snpar{1} \times \tnbz{3}$
\begin{equation}
    \Ch_2 = \frac{1}{2} \int_{\snpar{1} \times \tnbz{3}} \tr\left(\frac{F}{2\pi} \wedge \frac{F}{2\pi}\right), \label{eq:Chern_3dQWZ}
\end{equation}
  where $\snpar{1}$ is a parametric two-sphere living within the gapped states and surrounding any of the three gapless diabolical points $m= 0,\pm2,~\delta=0$. The numerical calculation of $\Ch_2$ is shown in \cref{fig:Chern_QWZ_MC} for $\mu=0$ at various unit circles centered at $\{\delta = 0, m\}$. We see that $\Ch_2 = -1$ for circles that wrap $m=\pm2,\delta=0$ and $\Ch_2=2$ for circles that wrap $m=\delta=0$. When the wrapping circle touches the gap closing point, for $m = \pm 1,~\pm 3$ there is a change in the Chern number. Numerically, we get half integer values for $\Ch_2$ at these points, as expected from Ref.~\cite{Verresen20}, corresponding to the mean of the straddling values. 

This topological nature also results in a texture in the $m-\delta$ phase diagram that can be visualized via the higher Berry phase $\check{\gamma}$ computed for each point $m,\delta$ by integrating the Chern-Simons 3-form over the 3d Brillouin zone,
\begin{equation}
    \check{\gamma}(m,\delta) = \frac{1}{4\pi} \int_{\tnbz{3}} \tr\left(A \wedge dA +\frac{2}{3} A\wedge A \wedge A \right). \label{eq:CS_QWZ_3d}
\end{equation}
Plotting $\check{\gamma}$ results in the texture shown in \cref{fig:ChernPump}. We see the presence of vortices of various strengths whose cores are located at $m = 0, \pm 2,~\delta = 0$. For $\mu=0$, the vortex core is represented by $N_f=2 ~(m=0)$ and $N_f=1~(m=\pm2)$ relativistic massless Dirac fermions, whereas for small but finite $\mu \neq 0$, the vortex cores become gapless metallic phases, as the Dirac bands are partially occupied. 

We can study the edge modes by considering the system with open boundaries. We retain periodic boundary conditions in $x_1,x_2$ directions and open boundaries in the $x_3$ direction. For concreteness, take $x_3 \in [1,L]$ with $L$ being an even number. We begin in the limit of $\delta = 1$ and $\mu=0$. The 2d boundaries on $x_3 = 1,L$ are governed by the QWZ model \eqref{eq:H_QWZ}. As we tune $m$, this changes the boundary Chern insulating phases and the boundary is gapless for $m = 0, \pm 1$. The gapless boundary states are expected to survive for the range $\delta \in (0,1]$ and terminate at the bulk gapless points as shown in \cref{fig:ChernPump}(a). When we add a small chemical potential $0<|\mu|$,  edge modes become stable and extend over a range of parameter values with no additional fine-tuning as shown in  \cref{fig:ChernPump}(a). This is just like the 2d Rice mele ascendant \cref{fig:2dRiceMele_mu}, as the boundary bands are partially populated. However, the difference is that the edge modes in  \cref{fig:2dRiceMele_mu} are relativistic massless Dirac fermions whereas in  \cref{fig:ChernPump}(b), they are nonrelativistic metals with a finite Fermi surface. As $\delta$ is lowered, we expect that these edge modes terminate on the metallic phases and merge with bulk gapless modes as shown in \cref{fig:ChernPump}(a,b). Just as in the case of the 1d Berry ascendant, as we increase $|\mu|$ to large values, different bulk and boundary bands can get partially populated, and the bulk and boundary metallic phases can merge and restructure in complex and interesting ways. We will not explore these further in this work. 

\subsection{Field theory and proximate phase diagrams}
\begin{figure}[!h]
    \centering
    \makebox[\linewidth][c]{\includegraphics[width=\textwidth]{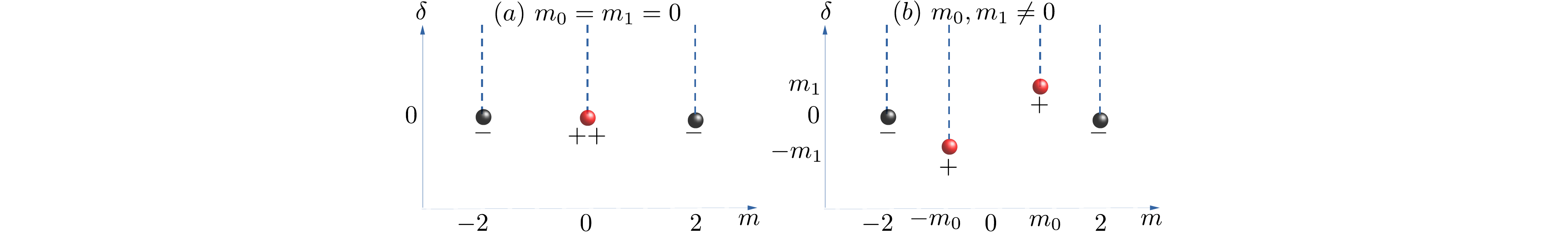}}
    \caption{The addition of singlet masses $m_0,~m_1$ shown in \cref{eq:QWZ_singlet_lagrangian,eq:QWZ_singlet_h} splits the charge $2$ vortex in the phase diagram at the origin into two charge $1$ vortices at $(m,\delta) = \pm (m_0,m_1)$}
    \label{fig:QWZ_singlet}
\end{figure}
We now study the 3d QWZ ascendant using effective field theory. We will consider a low-energy continuum description in the vicinity of the diabolical points. Let us begin with $m \approx \pm 2,\delta \approx 0$ where the continuum  Lagrangian density in Euclidean space can be written as
\begin{align}
    \cL = -\bar{\psi}\slashed{\partial} \psi - \delta \bar{\psi} \psi + i (m \pm 2) \bar{\psi} \gamma_c \psi\ . \label{eq:QWZ_mpm2}
\end{align}
Here, $\psi$ is a four component spinor field and \cref{eq:QWZ_mpm2} describes a relativistic Dirac fermion in 3+1 d. The couplings $\delta$ and $m$ correspond to the ordinary and chiral mass terms, which saturate the space of relevant operators. Thus, in the vicinity of $m \approx \pm 2,\delta \approx 0$, the phase diagram is qualitatively unchanged for arbitrary weak perturbations, including interactions. 

{At each of the two $N_f=1$ critical points $(m,\delta)=(\pm2,0)$, the continuum theory has an emergent axial $\mathrm{U}(1)$ symmetry that rotates the two mass perturbations---$\delta$ and the deviation of $m$ from its critical value---into one another~\cite{manjunath2026searchdiabolicalcriticalpoints}.}

Near $m\approx 0,~\delta \approx 0,$ the continuum Lagrangian density can be written as 
\begin{align}
    \cL = - \sum_{a=1}^2\bar{\psi}_a\slashed{\partial} \psi_a - \sum_{a,b=1}^2 \tau^3_{ab} \left(\delta \bar{\psi}_a \psi_b - i m \bar{\psi}_a \gamma_c \psi_b\right).\label{eq:QWZ_mzero}
\end{align}
\cref{eq:QWZ_mzero} is an $N_f = 2$ component Dirac fermion where $\delta$ and $m$ couple to one of the components of the ordinary and chiral triplet masses. {At $m=\delta=0$, the kinetic term has a flavor $\mathrm{SU}(2)$ symmetry that mixes these two flavors. It also has an axial $\mathrm{U}(1)$, generated by $\gamma_c/2$, that rotates the $\delta$ and $m$ mass terms into one another. This is the symmetry enhancement discussed in Ref.~\cite{manjunath2026searchdiabolicalcriticalpoints}.} The space of independent relevant perturbations for $N_f =2$ Dirac fermions is three dimensional and the diabolical point is not stable. We can add, for instance, the singlet ordinary and chiral mass
\begin{align}
    \delta \cL = \sum_{a=1}^2 \left(m_1 \bar{\psi}_a \psi_a -i m_0 \bar{\psi}_a \gamma_c \psi_a \right). \label{eq:QWZ_singlet_lagrangian}
\end{align}
For the microscopic Bloch Hamiltonian in \cref{eq:h_QWZ_3d}, this perturbation takes the form
\begin{align}
    \delta \h = \frac{1}{2}(\cos k_1 - \cos k_2) (m_1 \Gamma_3 - m_0 \Gamma_5).\label{eq:QWZ_singlet_h}
\end{align}
It is easy to check that this results in the $N_f=2$ gapless point at $m=\delta=0$ splitting into two $N_f =1$ points at $(m,\delta) = \pm (m_0,m_1)$. For the overall microscopic phase diagram, this splits the core of the charge 2 vortex into two charge 1 vortices as shown in \cref{fig:QWZ_singlet}. We can also consider adding in other triplet mass components. These merely move the location of various diabolical points and add smooth changes to the texture.

\subsection{QWZ ascendants in arbitrary dimensions}
The higher dimensional ascendants $\Ham_{\text{QWZ}}^{[d]}$ for $d \ge 3$ are constructed  from the lower-dimensional model $\Ham_{\text{QWZ}}^{[d-1]}$ using the suspension recipe,
\begin{multline}
    \Ham^{[d]}_\text{QZW}=\mu \sum_{\vec{x}\in \mathbb{Z}^d}n(\vec{x})+\sum_{x_d\in \mathbb{Z}}(-1)^{x_d}\Ham^{[d-1]}_{\text{QWZ},x_d}|_{\mu=0}\\
    +\sum_{x_d\in \mathbb{Z}}\left(\frac{1+(-1)^{x_d}\delta_{d-2}}{2}\right)(c^\dagger(\vec{x})c(\vec{x}+\hat{e}_d)+h.c)\ .
\end{multline}
An explicit form can be given by iteratively repeating the above, and expanding out the Hamiltonian to get
\begin{multline}
   \Ham^{[d]}_{\text{QWZ}} = \sum_{\vec{x} \in \bZ^d} \sum_{\alpha, \beta \in \{\uparrow,\downarrow \}}  \sum_{a=1}^d t^{a}_{\alpha \beta}(\vec{x}) (c^\dagger_\alpha(\vec{x})c_\beta(\vec{x}+\hat{e}_a)+h.c)\\+\sum_{\vec{x} \in \bZ^d} \sum_{\alpha, \beta \in \{\uparrow,\downarrow \}} \mu_{\alpha \beta}(\vec{x}) c^\dagger_\alpha(\vec{x})c_\beta(\vec{x})\ , \label{eq:H_QWZ_d}
\end{multline}
where 
\begin{align}
    &t^{1,2}_{\alpha \beta} =  (-1)^{1+\sum_{k=3}^d x_k} \left(\frac{\sigma^3_{\alpha \beta} + i\sigma^{1,2}_{\alpha \beta}}{2}\right),\nonumber\\
   &t^{a}_{\alpha \beta} = \delta_{\alpha \beta}(-1)^{\sum_{k=a+3}^d x_k} \left(\frac{1+(-1)^{x_a}\delta_a}{2}\right), a = 3,\ldots,d-1,\nonumber\\
    &t^{d}_{\alpha \beta} = \delta_{\alpha \beta} \left(\frac{1+(-1)^{x_d}\delta_d}{2}\right),~\nonumber\\
    &\mu_{\alpha \beta} =     \mu\delta_{\alpha \beta} + (-1)^{\sum_{k=3}^d x_k} m ~\sigma^3_{\alpha \beta}\ .
\end{align}
Choosing a $2^{d-2}$ site unit cell, we get a $2^{d-1}$ band model with the Bloch Hamiltonian as follows,
\begin{align}
    \h(\vec{k})=\mu \mathbb{1}+\sum_{m=1}^{2d-1}g_m\Gamma^m\ ,
\end{align}
where for a particular choice of the unit cell, we have 
\begin{align}
    g_1&=\sin k_1,\ g_2=\sin k_2, \nonumber\\
    g_{2m-3}&= \left(\frac{1-\delta_{m-2}}{2}\right)+\left(\frac{1+\delta_{m-2}}{2}\right) \sin k_{2m-1}, \nonumber\\
    g_{2m-2}&=\left(\frac{1+\delta_{m-2}}{2}\right) \cos k_{2m-1}, \ m=3,\ldots,d \nonumber\\ 
    g_{2d-1}&=(-1)^{d+1}(m+\cos k_1 +\cos k_2)\ .
\end{align}
The $\{\Gamma^m\}$ represent the $2^{d-1}$ dimensional irreducible representation of the Clifford algebra $\mathrm{Cl}(2d-1)$ satisfying \cref{eq:Clifford} whose matrix representation for the same choice of the unit cell is given by
\begin{align}
        &\Gamma^{2m-3} = \underbrace{\sigma^3 \otimes \sigma^3 \cdots \otimes \sigma^3}_{\text{$d-m$ terms}} \otimes \ \sigma^1 \otimes \underbrace{\mathbb{1} \otimes \mathbb{1} \otimes \cdots \otimes \mathbb{1}}_{\text{$m-2$ terms}} \nonumber,\\
        &\Gamma^{2m-2} = \underbrace{\sigma^3 \otimes \sigma^3 \cdots \otimes \sigma^3}_{\text{$d-m$ terms}} \otimes \ \sigma^2 \otimes \underbrace{\mathbb{1} \otimes \mathbb{1} \otimes \cdots \otimes \mathbb{1}}_{\text{$m-2$ terms}}. \nonumber \\  
        &\Gamma^{2d-1} = \underbrace{\sigma^3 \otimes \sigma^3 \cdots \otimes \sigma^3}_{\text{$d-1$ terms}}\ m=2,\ldots,d\ .
\end{align}
The $d$-dimensional ascendant gives us a non-trivial topological family over a parametric $\snpar{d-2}$ which can be detected by computing the Chern number $\Ch_{d-1}$ evaluated over $\cM_{2d-2}=\snpar{d-2}\times \tnbz{d}$. A non-trivial texture in the phase diagram can be extracted by evaluating the higher Berry phase $\check{\gamma}$ defined over a parametric $\snpar{d-3}$ as an integral over the Chern-Simons form $\CS_{2d-3}$ over $\cM_{2d-3}=\snpar{d-3}\times \tnbz{d}$.

\section{Conclusion and outlook}
In this work, we have comprehensively analyzed the topological structures of charge conserving systems in Class A using free-fermion representatives. Using topological and geometric invariants, we show that phase diagrams can have topological textures and associated gap closing loci in the bulk and boundary. We have studied three series of textures using microscopic models in arbitrary dimensions constructed using suspension, and analyzed them using band theory and effective field theory. We also make connections between interacting invariants and our free-fermion invariants. Altogether, we show that trivial insulators are not featureless, but have interesting topological structures hidden in phase diagrams. 

There are various potential future directions following this work. First, it would be nice to extend our analysis to other symmetry classes of insulators and superconductors and study how topological textures can arise in their phase diagrams. It would also be useful to consider disordered systems and understand how the topological textures can be extracted without access to band theory~\cite{KITAEV20062,RealSpaceSecondChern}, possibly through appropriate local observables. Related to this, it would be interesting to see how topological and geometric invariants that result in phase diagram textures can be measured experimentally. For the 1d Rice-Mele model, the answer to this question follows from the well-known connection between Berry's phase and charge polarization~\cite{KANE_topbandtheory}. For higher dimensions, Berry phases and Chern numbers are expected to be related to nonlinear response coefficients~\cite{KapustinSpodyneiko2020higherdimensionalgeneralizationsthoulesscharge,KapustinSpodyneikoo_HigherBerry_PhysRevB.101.235130} which would be interesting to study in detail. The main focus of our work has been on invertible phases, particularly the trivial phase, where the conjectured generalized cohomology framework allows us to determine the structure of phase diagrams across dimensions from information about one particular dimension. This advantage is lost for non-invertible phases of matter such as symmetry breaking, gapless, critical and topologically ordered phases. It would be interesting to repeat our study  how non-trivial families in such phases~\cite{Sharon_GlobalAspectsVacua_2020,Ohyama2026parameterizedfamiliestoriccode} leaves an imprint on phase diagrams. It would also be illuminating to know if textured phase diagrams can occur for classical systems~\cite{APJones_ClassicalDQC_PhysRevLett.134.097103,AP_Titas_J1J2}. We leave these and other questions for future work.

 \section*{Acknowledgments} 
 The authors thank Dileep Jatkar, Ashoke Sen,  R. Loganayagam, Bhandaru Phani Parasar, Sayantan Mandal, Abhijit Gadde, Shiraz Minwalla, and Onkar Parrikar for helpful discussions. 

 A.P. acknowledges support from the Anusandhan National Research Foundation (ANRF), Government of India, under the Prime Minister's Early Career Research Grant (Grant No.~ANRF/ECRG/2025/000520/PMS).

 The authors acknowledge the use of A.I. in this work (GPT 5.4 via ChatGPT and Codex) to clarify certain conceptual points, search for relevant literature, and develop scripts used for numerical analysis. All code outputs were manually verified for physical accuracy. A copy of the scripts used to produce the numerical plots, as well as the data files, can be found at \href{https://doi.org/10.5281/zenodo.20288955}{https://doi.org/10.5281/zenodo.20288955}.

\bibliography{references}{}

\end{document}